\documentclass[twocolumn,usenatbib]{aastex62}

\usepackage{newtxtext,newtxmath}
\usepackage[T1]{fontenc}
\usepackage{ae,aecompl}
\usepackage{url}
\usepackage{graphicx}	
\usepackage{amsmath}	
\usepackage{amssymb}	
\usepackage{appendix}

\defcitealias{martin_2013}{M13}
\defcitealias{ibata_2014b}{I14}

\hypersetup{
  colorlinks=true,
  urlcolor=blue,
  citecolor= blue,
  linkcolor = blue
}

\graphicspath{{./}{figures/}}

\shorttitle{$Andromeda\; IX$}
\shortauthors{$Abdollahi\; et \;al.$}
\usepackage[utf8]{inputenc}
\DeclareUnicodeCharacter{2212}{-}

\begin{document}

\title {THE ISAAC NEWTON TELESCOPE MONITORING SURVEY OF LOCAL GROUP DWARF GALAXIES. VI. THE STAR FORMATION HISTORY AND DUST PRODUCTION IN ANDROMEDA IX}

\author{Hedieh Abdollahi}
\affiliation{Physics Department, Faculty of Physics and Chemistry, Alzahra University, Vanak, 1993891176, Tehran, Iran}

\author{Atefeh Javadi}
\affiliation{School of Astronomy, Institute for Research in Fundamental Sciences (IPM), PO Box $19395-5531$, Tehran, Iran}

\author{Mohammad Taghi Mirtorabi}
\affiliation{Physics Department, Faculty of Physics and Chemistry, Alzahra University, Vanak, 1993891176, Tehran, Iran}
\affiliation{ICRANet, Piazza della Repubblica 10, I-65122 Pescara, Italy}

\author{Elham Saremi}
\affiliation{School of Astronomy, Institute for Research in Fundamental Sciences (IPM), PO Box $19395-5531$, Tehran, Iran}
\affiliation{Instituto de Astrofsica de Canarias, VA Lctea s/n, 38205 La Laguna, Tenerife, Spain}
\affiliation{Departamento de Astrofsica, Universidad de La Laguna, 38205 La Laguna, Tenerife, Spain}

\author{Jacco Th. van Loon}
\affiliation{Lennard-Jones Laboratories, Keele University, ST5 5BG, UK}

\author{Habib G. Khosroshahi}
\affiliation{School of Astronomy, Institute for Research in Fundamental Sciences (IPM), PO Box $19395-5531$, Tehran, Iran}
\affiliation{Iranian National Observatory, Institute for Research in Fundamental Sciences (IPM), Tehran, Iran}

\author{Iain McDonald}
\affiliation{Jodrell Bank Centre for Astrophysics, Alan Turing Building, University of Manchester, M13 9PL, UK}
\affiliation{Department of Physical Sciences, The Open University,
Walton Hall, Milton Keynes, MK7 6AA, UK}

\author{Elahe Khalouei}
\affiliation{School of Astronomy, Institute for Research in Fundamental Sciences (IPM), PO Box $19395-5531$, Tehran, Iran}

\author{Hamidreza Mahani}
\affiliation{School of Astronomy, Institute for Research in Fundamental Sciences (IPM), PO Box $19395-5531$, Tehran, Iran}

\author{Sima Taefi Aghdam}
\affiliation{School of Astronomy, Institute for Research in Fundamental Sciences (IPM), PO Box $19395-5531$, Tehran, Iran}

\author{Maryam Saberi}
\affiliation{Rosseland Centre for Solar Physics, University of Oslo, P.O. Box 1029 Blindern, NO-0315 Oslo, Norway}
\affiliation{Institute of Theoretical Astrophysics, University of Oslo, P.O. Box 1029 Blindern, NO-0315 Oslo, Norway}

\author{Maryam Torki}
\affiliation{School of Astronomy, Institute for Research in Fundamental Sciences (IPM), PO Box $19395-5531$, Tehran, Iran}

\email{atefeh@ipm.ir}

\begin{abstract}

We present a photometric study of the resolved stellar populations in And\,IX, the closest satellite to the M31, a metal-poor and low-mass dwarf spheroidal galaxy. We estimate a distance modulus of $24.56_{-0.15}^{+0.05}$ mag based on the tip of the red giant branch (TRGB). By probing the variability of asymptotic giant branch stars (AGB), we study the star formation history of And\,IX. We identified 50 long period variables (LPVs) in And\,IX using the Isaac Newton Telescope (INT) in two filters, Sloan $i'$ and Harris $V$. In this study, we selected LPVs within two half-light radii with amplitudes in the range of $0.2-2.20$ mag. It is found that the peak of star formation reaches $\sim$ $8.2\pm3.1\times10^{-4}$ M$\textsubscript{\(\odot\)}$ yr$^{-1}$ at $\approx 6$ Gyr ago. Our findings suggest an outside-in galaxy formation scenario for And\,IX with a quenching occurring $3.65_{-1.52}^{+0.13}$ Gyr ago with the SFR in the order of $2.0\times10^{-4}$ M$\textsubscript{\(\odot\)}$ yr$^{-1}$ at redshift < $0.5$. We calculate the total stellar mass by integrating the star formation rate (SFR) within two half-light radii $\sim$ $3.0\times10^5$ M$\textsubscript{\(\odot\)}$. By employing the spectral energy distribution (SED) fitting for observed LPVs in And\,IX, we evaluate the mass-loss rate in the range of $10^{-7}$ $\leq$ $\dot{M}$ $\leq$ $10^{-5}$ M$\textsubscript{\(\odot\)}$ yr$^{-1}$. Finally, we show that the total mass deposition to the interstellar medium (ISM) is $\sim$ $2.4\times10^{-4}$ M$\textsubscript{\(\odot\)}$ yr$^{-1}$ from the C- and O-rich type of dust-enshrouded LPVs. The ratio of the total mass returned to the ISM by LPVs to the total stellar mass is $\sim 8.0\times10^{-10}$ yr$^{-1}$, and so at this rate, it would take $\sim$ 1 Gyr to reproduce this galaxy.

\end{abstract}


\section{Introduction}

Dwarf galaxies are the most abundant type of galaxies in the Universe. They contain valuable information about the early Universe and its evolution \citep{2010AdAst2010E...3C}. Dwarf galaxies also continue to play an important role in our understanding of galaxy formation and stellar evolution \citep{2003Ap&SS.284..579T}. Dwarf galaxies are divided into two main categories based on their gas content: gas-rich dwarfs: dwarf irregulars (dIrrs), blue compact dwarfs (BCDs), and gas-poor dwarfs: dwarf ellipticals (dEs), dwarf spheroidals (dSphs), and ultra-faint dwarfs (UFDs) \citep{2021A&A...645A..92M}. The Local Group hosts various kinds of dwarf galaxies, and it is possible to study their resolved populations with space- or ground-based telescopes \citep{2014ApJ...789..147W}.

$\Lambda$CDM, which is known as the standard model of cosmology, is compatible with most observations. In the $\Lambda$CDM, cold dark matter is assumed to be the dominant matter content of the Universe \citep{2018dg}.
In the hierarchical $\Lambda$CDM paradigm, galaxies evolve in dark matter halos formed in the early Universe by the gravitational collapse of overdense regions. Moreover, the halos grow hierarchically through the accretion of subhalos. The galaxies are surrounded by these subhalos which are called satellite galaxies \citep{2020SFH}. 

One of the main challenges for the cold dark matter scenario is the missing satellite problem (MSP) \citep{1999missing, 2017CDM}. There is a discrepancy between the number of satellites predicted in the CDM simulation and the observed satellites. There is also another critical challenge called Too Big To Fail. The N-body simulation, based on the $\Lambda$CDM, predicts that the satellites are too massive and dense compared to those observed \citep{2011toobigtoofail, 2012MW, 2017CDM}. Thus, observing and studying dwarf galaxies enables us to understand the Universe and solve the aforementioned problems. 

Several mechanisms (internal and external) could influence the evolution of dwarf galaxies. Compared to massive galaxies, dwarf galaxies have a smaller stellar population and a simple star formation history (SFH). There are internal processes, such as stellar feedback and depletion gas, and also environmental processes, such as tides and ram pressure stripping, that influence star formation in dwarf galaxies
\citep{2014ApJ...793...29G, 2015ApJ...808L..27W, 2016ApJ...833...84X, 2018quenching, 2019MNRAS.490.4447W, fillingham2019characterizing, 2021ApJ...906...96A}.

An optical monitoring survey was initiated using the Isaac Newton Telescope (INT) to study the SFH of dwarf galaxies in the Local Group by probing the asymptotic giant branch (AGB) stars \citep{2020ApJ...894..135S}. Mass-loss is one of the notable features of AGB stars with large amplitude pulsations. These pulsations lead to long period photometric variability that can be  observed in durable photometric campaigns \citep{javadia}. The INT survey was launched to estimate the dust ejected into the interstellar medium (ISM) and to determine the mass-loss rate in $55$ nearby galaxies, spanning an order of magnitude in metallicity. The long period variables (LPVs) can be used to trace the stellar evolution and history of the galaxy. This survey allows us to compare the SFH in different galaxy types and to study the evolution and quenching times of dwarf galaxies. We used a method introduced by \cite{javadib, 2017MNRAS.464.2103J} to build the SFH of the galaxy M33 and to estimate the dust ejection into the ISM.

AGBs and red supergiants (RSGs) are powerful tools to study SFH, as they have been presented in most of the history of the Universe ($\sim$ $10$ Myr to $10$ Gyr) and are in their final evolution stage, where their luminosity relates directly to their birth mass \citep{javadib}. Various surveys in the IR-wavelength have widely studied AGBs and RSGs \citep{javadia, 2015MNRAS.447.3973J, 2013POBeo..92..117B, boyer2009spitzer, 2015ApJS..216...10B, 2015ApJ...800...51B}. Dust-producing LPVs are more easily detected in the optical bands because their amplitude is larger compared to the IR-wavelength, though there has not yet been a comprehensive survey of long period variable stars in Local Group dwarf galaxies. In addition, the inclusion of IR bands in the spectral energy distributions (SED) makes the calculation of dust density more accurate, since optical color alone is not a sufficient indicator of dust density. The INT monitoring survey, thanks to a comprehensive sampling (to date), could address open questions about the evolution of dwarfs and shutting-down the star formation by various effects such as environmental processes and tidal effects.

AGB stars can reach a luminosity of $10^4$ L$\textsubscript{\(\odot\)}$ \citep{2005A&A...438..273V} at their most luminous stage. These luminous populations are easily distinguished from the background, especially in the infrared, where the difference in brightness is very pronounced. These types of stars are cool with temperatures of $\lesssim$ 4000 K \citep{2005A&A...438..273V}, and have a birth mass of 0.8 M$\textsubscript{\(\odot\)}$ $\leq M \leq$ 8 M$\textsubscript{\(\odot\)}$ \citep{2018A&ARv..26....1H}. AGB stars consist of an electron-degenerate (C-O) core and two shells around it. Burning helium and hydrogen in the shells generates energy for evolution in AGBs. Helium shell burning releases considerable energy in a flash-like process, resulting in a long series of thermal pulses \citep{2014ApJ...790...22R}. Periodic expansions and contractions of the outer layers lead to radial pulses, usually on the order of $10^2$ to $10^3$ days \citep{2018A&ARv..26....1H}. Most stars (especially AGBs) experience mass-loss at the end of their evolution. Dust can be produced mainly in two environments: during the thermal pulsation phase in the cool and dense atmosphere of AGBs with low to intermediate stellar mass (0.8-8 M$\textsubscript{\(\odot\)}$), and during the core-collapse phase of stars with enough heavy mass ($M$ > 8 M$\textsubscript{\(\odot\)}$) to end their lives with supernova explosions \citep{2009MNRAS.397.1661V}.
AGB stars deposit some of their mass into the ISM through radial mechanical pulsation and thus play a crucial role in enriching the ISM. \cite{1999A&A...351..559V} estimated the mass-loss rate in the range of $10^{-7}$ < $\dot{M}$ $\leq$ $10^{-3}$ M$\textsubscript{\(\odot\)}$ yr$^{-1}$ based on a sample of AGBs and RSGs in the Large Magellanic Cloud (LMC). 

In this paper, we focus on a spheroidal dwarf satellite along the major axis of M31 galaxy, And\,IX, which is closest to the host, and one of the least luminous satellites. And\,IX was discovered using the resolved stellar photometry of the Sloan Digital Sky Survey (SDSS) by \cite{2004ApJ...612L.121Z} and categorized as an old dwarf galaxy with little baryonic matter. The term of old was due to the fact that no significant population of intermediate-age carbon and main sequence stars were observed with the WIYN 3.5-m telescope by \cite{2005ApJ...623..159H}. Table $1$ shows in detail the apparent characteristics of And\,IX, which was selected for this study for the following reasons:

\begin{itemize}
\item Studying one of the Andromeda satellites allows us to investigate whether the star-forming pattern and quenching time are consistent with those of the Milky Way. It also helps to understand whether this dwarf galaxy follows other M31 satellites in its formation scenario, as studied by \cite{2019ApJ...885L...8W}.
\item And\,IX is an important case study because of its proximity to the host $\sim$ $39_{-2}^{+5}$ kpc \citep{2019MNRAS.489..763W}, in which quenching time and galaxy evolution may have been affected by environmental effects such as the strong tidal effect of M31.
\item With a mass-to-light ratio of $1$ M$\textsubscript{\(\odot\)}$/ L$\textsubscript{\(\odot\)}$, \cite{2012AJ....144....4M} estimated the dynamical mass ($6.5\times10^6$ M$\textsubscript{\(\odot\)}$) and stellar mass ($0.15\times10^6$ M$\textsubscript{\(\odot\)}$), which could be a sign of the existence of large amounts of dark matter with a very low surface brightness of $\Sigma_V$ = $28.0\pm1.2$ mag\;arcsec$^{-2}$.

\end{itemize}
\begin{table}
\begin{center}
\caption{Properties of And\,IX dwarf galaxy}
\begin{tabular}{c c c}
\tableline
\tableline
RA.(J$2000$)& $00\, 52 \,53$ &\cite{2004ApJ...612L.121Z}\\[1ex]
DEC.(J$2000$)  & $+43\, 11\, 45$ &\cite{2004ApJ...612L.121Z}\\[1ex]
M$_{V}$ (mag)  &  $-9.0\pm0.3$&\cite{2019MNRAS.489..763W}\\[1ex]
r$_{h}$ (pc) & $444_{-53}^{+68}$&\cite{2019MNRAS.489..763W}\\[1ex]
[Fe/H] (dex) & $-2.03\pm0.01$&\cite{2020ApJ...895...78W}\\[1ex]
$\sigma$ (km s$^{-1}$) \footnote{Observed velocity dispersion}&$4.5_{-3.4}^{+3.6}$& \cite{2010MNRAS.407.2411C} \\ [1ex] 
$\Sigma_V$ (mag\;arcsec$^{-2}$)\footnote{Surface brightness} & $28.0\pm1.2$ &\cite{2012AJ....144....4M} \\ [1ex]
V (mag) \footnote{Apparent magnitude in Vega magnitude system} & $16.3\pm1.1$ &\cite{2012ApJ...758...11C} \\ [1ex]
\tableline
\end{tabular}

\end{center}
\end{table} 

Different values have been reported for the distance modulus using two methods, the horizontal branch (HB) and the tip of the red giant branch (TRGB). In Table 2, we summarize the distance modulus calculated in other works. In this paper, we calculate a distance modulus of $(m-M) = 24.56_{-0.15}^{+0.05}$ mag using the TRGB method (see Section 5).

\begin{table}
\begin{center}
\setlength{\tabcolsep}{0.5pt}
\caption{Distance modulus reported in the literature.}
\begin{tabular}{ccc}
\tableline
\tableline
Distance modulus (mag) & Method & Reference\\
\hline
$24.42\pm0.07$ &  TRGB & \cite{2005MNRAS.356..979M}\\[1ex]
$24.48\pm0.20$&  TRGB & \cite{2004ApJ...612L.121Z}\\[1ex]
$24.33\pm0.10$ &  TRGB & \cite{2005ApJ...623..159H}\\[1ex]
$24.42\pm0.39$&  TRGB & \cite{2010MNRAS.407.2411C}\\[1ex]
$24.4$ & HB &\cite{2010MNRAS.407.2411C}\\[1ex]
$24.46_{-0.15}^{+0.28}$& TRGB, HST \footnote{HST-based TRGB distance modulus} & \cite{2019MNRAS.489..763W} \\[1ex]
$24.43_{-0.03}^{+0.06}$ & HB, HST \footnote{HST-based HB distance modulus} & \cite{2019MNRAS.489..763W} \\[1ex]
$23.89_{-0.08}^{+0.31}$ & TRGB, Ground \footnote{Ground-based TRGB distance modulus } &  \cite{2019MNRAS.489..763W}\\[1ex]

\tableline
\end{tabular}
\end{center}
\end{table}

This paper is organized as follows. In Section $2$, we present the photometry results of And\,IX. Section $3$ deals with the study of the variable candidates and their amplitudes. The cross-correlation of the INT catalog with the {\it Spitzer}, WISE, and SDSS catalogs is discussed in Section $4$. A description of the physical parameters of And\,IX is presented in Section $5$. We focus on the estimation of the SFH based on the LPVs, specifically in And\,IX in Section $6$. Estimation of the mass-loss rate by modelling the SEDs based on the dust of the LPVs is discussed in Section $7$. Finally, a summary of this work is presented in Section $8$.

\section{Photometry}

\begin{figure*}[ht]
\centering
\includegraphics[width=18cm]{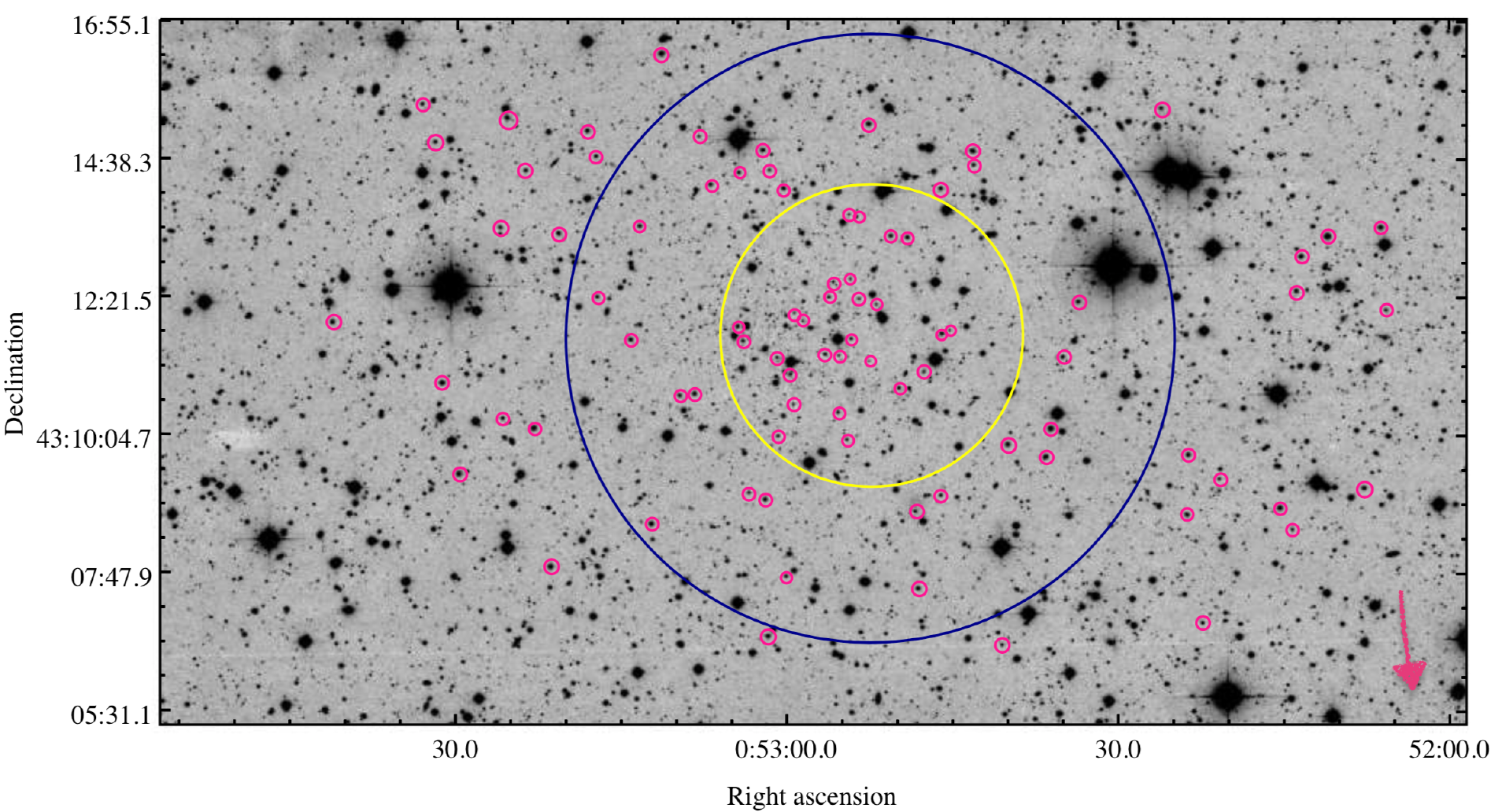}
\centering
\advance\leftskip-0.5cm
\caption{Montage of all frames of And\,IX dSph galaxy with variable candidates in pink circles. Half-light radius of $\sim$ $2.5$ arcmin (yellow circle) and two half-light radii (blue circle) of And\,IX are approximately noted. The pink arrow points toward the center of the M31 galaxy.\centering}
\end{figure*}

Nine observations were made from June $21$, $2015$, to October $6$, $2017$ (Table $3$), to determine the photometric variability of the stars. The observations were made with the 2.5-m wide field camera (WFC) at INT in the Sloan $i'$, Harris $V$, and RGO $I$ filters. The Sloan $i'$ was used to observe the minimal effects of dust attenuation among the visible wavelengths in addition to the most significant magnitude differences of LPVs compared to other populations. The Harris $V$ was also used to check the color, temperature, and radius. {\sc Theli} (Transforming HEavenly Light into Image) pipeline was used to process each night's observations and create a comprehensive mosaic image by removing noises and tool errors \citep{2020ApJ...894..135S}.

We performed photometric measurements of And\,IX using the {\sc daophot/allstar} package \citep{1987PASP...99..191S} on a charge-coupled device (CCD). 
Because the CCD4 covered more than two half-light radii ($\sim$ $5$ arcmin) of And\,IX, we focused our study on the CCD4 ($11.26\times22.55$ arcmin$^2$). {\sc daophot} distinguished stars from background noise and measured stellar brightness by aperture photometry. To obtain a more accurate magnitude and astrometry, a point spread function (PSF) was created by selecting a number of isolated and unsaturated stars in each image. The PSF-fitting photometry was performed with the {\sc allstar} by subtracting all stars in each image based on the constructed PSF model. The {\sc daomaster} combined the output of the {\sc allstar} from multiple individual images. Individual images combined to create a master mosaic image of the galaxy by {\sc montage2}. Simultaneous reduction of all images by PSF-fitting photometry was performed by {\sc allframe} \citep{stetson1994center}. The master image of CCD4 with a half-light radius (yellow circle) and two half-light radii (blue circle) of And\,IX are shown in Fig.\ 1. A total of 8653 stellar sources were detected in CCD4, of which $\sim$ 4030 are within the two-half light radius. The photometric calibrations were performed as follows:
\begin{itemize}
\item Aperture correction: aperture growth curves were generated for a sample of stars ($\sim$ $40$ bright and isolated stars) using the {\sc daogrow} routine. The difference between the PSF-fitting and a large-aperture magnitude was derived from the {\sc collect} routine and added to all stellar sources using the {\sc newtrial} routine \citep{1996PASP..108..851S, 1990PASP..102..932S}.
\item Zero point derivation: the transformation equations are derived by comparing Landolt standard stars and SDSS photometry based on the zero point and atmospheric extinction \citep{2006A&A...460..339J}. The average of the zero points is used for nights without a standard field. The transformation equations are applied by the {\sc ccdave} routine and all images are calibrated with the {\sc newtrial} \citep{1996PASP..108..851S}.

\item Relative calibration: we selected a sample of $1000$ common stars in all images with magnitudes ranging from $17$ to $21$ mag. The mean magnitude was estimated at all epochs for each star, taking into account the  photometric errors. We then applied the average value of the deviation of the mean magnitude of each star to all epochs. This step ensures that we have calibrated the magnitude of the LPVs. Fully described details of the photometric procedure can be found in \cite{2020ApJ...894..135S}.
\end{itemize}

\begin{table}
\begin{center}
\caption{Observations of And\,IX}
\begin{tabular}{c c c c c}

\tableline
\tableline
Date (y m d) & Epoch & Filter & $t_{exp}$(sec) & Airmass  \\

\tableline
$2015$ 06 $21$ & $1$ &  I & $45$ & $1.265$ \\[1ex]
$2015$ 06 $21$ & $1$ &  V & $72$ & $1.253$ \\[1ex]
$2016$ 02 $09$  & $2$ & i &  $540$ & $1.854$\\[1ex]
$2016$ 06 $15$  & $3$  & i & $555$ & $1.409$\\[1ex]
$2016$ 08 $11$  & $4$ & i & $555$ & $1.045$ \\[1ex]
$2016$ 08 $13$  & $4$ &  V & $735$ & $1.033$ \\[1ex]
$2016$ 10 $21$  & $5$ & i & $555$ & $1.320$ \\[1ex]
$2017$ 01 $29$  & $6$ & i & $435$ & $1.655$ \\[1ex]
$2017$ 08 $01$  & $7$ & i & $555$ & $1.306$ \\[1ex]
$2017$ 08 $01$  & $7$ &  V & $735$ & $1.227$ \\[1ex]
$2017$ 09 $02$  & $8$ & i & $555$ & $1.070$ \\[1ex]
$2017$ 09 $02$  & $8$ & V & $735$ & $1.047$ \\[1ex]
$2017$ 10  $06$  & $9$ & i & $555$ & $1.341$ \\[1ex]

\tableline
\end{tabular}

\end{center}
\end{table}

\begin{figure}
\centering
\includegraphics[width=.52\textwidth]{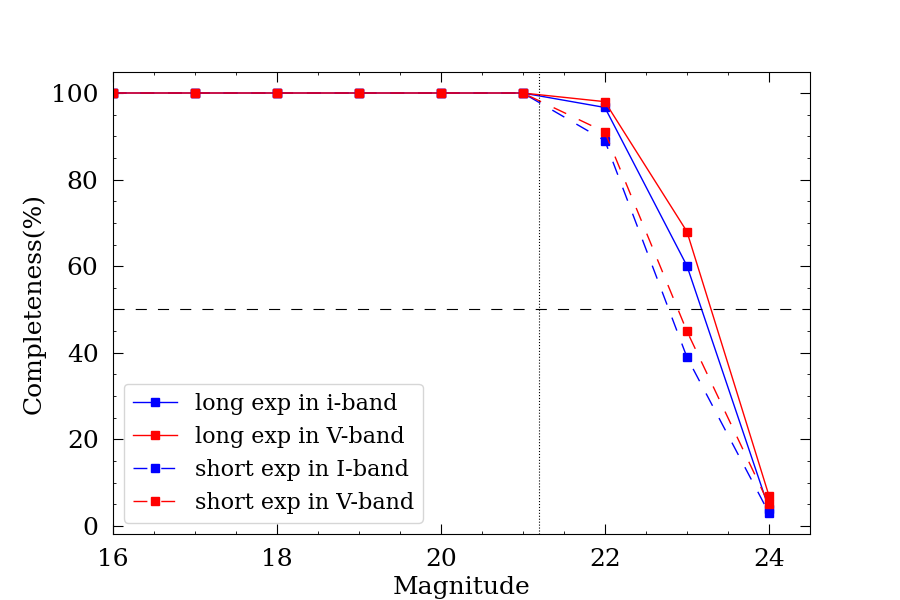}

\caption{Completeness limit {\it vs}.\ magnitude for the long exposure times in the $i$-band (blue solid line) and in the $V$-band (red solid line). The short exposure times are also in the $I$-band (blue dashed line) and $V$-band (red dashed line). The dotted vertical line shows the tip of the RGB.\centering}
\centering
\end{figure}


To investigate photometric completeness, we performed the {\sc addstar} task in the {\sc daophot} package. 1050 artificial stars in a range from $17$ to $24$ mag were added at random positions. The completeness limits as a function of magnitude are shown for long and short exposure time images in the $i$-, $I$-, and $V$-bands in Fig.\ 2. The completeness limit is at $100\%$ above the peak of the RGB around $\sim$ 21.20 mag. The extracted catalog is more than $90\%$ complete up to the $\sim$ $22$ mag for long exposure frames and more than $85\%$ complete in frames taken in 2015. The completeness limit as a criterion of recovered stars has dropped to $50\%$ for fainter stars (magnitude > 22.8 mag). The catalog covers all AGB stars in the observed region, as we search for AGB stars between the tip of the AGB at $i$ = $17.29$ mag and the tip of the RGB at $i$ = $21.20$ mag (see Section 5 for more information on how tips of the AGB and RGB are calculated). The magnitude differences between the recovered (output of photometry) and assigned magnitude (by {\sc addstar}) of the artificial stars {\it vs}.\ magnitude in the $i$-band are shown in Fig.\ 3. These differences range from $-1 < \Delta i < 0.5$ mag. Despite the acceptable completeness in the $I$-band magnitude, we eliminated it in the later steps such as the search for variables and SFH. The calibration procedures were not applied to this filter because there was no transformation equation between the $I$-band magnitude and two other bands (refer to Section 4.3 for more details). 
\begin{figure}
\centering
\includegraphics[width=.52\textwidth]{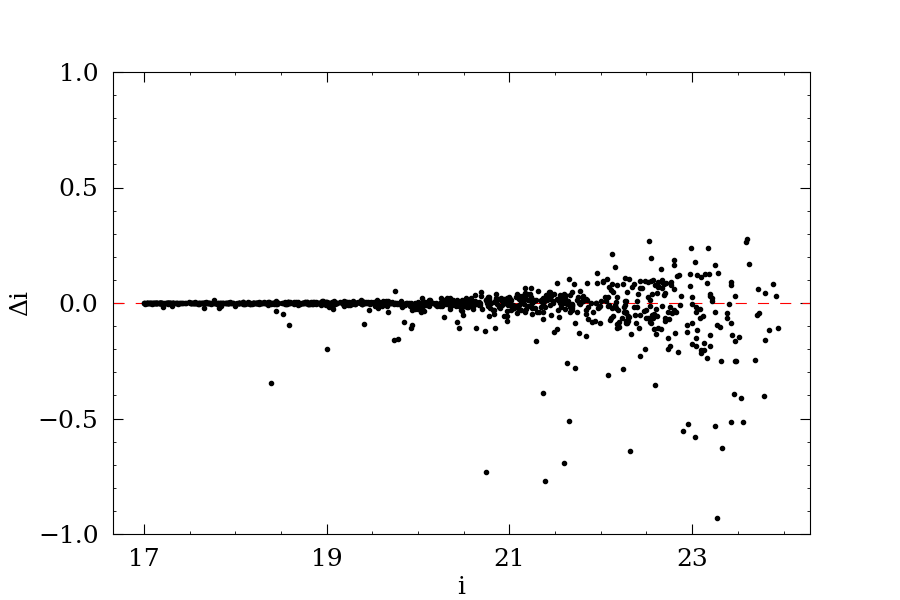}
\caption{Differences between the recovered and assigned magnitude of artificial stars by {\sc addstar} {\it vs}.\ magnitude in the $i$-band.\centering}
\centering
\end{figure}

\section{Probing variable candidates}

\subsection{Evaluation of variables}

One of the most reliable methods for determining variability in a sample of stars was introduced by \cite{1993AJ....105.1813W}, which was further developed in \cite{1996PASP..108..851S}. The decisive variable index, $L$, is determined by combining two indexes $J$ and $K$ as follows:

\begin{equation}
L=\frac{J\times K}{0.798} {\frac{\displaystyle\sum_{i=1}^{n}\omega_i}{\omega_{all}}}.
\end{equation}

\begin{figure}
\centering
\includegraphics[width=.51\textwidth]{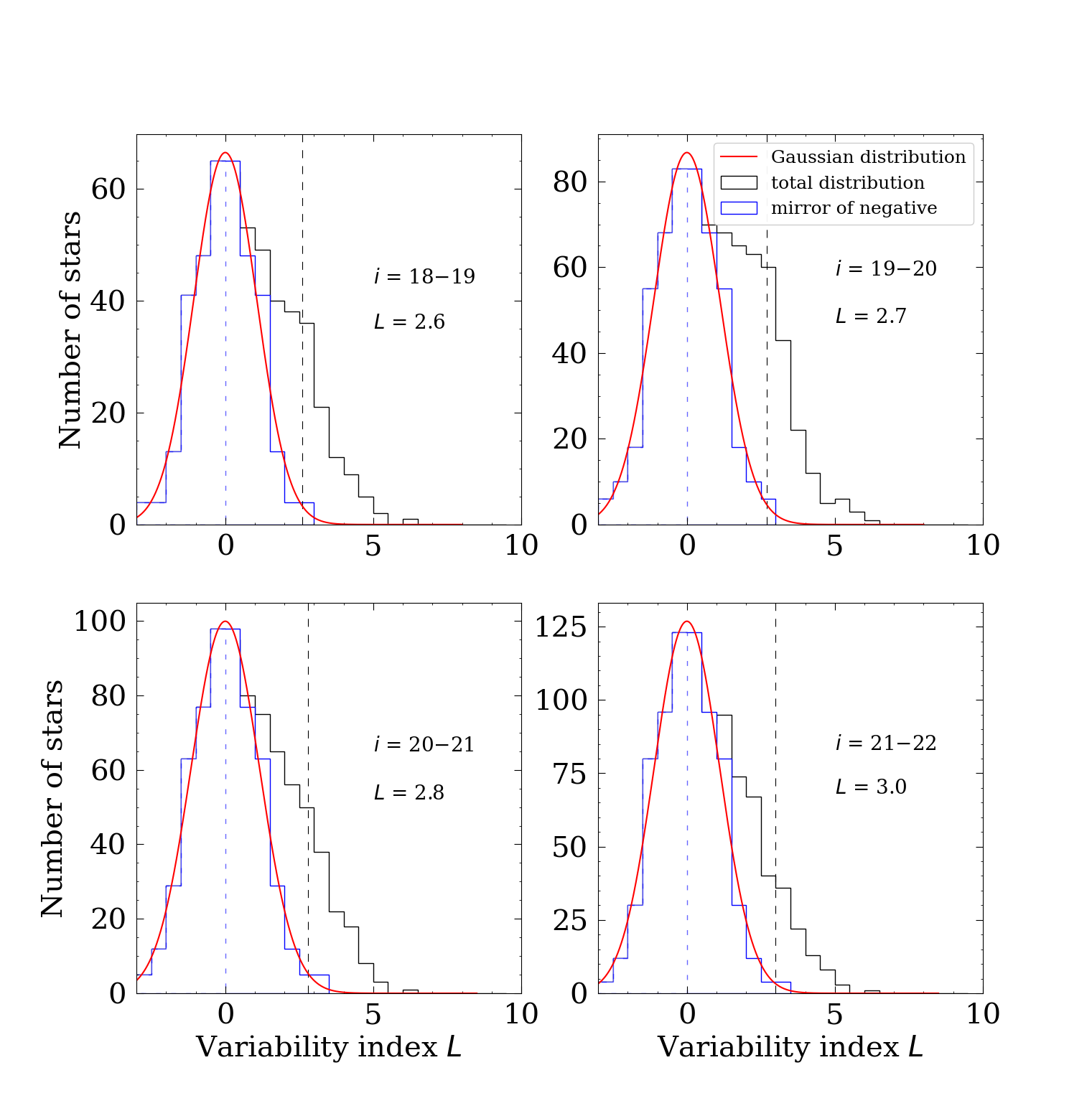}
\centering
\caption{The red curves are the predicted Gaussian functions fitted to the histograms of variability indexes of And\,IX population. The blue dashed lines are with $L$ = 0 and the negative part of the blue histograms ($L$ < 0) are mirrored. Vertical black dashed lines represent the thresholds of the variability indexes in each magnitude bin.}
\end{figure}

With random noise, the index $J$ scatters around zero and has a large positive value for variable stars. The shape of the light-curve is also influenced by the index $K$. If the time difference between two frames is closer than the period of variability, the weight of each star, $\omega_{i}$, is equal to 1, whereas in a single frame, $\omega_{i}$ = 0.5. $\sum \omega_{i}$ is the total weight assigned to a star based on the number of detections and $\omega_{all}$ is the total weight assigned to a star when it has been observed at all epochs.

Fig.\ 4 can be considered as a Gaussian function (red curve), mirrored around $L = 0$, with an excess at high $L$, representing statistically significant variable stars. We fit a Gaussian function to the negative distribution of $L$ in each magnitude bin. If the number of stars in a given (positive) $L$ and a given magnitude bin exceeds the Gaussian fit by a factor of 10 (indicating a $90\%$ chance that it is a true variable), we set a threshold for candidate variability (vertical black dotted line). This threshold is indicated by the red dots in Fig.\ 5.

\begin{figure}
\includegraphics[width=.49\textwidth]{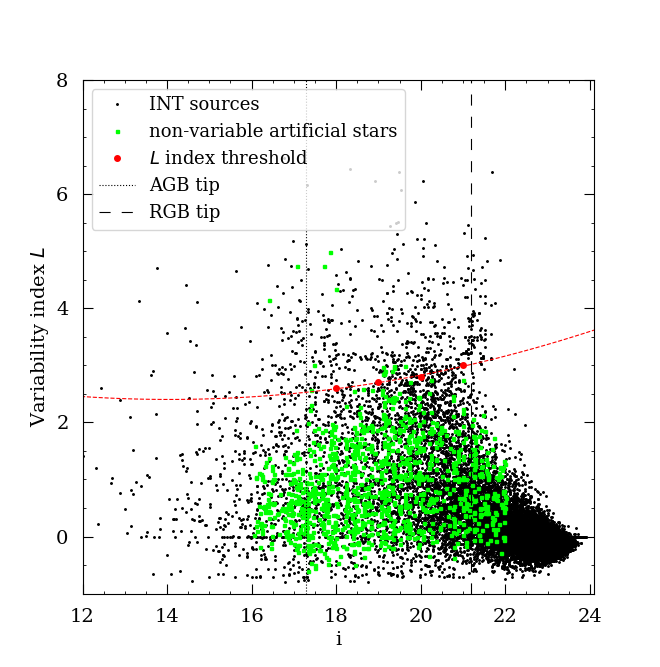}
\centering
\caption{Distribution of variability indexes as a function of magnitude with the best fit of the polynomial function (red dashed line) to the indexes threshold in $i$ $\in$ [$18$,$19$,$20$,$21$] mag. Variability indexes of artificial stars in the range of 16 $\leq i \leq$ 22 mag marked in green. The tips of the AGB and RGB are shown with vertical black lines at $17.29$ and $21.20$ mag, respectively.}
\end{figure}

Fig.\ 5 shows a scatter plot of variability index $L$ for different magnitude intervals. To detect variable candidates in all magnitude ranges, a polynomial function is fitted to the variability index threshold which varies with magnitude. A total of $411$ variable candidates were detected over an area of about $0.07$ deg$^2$ ($\sim$ area of CCD4).

The accuracy of the variability index estimate was evaluated using the task {\sc addstar} of the package {\sc daophot} \citep{1987PASP...99..191S}. In this way, 1200 artificial stars with magnitudes in the range of $16-22$ mag were added to all frames in $6$ steps with random positions and constant light-curves. Photometric procedures and assessment of variability index $L$ for the synthetic population were performed as previously described. The distribution of the variability indexes of the artificial stars, based on their magnitude is overplotted in Fig.\ 5. It can be seen that no more than $1.16\%$ of these artificial constant stars have a variability index above the threshold value, indicating a fairly reliable estimate of the threshold value.

\begin{figure}
\centering
\includegraphics[width=.5\textwidth]{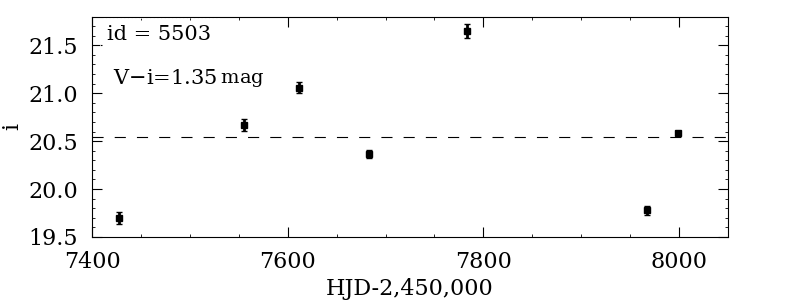}
\includegraphics[width=.5\textwidth]{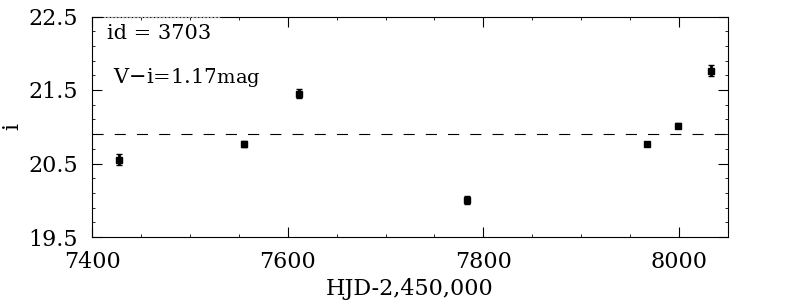}
\includegraphics[width=.5\textwidth]{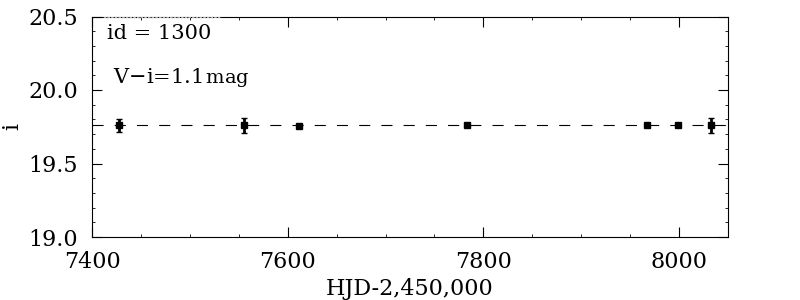}
\caption{Examples of light-curves, two variable candidates in the first two plots, and a non-variable star in the third plot. Vertical error bars represent magnitude error.\centering}
\centering
\end{figure}

Fig.\ 6 shows the light-curves of the variable candidates and a non-variable star to highlight the distinction between their light-curves. The magnitudes of the variable ones show remarkable changes around the mean magnitude (the first two plots), while the non-variable star (the third plot) shows only small changes in magnitude around its mean, shown by horizontal dashed lines. Variable candidates (the first two plots) were detected in {\it Spitzer} and SDSS surveys \citep{2014A&A...571A..16P, 2015ApJS..216...10B}. In addition to these surveys, WISE also discovered the candidate \#5503 \citep{2014yCat.2328....0C}. Table 6 provides more details about variable stars.

\begin{figure}

\includegraphics[width=.51\textwidth]{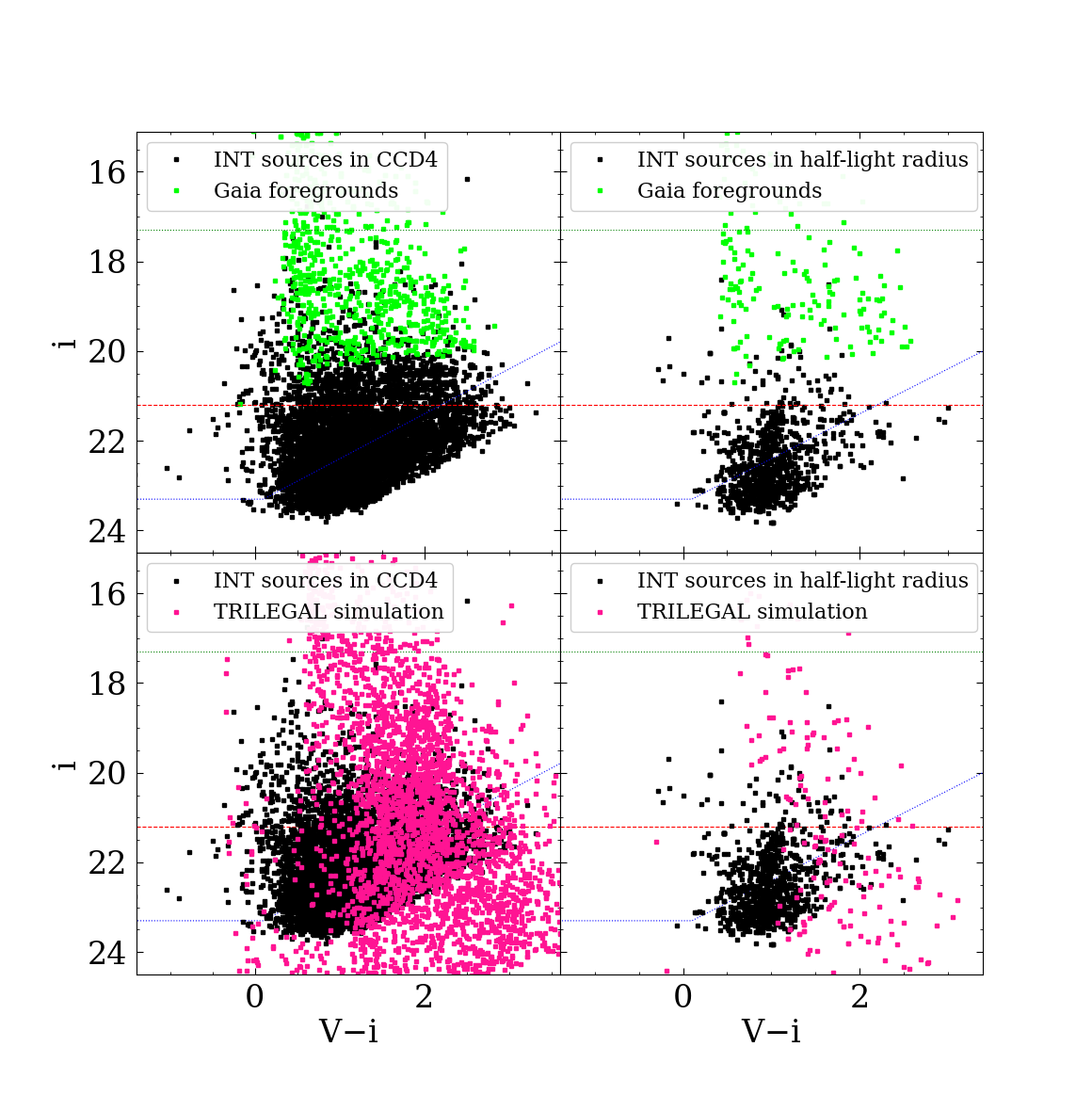}
\centering
\caption{\emph{Gaia} sources (in green) and {\sc trilegal} simulation of foreground contamination (in pink) are presented as a function of color. Stellar population confined in a field of $\sim$ $0.07$ deg$^2$ (area of CCD4) in the left panels, and in $\sim$ $0.005$ deg$^2$ corresponding to the half-light radius in the right panels. Tips of the AGB and RGB are marked in green and red dashed lines, respectively. The completeness limit of our photometry is marked in blue.}
\end{figure}
 
\subsection{And$\;$IX Contamination}
Foreground Milky Way contamination must be removed from And\,IX to distinguish dwarf galaxy stellar  populations, especially before SFH are constructed from LPVs. To detect foreground contamination, we cross-correlated our catalog with \emph{Gaia} Data Release 3 (DR$3$; \citep{GaiaDR3}). We also considered {\sc trilegal}  simulations of the Milky Way population to account for the level of contamination \citep{2005A&A...436..895G}. The \emph{Gaia} sources in green are overplotted on the INT populations in the upper panels of Fig.\ 7. We included the entire field observed in CCD4 ($0.07$ deg$^2$, left panels of Fig.\ 7) and a region within a half-light radius ($0.005$ deg$^2$, right panels of Fig.\ 7). 
Populations of the Milky Way must either satisfy the proper motion criteria $\sqrt{(\mu_{RA})^2+(\mu_{DEC})^2} > 0.28$ mas yr$^{-1}$ + 2.0 error for $\mu$ as proper motion or Pa / (error of Pa) $\geq 2\sigma$ (Pa as parallax) \citep{2020ApJ...894..135S}. Based on the adopted criteria, the total number of \emph{Gaia} sources detected in our observation is estimated to be $890$ in the region of CCD4 and $171$ within the half-light radius of And\,IX. About $80\%$ of the variable candidates in CCD4, $\sim$ $55\%$ in two half-light radii, and less than $27\%$ within the half-light radius are detected by \emph{Gaia} as foreground contamination. \emph{Gaia}'s faint source limit is $G \sim 20.5$ mag and its completeness limit is $G \sim 17$ mag \citep{GaiaDR3}.

\begin{figure}
\centering
\includegraphics[width=.5\textwidth]{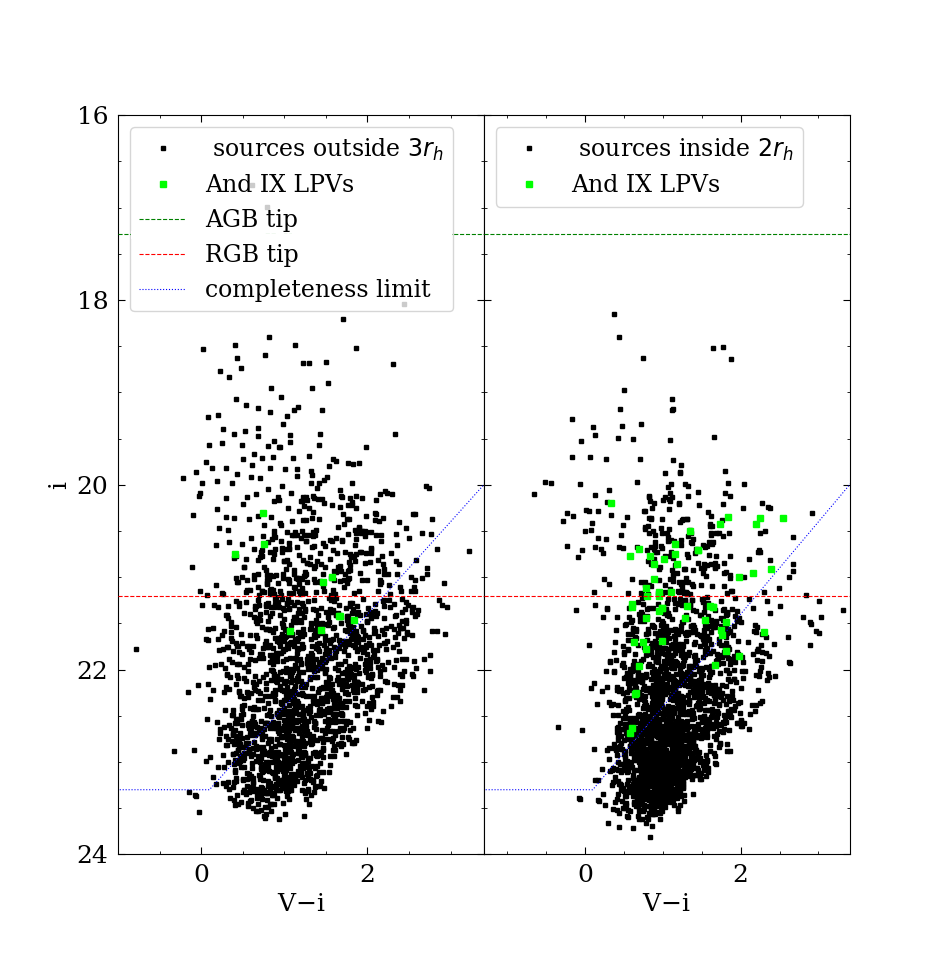}
\centering
\caption{CMDs of And\,IX in the $i$ {\it vs}. $V-i$ color showing our identified LPV candidates in green for two regions of a similar area within two half-light radii (right panel) and outside three half-light radii (left panel).}
\end{figure}

Foreground contamination of And\,IX is simulated using the {\sc trilegal} tool, at galactic coordinates $\ell$ = $123.212^\circ$, $b$ = $-19.675^\circ$, and Galactic extinction A$_V = 0.206$ mag. The stellar population predicted by {\sc trilegal} along the line of sight of And\,IX in two panels is shown in pink in the lower panels of Fig.\ 7. The synthetic stars predicted by {\sc trilegal} are $1990$ in CCD4 and $190$ within the half-light radius. There is additional foreground contamination from the {\sc trilegal} simulation within CCD4 and the half-light radius compared to \emph{Gaia} DR$3$ due to the \emph{Gaia} limitation in the detection of very faint stars ($\sim i = 20.5$ mag). Comparing the foreground contamination in the two regions in Fig.\ 7, we see that foreground stars have less impact on the half-light radius despite a large number of foreground stars in CCD4. 

In addition to the foreground stars of the Milky Way, our sample could be contaminated by stars in the halo of M31 (due to its proximity to M31).
A plausible contamination by a giant stellar stream in the northeast of M31 has been suggested based on the study of velocity dispersion in the SPLASH survey \citep{2012ApJ...752...45T}. In addition, And\,IX is located in a halo substructure known as the Triangulum-Andromeda (Tri-And) region, and stars in this region could represent possible contamination \citep{2004ApJ...615..732R}. 

The distribution of LPVs as a function of color ($V-i$) is compared in two regions with the same area to reveal the degree of contamination by the M31 halo in Fig.\ 8. Contamination is estimated by comparing the population within two half-light radii (right panel) and outside of the three half-light radii (left panel). About 28\% of the stellar population and $13\%$ of the LPV candidates within two half-light radii are estimated as background contamination. Possibly they are stars in M31 and/or background active galactic nuclei (AGN) and/or foregrounds that are below \emph{Gaia}'s limit of completeness. Fig.\ 8, left panel, shows no clear, curved RGB in the control field, suggesting that M31 contamination is negligible and most of the contamination comes from the Galactic foreground. It is noted that background galaxies and AGN contamination mainly affect the region around $V - i \sim  2$ mag and $i \gtrsim 21.5$ mag, not much affecting the AGB (or RGB) portion of the And\,IX CMD. In the following, foreground contamination of \emph{Gaia} is excluded from the And\,IX population.

\subsection {Amplitude of variable candidates}

\begin{figure}
\centering
\includegraphics[width=.5\textwidth]{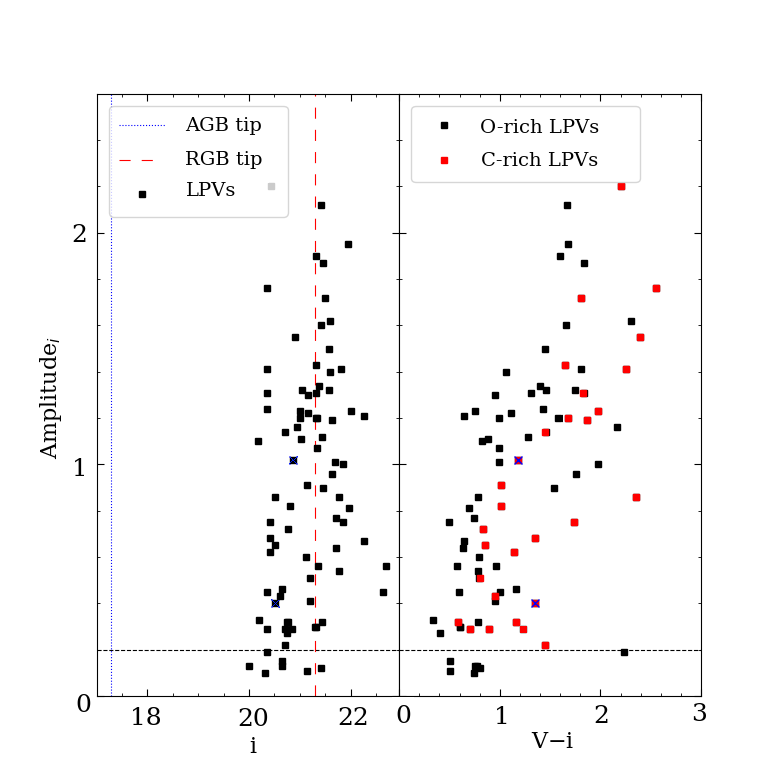}
\centering
\caption{Amplitude of variability {\it vs}.\ magnitude in the $i$-band (left panel) and {\it vs}.\ color (right panel). Vertical blue and red dashed lines represent the tips of the AGB and RGB, respectively. $0.2$ mag amplitude (horizontal black dashed line) is the threshold for distinguishing LPVs from other candidates.}

\end{figure}

The shape of the light-curve is assumed to be sinusoidal to evaluate the amplitude of variability. The light-curves of the variable candidates in Fig.\ 6 are considered as examples in our data that show sinusoidal variability due to pulsations (blue signs in Fig.\ 9). Considering a value of 0.707 for the standard deviation of the unit sine function and a standard deviation in our data, $\sigma$ \citep{javadia}, the amplitude is estimated by:
 
\begin{equation}
Amplitude =\frac{2\times\sigma}{0.707}.    
\end{equation}

The amplitude of the variable candidates in the range $0.1-2.20$ mag, as a function of magnitude in the $i$-band, is shown in the left panel of Fig.\ 9. The horizontal black dashed line ($0.2$ mag) illustrates the LPV amplitude threshold \citep{2020ApJ...894..135S}, and the blue dotted and red dashed lines are the tips of the AGB and RGB, respectively (see Section 5). It can be seen that LPV stars with larger amplitudes tend to be redder than those with smaller amplitudes. The right panel of Fig.\ 9 shows the amplitude in the $i$-band as a function of $(V-i)$ color. The amplitude almost increases with color for more candidates. The stars become redder and fainter as they evolve in the giant star branch. During the evolution of stars along the AGB, they become more luminous, so dust must attenuate their light if they appear fainter. Additionally, the amplitude of the variability increases in the AGB phase as they evolve \citep{1992ApJ...397..552W, 2016ApJ...823L..38M, 2019MNRAS.484.4678M}. The larger amplitude can be explained by the lower luminosity of lower-mass AGB stars; thus, RSGs typically have a lower amplitude \citep{1992ApJ...397..552W, 1998A&A...338..592W, whitelock2003obscured, 2008A&A...487.1055V}.

C-rich variable candidates of the INT survey which are classified on the basis of birth mass (see Section 6), are shown in red squares in the right panel of Fig.\ 9. The amplitudes of this sample almost increase with increasing reddening, just as in the {\it Spitzer} observations of C-rich variables in LMC and IC\,1631 \citep{2017}. Dust-enshrouded variables that have larger amplitudes experience more mass-loss and become redder \citep{1991MNRAS.248..276W}. As well, O-rich variable candidates in the INT survey (black squares in the right panel of Fig.\ 9) with a redder color have a larger amplitude.
 
A total of seven stars with amplitudes less than 0.2 mag in the $i$-band in CCD4 and four within two half-light radii were excluded from the variability analysis. The LPVs have variability amplitudes greater than 0.2 mag, and we are also unsure of the nature of the variable candidates with amplitudes less than 0.2 mag.

\begin{figure}

\includegraphics[width=.525\textwidth]{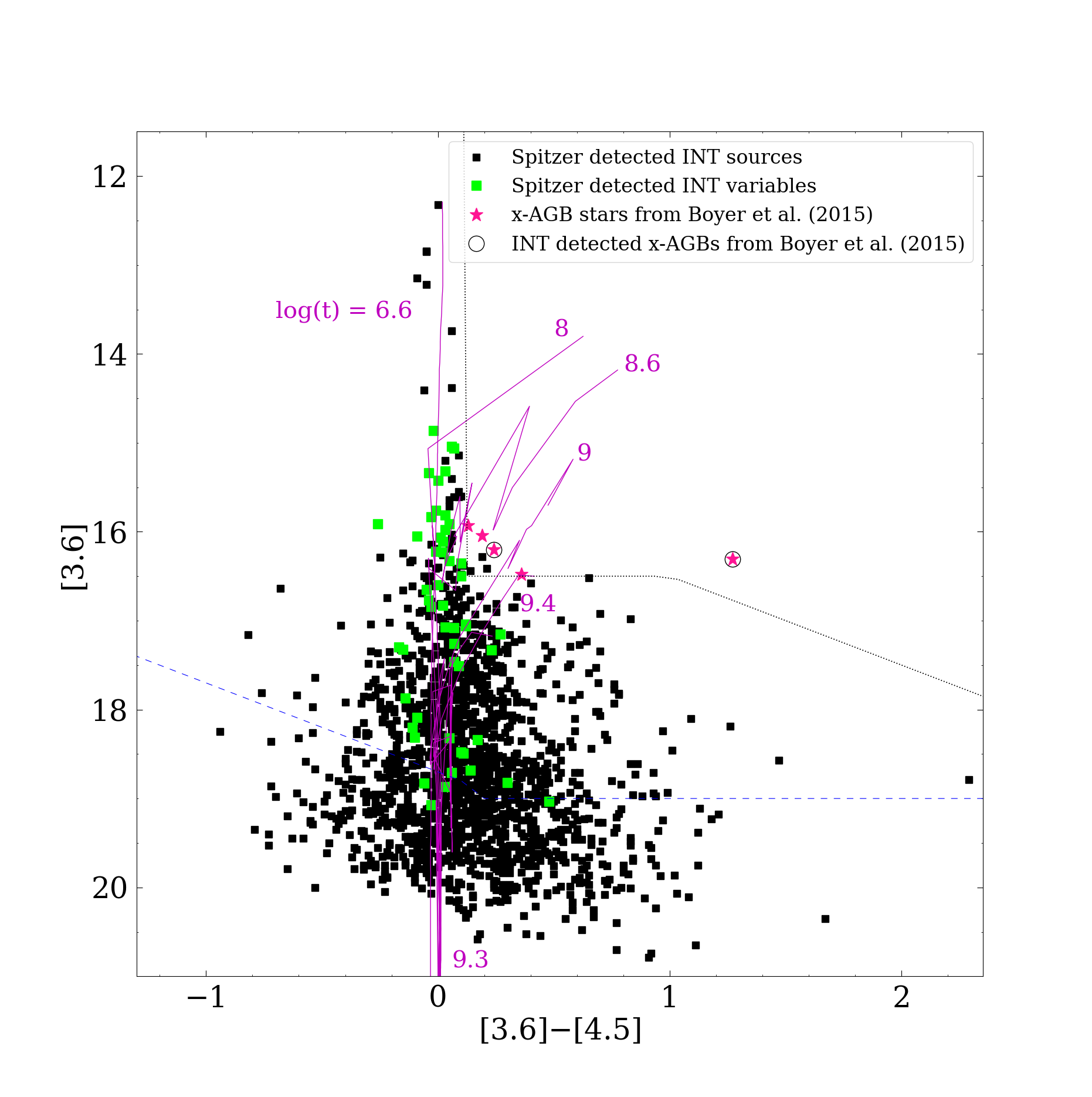}
\centering

\caption{CMD of {\it Spitzer} detected INT sources. {\it Spitzer} detected LPVs are highlighted in green squares, and PADOVA isochrones from \cite{2017ApJ...835...77M} are marked in magenta. The blue dashed line represents the $75\%$ completeness limit of {\it Spitzer} data, and the dotted black line separates the plausible region of x-AGB stars \citep{2015ApJ...800...51B}.\centering}
\end{figure}

\section{Cross-match with other catalogs}
\subsection{{\it Spitzer} catalog cross-identification}

A cross-correlation was carried out between the And\,IX INT catalog and mid-IR data from the {\it Spitzer} Space Telescope, IRAC, in $3.6$- and $4.5$- $\mu$m bands as part of the DUSTiNGS survey (DUST in Nearby Galaxies with {\it Spitzer}) \citep{2015ApJS..216...10B}. The two catalogs have 1638 common sources within two half-light radii of the center, out of which $50$ ($\sim$ $3\%$) are INT LPV candidates. DUSTiNGS reported five extreme AGB (x-AGB) candidates \citep{2015ApJ...800...51B}, only two detected in our survey (black open circles in Fig.\ 10) as LPV candidates (\#5671 and \#4433) and two others as non-variable population (\#3770 and \#5025). As in \cite{2015ApJ...800...51B} explained because of imaging artifacts some stars may appear artificially variable in their survey. We estimated the variability index of these two non-variable stars less than the variability threshold in their magnitude range and also their light-curves do not show apparent variability. Our survey did not observe the last one because it is located outside CCD4; therefore, it is unlikely to be And IX's population, as it lies outside the two half-light radii.
In \cite{goldman2019infrared}, just one of these LPV candidates was introduced with clear variability. This x-AGB (\#5671) is also detected in our survey within the half-light radius of And\,IX.
According to \cite{goldman2019infrared}, this candidate has a period of $467$ days and an amplitude of $0.8$ mag. There is also a huge difference in color between it and the rest of the x-AGBs, with a color of  $[3.6]-[4.5] = 1.26$ mag, in Fig.\ 10. Two x-AGBs sources detected as LPVs in our survey with sinusoidal fits are presented in Fig.\ 11, along with their mass-loss rate ($\dot{M}$) (see Section 7.2) and amplitude in the $i$-band (amplitude$_i$) based on our calculations.
In our estimation, LPVs \#5671 and \#4433 have periods of 530 (which is very close to \cite{goldman2019infrared}'s estimate of 467 days) and 585 days. 

\begin{figure}
    \centering
    \includegraphics[width=.5\textwidth]{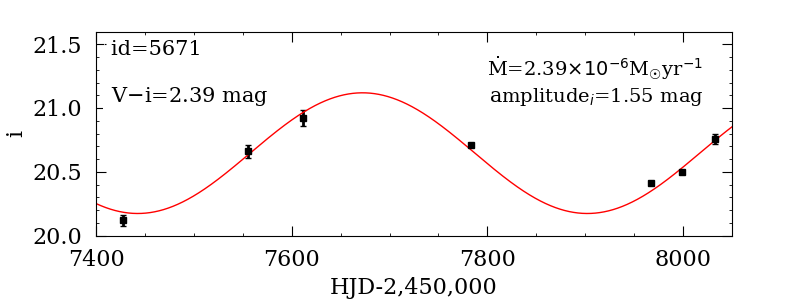}
     \includegraphics[width=.5\textwidth]{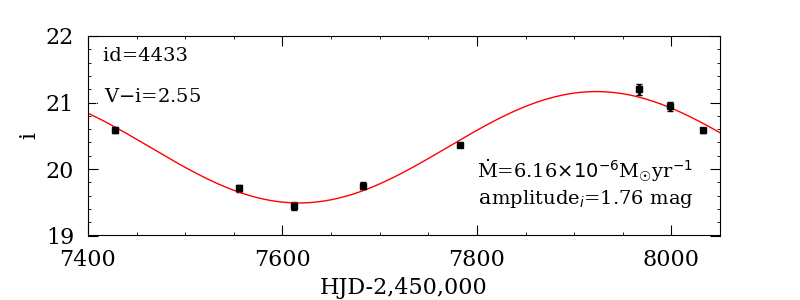}
    \caption{Two light-curves of mutual x-AGBs in INT and the DUSTiNGS catalog of variable {\it Spitzer} sources \citep{2015ApJ...800...51B} with sinusoidal fits in red curves.}
    \label{fig:my_label}
\end{figure}

\begin{figure}
\includegraphics[width=.51\textwidth]{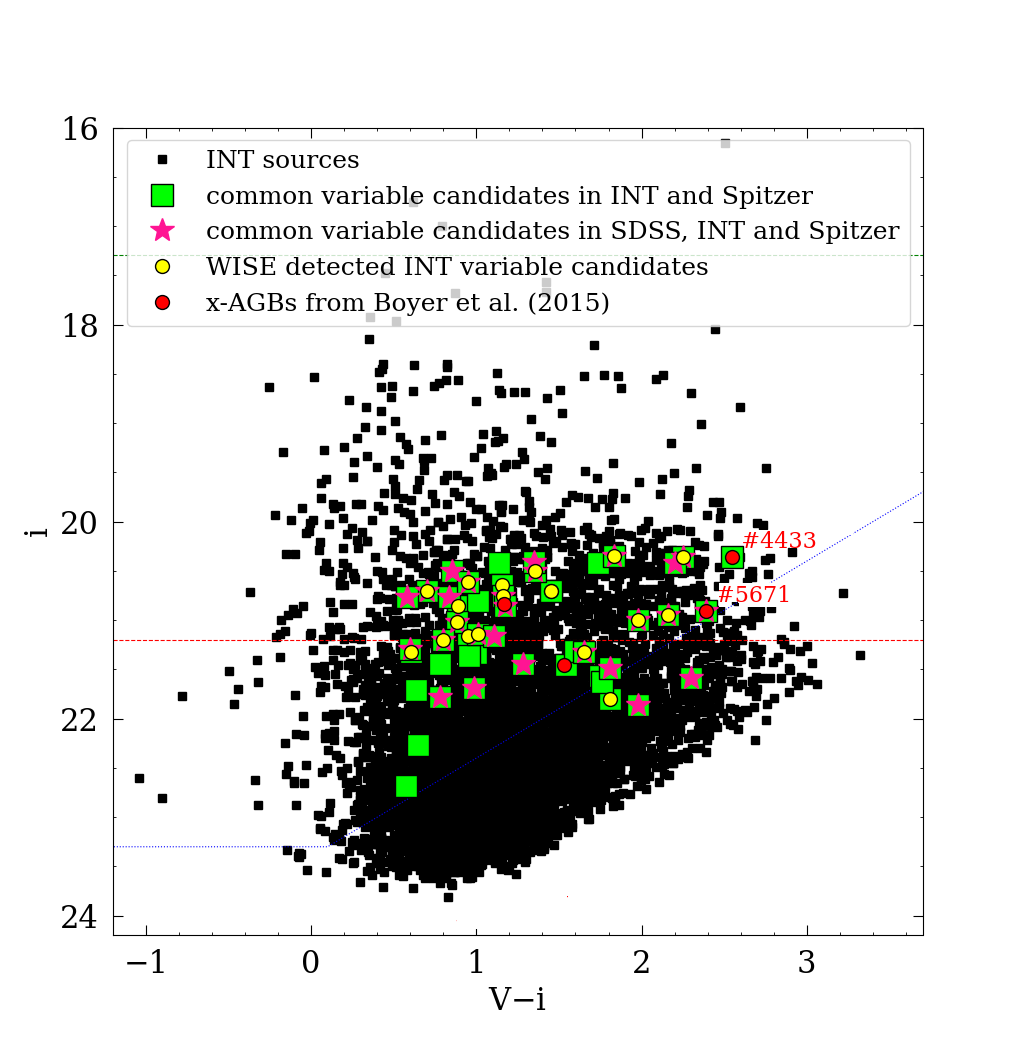}

\centering
\caption{CMD of mutual sources from the INT, {\it Spitzer}, WISE, and SDSS surveys within two half-light radii of And\,IX. Four x-AGBs from \cite{2015ApJ...800...51B} are marked with red circles. The AGB-tip and RGB-tip are illustrated by the green
and red dashed lines, respectively. The blue dashed line represents the estimated completeness limit. \#5561 and \#4433 are mutual x-AGBs between the INT and the {\it Spitzer}.}
\end{figure}

\subsection{WISE catalog cross-identifications} 
\begin{figure}
\includegraphics[width=.48\textwidth]{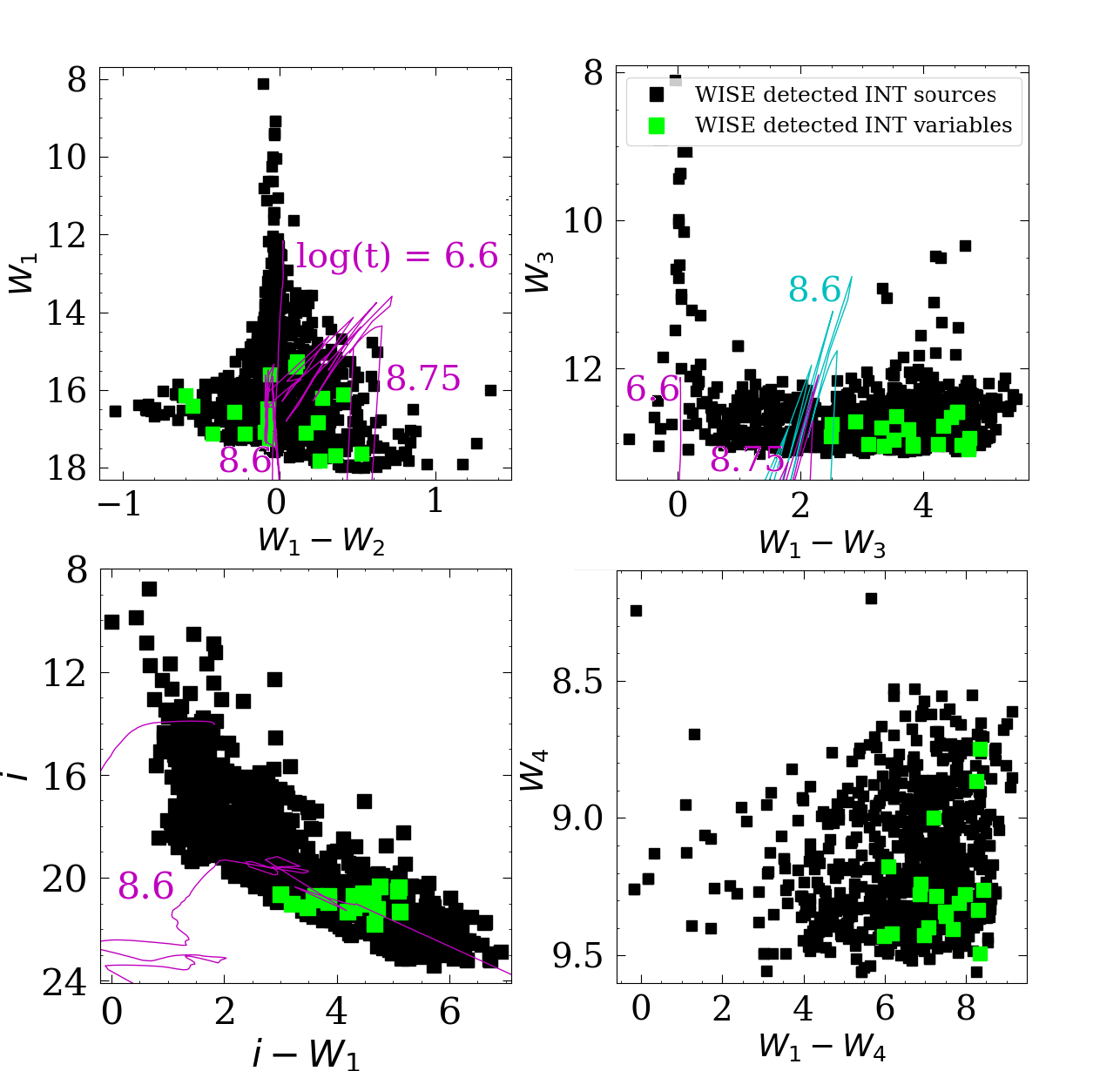}

\centering
\caption{CMD of mutual sources from INT and WISE surveys. Green squares represent mutual LPVs between INT and WISE. The Padova isochrones \citep{2017ApJ...835...77M} are also marked in magenta and blue.}
\end{figure}

The WISE (Wide-field Infrared Survey Explorer) ALL-SKY data release $2013$ \citep{2014yCat.2328....0C} in $W_1$ = $3.35$, $W_2$ = $4.60$, $W_3$ = $11.56$, and $W_4$ = $22.09$ $\mu$m bands was cross-correlated with the INT catalog of And\,IX. A total of $864$ stellar sources are identified between the INT and the WISE, of which $19$ are among the INT variable candidates. The mutual LPVs between the INT and the WISE are low because most of them are in dense regions where we do not get any WISE matches because of the limited angular resolution of WISE. Fig.\ 12 shows a comprehensive CMD of the And\,IX population with common LPVs in INT, {\it Spitzer}, WISE, and SDSS. Matched LPVs between WISE and INT are marked with yellow circles. Fig.\ 13 shows the mutual stars of the WISE and INT surveys in four subplots. The {\it Spitzer} and SDSS surveys have a better photometric quality than WISE and have more stars in common with the INT survey. 
According to the SEDs (Fig.\ 28), the $W_4$ and some $W_3$ data are too bright compared to the other bands. A poor spatial resolution (6.5" and 12.0" for $W_3$ and $W_4$, respectively) could have contributed to this.

\begin{figure}
\includegraphics[width=.5\textwidth]{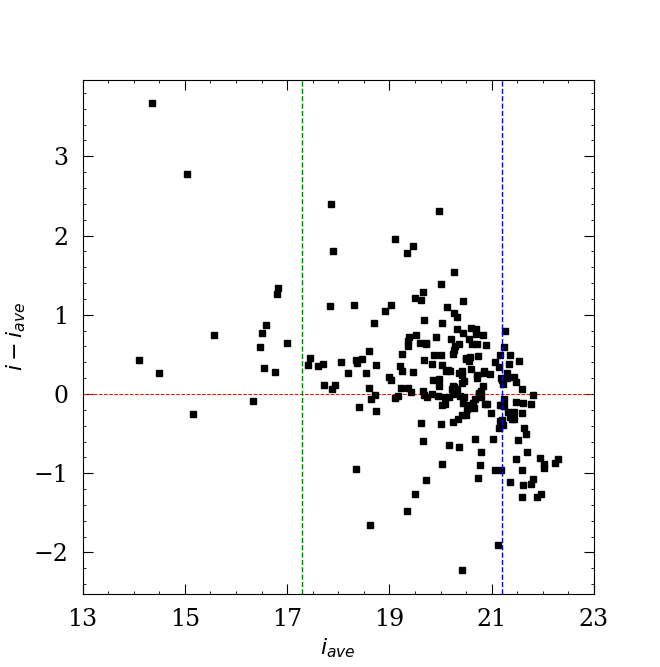}

\centering
\caption{The difference of the calculated magnitude in the Sloan $i$ filter based on the transformation equation \citep{2006A&A...460..339J} with the average magnitude of all frames except 21 Oct $vs.$ average magnitude for non-variable stars. The tips of the AGB and RGB are shown with vertical green and
blue lines at 17.29 and 21.20 mag, respectively.}
\end{figure}

\subsection{SDSS catalog cross-identifications}

We cross-matched the INT catalog with the SDSS data release 12 (DR12) \citep{2014A&A...571A..16P}, the last data released from SDSS-III, in five bands $u, g, r, i, and \,z$ from $2008$ to $2014$. The number of common sources between the SDSS catalog and master catalog of And\,IX is $\sim$ 2680, of which $27$ are LPVs. More than $80\%$ of matched sources have magnitude differences less than $\Delta i < 1$ mag, as we used SDSS photometry to transform our magnitude to the photometric standard system (Landolt). {\it Spitzer} data are available for all 27 LPVs that are mutual between the SDSS and INT (pink stars in Fig.\ 12).
\begin{figure}
\includegraphics[width=.5\textwidth]{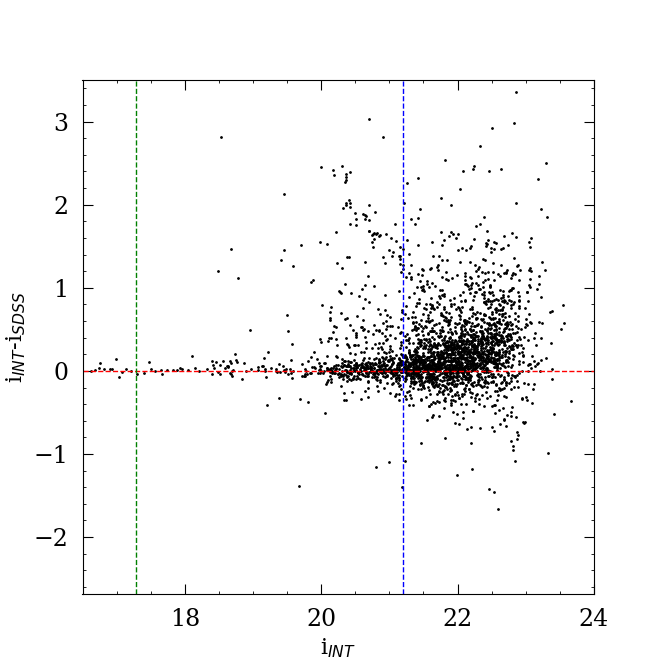}

\centering
\caption{Magnitude differences between INT catalog and SDSS of And\,IX, plotted against $i$ magnitude of our catalog. The tips of the AGB and RGB are shown with vertical green and
blue lines at 17.29 and 21.20 mag, respectively.}
\end{figure}

In the INT survey, an observation was made on 21 October in Landolt $I$ filter. The transformation equation \citep{2006A&A...460..339J} must be used to convert Landolt $I$ filter to Sloan $i$ filter for the 230 mutual stars between the SDSS catalog and frame in Landolt $I$ filter, but since we do not have color in the Landolt system, we use the color ($i-z$) in the SDSS catalog which has observation in Landolt $I$ filter. Due to the short exposure time of this frame (Table 3), fewer stars were identified than between the master catalog and the SDSS catalog.
Fig.\ 14 shows the difference of the magnitude in the Sloan $i$ filter on 21 October with the average magnitude ($i_{ave}$) in all observation (except 21 October) frames for non-variable stars $vs.$ the average magnitude.

There is a large scatter for the magnitude range of interest (between RGB-tip and AGB-tip), which makes the filter transformation inaccurate. 
As a further illustration, in Fig.\ 15, we plot the magnitude differences between the SDSS and INT catalogs ($i_{INT}-i_{SDSS}$ $vs.$ $i_{INT}$). The magnitude difference between the two surveys is distributed around zero, which indicates the accuracy of magnitude calibration, but the scatter is still quite large. As a result, SDSS colors cannot be used in transformation equations to estimate $i$-band magnitudes for LPV identification which requires high accuracy in magnitude estimation of any epoch.

\section{Surveying physical parameters of And\,IX}

We have presented And\,IX stellar density profile (in pink) and surface brightness (mag arcsec$^{-2}$) (in black) as a function of distance from the galaxy center in Fig.\ 16. A half-light radius of $2.50\pm0.26$ arcmin ($597_{-67}^{+62}$ pc) results from the calculation of the half area under the most optimal exponential fit (blue curve) to the number density and surface brightness data. 
Our calculation agreed well with the \cite{2012AJ....144....4M} estimate of about $2.5\pm0.1$ arcmin. In addition, radii of $2.5\pm0.1$, $2.6\pm0.1$, and $2.7\pm0.2$ arcmin were estimated for the half-light radius by \cite{2010MNRAS.407.2411C} considering the best fit of the exponential, Plummer, and King models, respectively, to the distribution of number density as a function of radius.

\begin{figure}
\centering
\includegraphics[width=.47\textwidth]{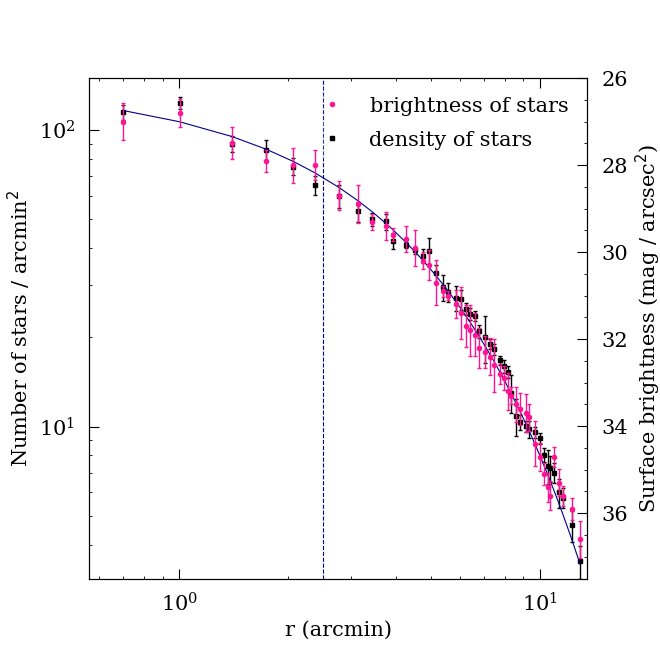}
\centering
\caption{The stellar number density and surface brightness of And\,IX with the best exponential fit to the data (blue curve) as a function of galactocentric distance. The blue dashed line represents the half-light radius ($2.50\pm0.26$ arcmin). The vertical error bar results from the Poisson uncertainty of the counts.}
 \end{figure}

The magnitude of the tip of the RGB, as a distance indicator, has been used to estimate the distance of the galaxy \citep{lee1993tip}. On reaching the end of the RGB, stars ignite helium in their cores. At the tip of the RGB, stars reach maximum luminosity through helium flashes. The TRGB magnitude in the $I$-band has the least dependence on a star's age and chemistry, making it the most reliable magnitude to use as a standard candle \citep{lee1993tip}. It was not possible to convert the photometry bands to $I$-band using the transformation equations in \cite{Lupton(2005)} via the Johnson-Cousins system due to the lack of a third filter. Therefore, we calculated the TRGB magnitude in the $i$-band.

\begin{figure}
\centering
\includegraphics[width=.52\textwidth]{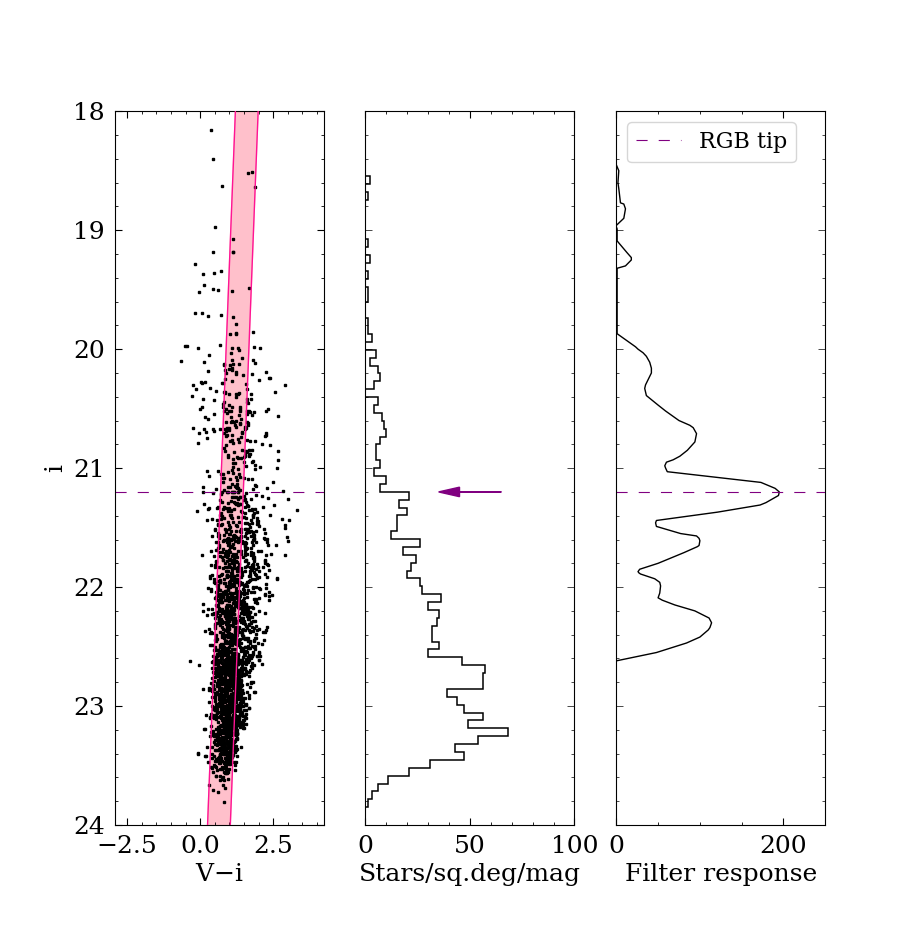}
\centering
\caption{The left panel shows And\,IX sources in two half-light radii. The middle panel represents the histogram of the luminosity function. The right panel shows the Sobel filter response for the tip of the RGB with edge detection. The tip of the RGB is detected at $i$ = $21.20_{-0.15}^{+0.05}$ mag and marked on the CMD by a horizontal purple dotted line and on the luminosity functions by an arrow.}
\end{figure}

A population of stellar sources in an area within two half-light radii ($\sim$ $0.022$ deg$^2$) is selected to estimate the distance of the galaxy, as used in \cite{2004MNRAS.350..243M}. This region is located between two pink lines (left panel in Fig.\ 17) of 0.28 < $V-i$ < 1.07 mag and 18.51 < $i$ < 23.81 mag to isolate the populations in the red giant branch. The magnitude of the TRGB can be calculated by constructing the $i$-band luminosity distribution as a binned histogram with $0.05$ mag (middle panel in Fig.\ 17). By convolving the smooth luminosity function with the summation of the normalized Gaussian distribution \citep{1996ApJ...461..713S} through a Sobel edge kernel [-2,-1,0,+1,+2], the position of the tip of the RGB at which the convolution is maximum is determined (right panel in Fig.\ 17) \citep{lee1993tip}. 

The distance modulus of $24.56_{-0.15}^{+0.05}$ mag ($\sim$ $816.58_{-54.50}^{+19.02}$ kpc) results from the tip of the RGB at $i$ = $21.20_{-0.15}^{+0.05}$ mag (highlighted in purple in Fig.\ 17). In this derivation, a correction of $2.086\times E(B-V)$ with $0.075$ mag as reddening \citep{1998ApJ...500..525S} is used for the Galactic extinction in $i$-band. Also, we adopt an absolute magnitude of $-3.52$ mag for the tip of the RGB in the $i$-band based on the PARSEC isochrones in the SDSS photometry system \citep{2012MNRAS.427..127B}.  Estimates of the distance modulus to And\,IX vary from $23.89_{-0.08}^{+0.31}$ mag \citep{2019MNRAS.489..763W} to $24.42\pm0.39$ mag
\citep{2010MNRAS.407.2411C}.

CMD of the stellar population of And\,IX within two half-light radii ($\sim5$ arcmin) is shown in Fig.\ 18. All LPV candidates and those with amplitude$_{i}$ < 0.2 mag are marked in green and orange squares, respectively. 
The spatial distribution of variable candidates within two half-light radii is shown in pink circles in Fig.\ 1. The overlaid Padova\footnote{http://stev.oapd.inaf.it} stellar evolutionary tracks illustrated in magenta, range from $\sim$ $31.62$ Myr to $10$ Gyr \citep{2017ApJ...835...77M}. A distance modulus of $24.56_{-0.15}^{+0.05}$ mag and metallicity $Z = 0.0001$ are used for all stellar tracks in this paper.
A total of 8653 stellar sources and $84$ variable candidates were detected in an area of $11.26\times22.55$ arcmin $^2$ ($2.69\times5.39$ kpc$^2$), which corresponds to the CCD4 of the WFC.

AGB stars at the tip of the AGB are optically obscured by dust due to high mass-loss. The Chandrasekhar core mass for the classical AGB limit is obtained in M$_{bol}$ = 7.1 mag (M$_{bol}$ < $-8$ mag for supergiants) considering the classical core luminosity relation \citep{1996MNRAS.279...32Z}. From Padova evolutionary model, the classical AGB limit is estimated to be $log (t) = 7.5$ (t = 31.62 Myr) for the tip of the AGB (green dashed line) with $i = 17.29$ mag. Fig.\ 18 also shows a red dashed line representing the tip of the RGB at $i = 21.20$ mag, and the completeness limit of the photometry in blue.

\begin{figure}
\includegraphics[width=.5\textwidth]{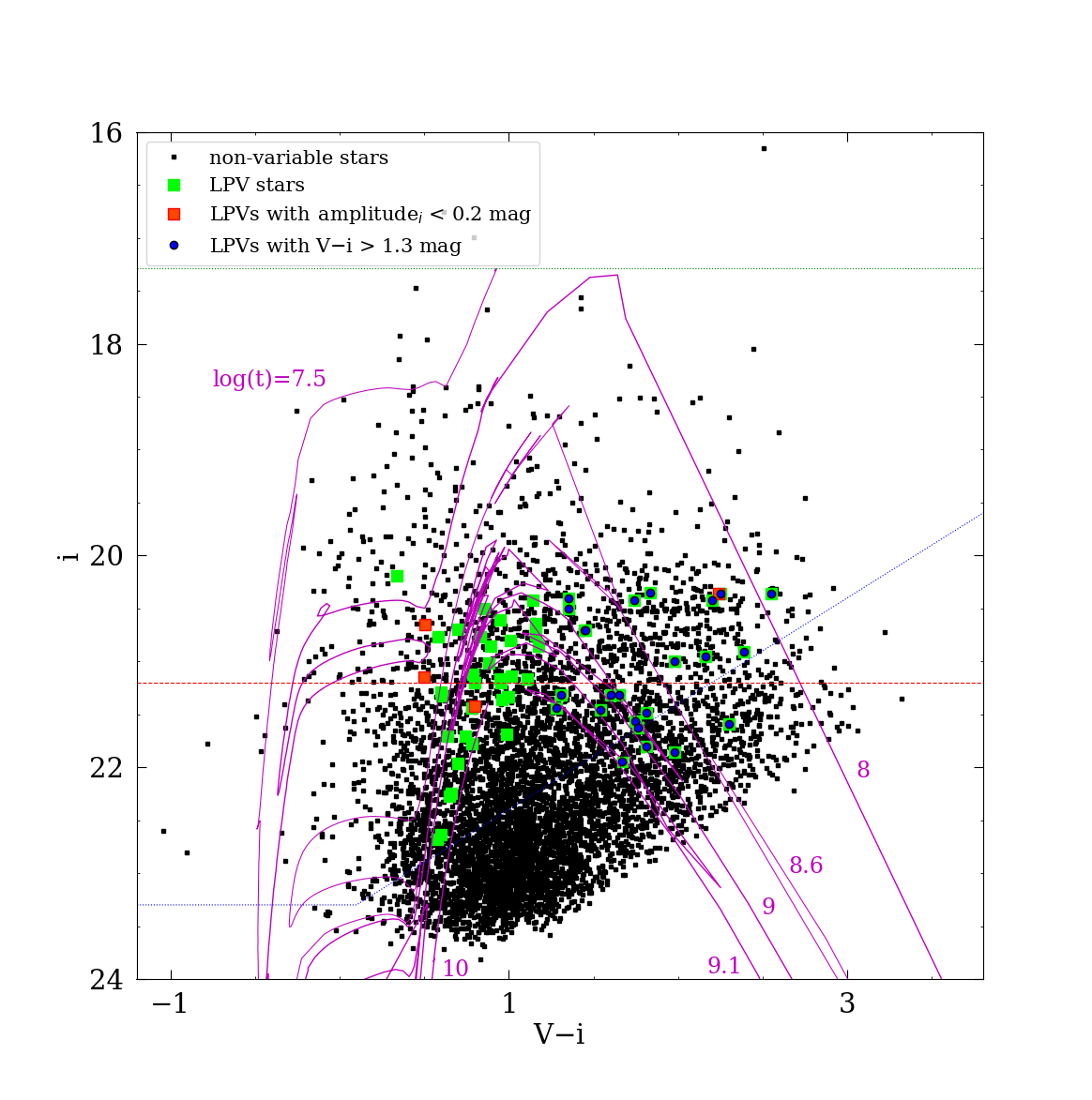}
\centering
\caption{CMD of And\,IX sources in an area within two half-light radii. The variable candidates are marked with green squares and those with amplitude$_i$ < 0.2 mag with orange squares. The completeness limit of the photometry (blue), the AGB-tip (green), and the RGB-tip (red) are shown in the graph. The Padova isochrones \citep{2017ApJ...835...77M} are also marked in magenta. Dust correction will be applied to the LPVs shown in blue.}
\end{figure}

\begin{figure}
\includegraphics[width=.5\textwidth]{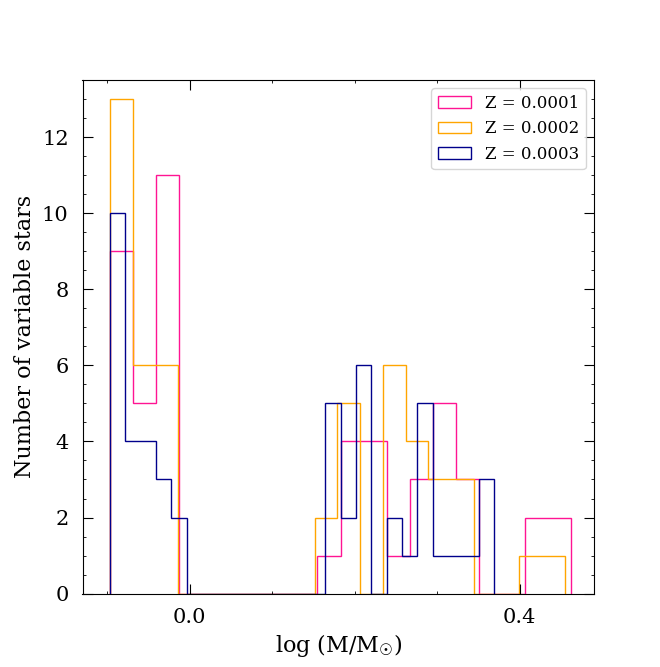}
\centering
\caption{The histogram of the number of variable stars within two half-light radii in metallicities $Z = 0.0001$ (pink), $Z = 0.0002$ (orange), and $Z = 0.0003$ (blue).}
\end{figure}

\section{From variable candidates to Star formation History}

As mentioned, to reconstruct the SFH of And\,IX we will use the candidates of LPVs, since there is a relation between their luminosity and their birth mass. This is possible by Padova stellar evolutionary tracks which link the luminosity of LPVs to their birth mass. The Padova tracks are a comprehensive description of stellar evolution from the first thermal pulse of the AGBs to the post-AGB phase. The effects of circumstellar extinction and different chemical composition of dust are also considered \citep{2017ApJ...835...77M}. The age and pulsation duration of LPVs are derived from the birth mass-age and birth mass-pulsation duration relationships. Suitable coefficients for these relations with a distance modulus of $24.56_{-0.15}^{+0.05}$ mag and metallicities of $Z = 0.0001$, $Z = 0.0002$, and $Z = 0.0003$ were derived by \cite{2021ApJ...923..164S}.

As AGB stars evolve, their bolometric luminosity should increase, but as they become cooler and more dust-enshrouded, their optical brightness will decrease. In other words, AGB stars re-emit the absorbed IR radiation at larger wavelengths. The AGB stars become fainter and redder due to extinction, both from interstellar dust and circumstellar dust. It is therefore necessary to apply a magnitude correction to stars enveloped by dust in addition to the Galactic extinction correction.

We have a magnitude correction for the LPVs with $V-i > 1.3$ mag to bring them back to $V-i = 1$ mag. The slopes of the isochrones for O-rich evolutionary tracks tend to redden faster compared to C-rich tracks. In our sample, O-rich and C-rich stars have average slopes of $3.31$ and $2.37$ mag mag$^{-1}$, respectively. The calculations were performed using the isochrones from Fig.\ 18. Blue points in Fig.\ 18 indicate LPV candidates affected by circumstellar dust, so the magnitude correction will be applied. The correction equation with "a" as the slope of the isochrones is:

\begin{equation}
    i_0 = i + a [(V-i)_0-(V-i)]
\end{equation}

First, a carbon correction equation is applied to our sample by assuming that our LPVs are C-rich. The birth mass is derived using the corrected magnitude in the $i$-band and the relation between birth mass and luminosity. This assumption is correct if $1.5\leq M/M\textsubscript{\(\odot\)}\leq 4$ \citep{2021ApJ...923..164S}. Otherwise, our sample should be de-reddened by the oxygen correction equation (see Section 7 for more details on the mass range). The histogram in Fig.\ 19 shows the number of variable stars in different mass ranges for three metallicities. There are no variable stars with a mass of 1-1.5 M$\textsubscript{\(\odot\)}$ in this galaxy, according to the histogram.

\subsection{Calculation method for star formation rate}

Based on the mass, age, and pulsation duration of the LPVs, the star formation rate (SFR) is calculated. We use the mass-luminosity relation to convert the $i$-band magnitudes of the LPV stars to their masses. There is a correlation between the most luminous point in each isochron and its associated mass. Using a function fitted to all points derived from different isochrones, constant coefficients can be obtained for different luminosity intervals at different magnitude intervals
to calculate the birth mass. Here, we used the coefficients of the best fit of the function reported by \cite{2021ApJ...923..164S} to derive the relationships between mass and luminosity, mass and age, and mass and pulsation duration.

The SFR, $\xi(t)$ (M$\textsubscript{\(\odot\)}$ yr$^{-1}$), as a function of time, is used to derive the SFH. The method used in this paper was adapted from \cite{javadib, 2017MNRAS.464.2103J, 2017MNRAS.466.1764H}, and has also been used to reconstruct the SFH of other dwarf galaxies in the Local Group by \cite{2014MNRAS.445.2214R, 2017MNRAS.466.1764H, 2019MNRAS.483.4751H, 2021ApJ...923..164S, 2021, 2023ApJ...942...33P}. In this method, the SFR is derived based on the initial mass function (IMF) of \cite{2001MNRAS.322..231K} to describe the initial mass distribution of stars, rather than the number of stars. The SFR is calculated by considering the LPV mass in the range from $m(t)$ to $m(t+dt)$, $\delta t$ as the pulsation duration (the total amount of time a star is a LPV), and $dn^\prime$ as the number of stars in each period. The SFR is given by:

\begin{equation}
\xi(t)=\frac{dn^\prime(t)}{\delta t} \frac{\displaystyle\int_{min}^{max} f_{IMF}(m) m\; dm}{\displaystyle\int_{m(t)}^{m(t+dt)} f_{IMF}(m) dm}
\end{equation}

Massive stars evolve more quickly, and their pulsating phases last only a short time. However, low-mass stars spend more time in this phase and are more likely to appear in the pulsation phase. So we use pulsation duration in this formula as a correction factor. By considering a Poisson statistic distribution of the number of stars in each bin as $N$, the statistical error is calculated as, 

\begin{equation}
    \sigma = \frac{\sqrt{N}} {N} \xi(t)
\end{equation}

\subsection{The star formation history in And\,IX}

The birth mass, age, and pulsation duration are calculated from the brightness of LPVs. Stars are sorted by mass and classified into bins with the same number of stars, so the mass span in each interval is specified for IMF integration. 

Metallicity is an important factor in stellar evolutionary tracks, so it could affect the determined SFH as well. According to the estimates, And\,IX's metallicity is [Fe/H] = $-2.2\pm0.2$ dex ($\sim$ $Z = 0.0001$) in \cite{2010MNRAS.407.2411C, 2012AJ....144....4M}, and also \cite{2020ApJ...895...78W} calculated metallicity [Fe/H] = $-2.03\pm0.01$ dex ($\sim$ $Z = 0.0002$). \cite{2013ApJ...779..102K} reported a mean metallicity [Fe/H] = $-1.93\pm0.20$ dex ($\sim$ $Z = 0.0002$), assuming a solar abundance of 12 + log (Fe/H) = $7.52$.
The metallicity of galaxies varies during their evolution, so in this study, we consider metallicity $Z = 0.0002$ as well as a more metal-rich estimate $Z = 0.0003$ in addition to the main metallicity $Z = 0.0001$.

\begin{figure}
\includegraphics[width=.5\textwidth]{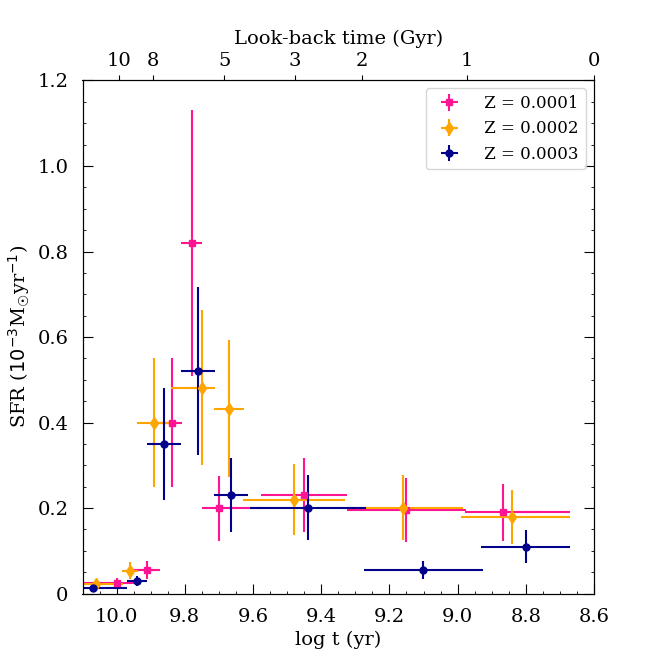}
\centering
\caption{SFHs of And\,IX for metallicities of $Z = 0.0001$ (pink), $Z = 0.0002$ (orange), and $Z = 0.0003$ (blue) within two half-light radii ($\sim$ 0.022 deg$^2$).}
\end{figure}

Fig.\ 20 illustrates SFH in two half-light radii corresponding to metallicities $Z = 0.0001$, $Z = 0.0002$, and $Z = 0.0003$. Each age bin length and the statistical error on the SFR are represented by horizontal and vertical error bars, respectively. In this paper, we use the $\Lambda$CDM standard model with the cosmological parameters, the Hubble constant $H_{0} = 67.3\pm1.2$ km s$^{-1}$ Mpc$^{-1}$, the matter density parameter $\Omega_{m} = 0.315\pm0.017$, and the physical densities of baryons and cold dark matter $\Omega_{b}h^2 = 0.02205\pm0.00028$ and $\Omega_{c}h^2 = 0.1199\pm0.0027$, respectively, to calculate the redshift \citep{2014A&A...571A..16P}.

The star formation epochs occurred between 500 Myr ($log\ t = 8.67$) and $13$ Gyr ago ($log\ t = 10.1$). 
Assuming $Z = 0.0001$, the SFR reached a maximum of $8.2\pm3.1\times10^{-4}$ M$\textsubscript{\(\odot\)}$ yr$^{-1}$ in 6 Gyr ago. During this period, SFR peaked at $4.8\pm1.8\times10^{-4}$ M$\textsubscript{\(\odot\)}$ yr$^{-1}$ and $5.2\pm2.0\times10^{-4}$ M$\textsubscript{\(\odot\)}$ yr$^{-1}$ for metallicities $Z = 0.0002$ and $Z = 0.0003$, respectively.

Comparing the SFRs in different metallicities show that $Z = 0.0001$ produces the highest peak of SFR due to its more metal-poor environment. Generally, the SFR decreases with increasing metallicity except for $1.41$, $5.01$, and $6.02$ Gyr. SFR distributions for metallicities of $Z = 0.0001$ and $Z = 0.0003$ are used for this comparison.
Unlike two other metallicities, the SFR at $Z = 0.0003$ peaks at 630 Myr ago (corresponding the errors, 850$-$457 Myr ago) with a rate of $11.0\pm4.0\times10^{-5}$ M$\textsubscript{\(\odot\)}$ yr$^{-1}$.

Cross-correlation with the \citet{2015ApJ...800...51B} catalog of extreme AGB (x-AGB) stars provided the detection of two x-AGBs. Based on \citet{2015ApJS..216...10B} classification, x-AGB stars are variables with M$_{3.6}$ < 8 mag and colors [3.6] − [4.5] > 0.1 mag. They produce more than 75\% of the dust produced by cool evolved stars, but they represent less than 6\% of the total population of AGBs \citep{2012ApJ...748...40B}. Two stars with masses exceeding 1.5 M$_\odot$ and an estimated 1 Gyr to 1.58 Gyr of age have been identified as carbon stars. As a result of the existence of these C-rich AGB stars, there is a possibility that other C-rich AGBs are responsible for the revival of SFH. Also, these two carbon stars may be older AGBs that going through the bright part of their thermal-pulse cycle. However, more research is necessary to determine if dusty LPV mass-loss has recently increased SFR.
 
A total stellar mass over a specified time is calculated by aggregating $\xi(t)$ over that period. A total stellar mass of $3.0\times10^5$ M$\textsubscript{\(\odot\)}$ obtained for a metallicity of $Z = 0.0001$ within two half-light radii. In And\,IX, stellar mass is reduced by $20\%$ ($\sim 2.4\times10^5$ M$\textsubscript{\(\odot\)}$) and $23\%$ ($\sim 2.3\times10^5$ M$\textsubscript{\(\odot\)}$), in metallicities of $Z = 0.0002$ and $Z = 0.0003$, respectively. This study observed a decrease in total stellar mass due to increasing metallicity. Considering a metallicity of [Fe/H] = $-2.03\pm0.01$ ($\sim$ $Z = 0.0002$), \cite{2020ApJ...895...78W} estimated a stellar mass of $\sim$ $2.4\times10^5$ M$\textsubscript{\(\odot\)}$ ($log (M/M\textsubscript{\(\odot\)}$) = $5.38\pm0.44$). According to the paper mentioned, the estimated stellar mass is similar.
\begin{figure}
\includegraphics[width=.50\textwidth]{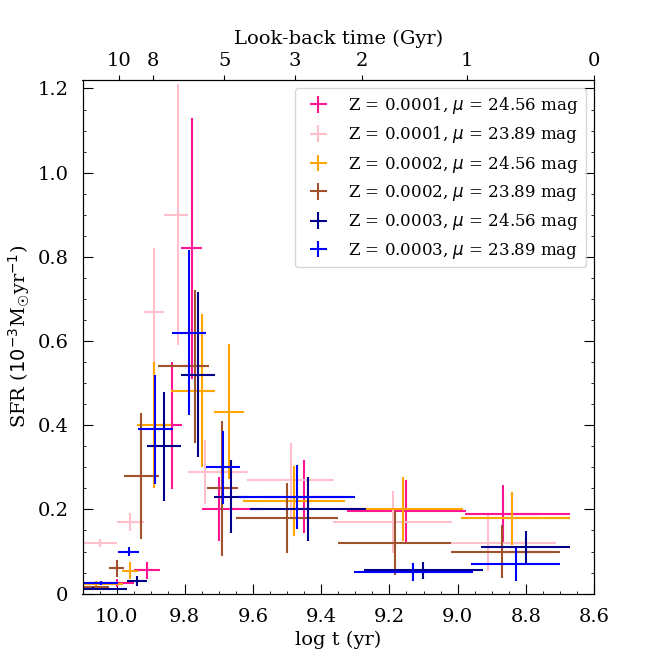}
\centering
\caption{SFHs of And\,IX for three metallicities of $Z = 0.0001$, $Z = 0.0002$, and $Z = 0.0003$ within two half-light radii ($\sim$ 0.022 deg$^2$) in distance modulus of $23.89_{-0.08}^{+0.31}$ mag \citep{2019MNRAS.489..763W} and $24.56_{-0.15}^{+0.05}$ mag.}
\end{figure}

While changing the distance modulus and Galactic extinction do not affect the age–mass, and pulsation-duration–mass relations, it will change the magnitude of LPV stars and hence their mass and the SFH of the galaxy. Fig.\ 21 and Fig.\ 22 are plotted separately in order to illustrate the potential differences between the effect of the distance modulus and Galactic extinction on SFRs. A comparison of the SFHs of And\,IX for distance modulus derived in this paper ($24.56_{-0.15}^{+0.05}$ mag) and the lowest distance modulus reported ($23.89_{-0.08}^{+0.31}$ mag \citep{2019MNRAS.489..763W}) is shown in Fig.\ 21. As can be seen, the SFHs exhibit similar behavior and there is only a shift towards recent times with increasing distance modulus. The Galactic extinctions for And\,IX are estimated A$_i$ = 0.127 mag (E(B-V) = 0.075 \citep{2012AJ....144....4M}) and A$_i$= 0.129 mag (E(B-V) = 0.076 \citep{2012ApJ...758...11C}) from the \citet{2011ApJ...737..103S}. In Fig.\ 22, we compare the SFH of the galaxy, assuming the average of the reported Galactic extinctions (A$_i$ = 0.128 mag), with the SFH of And\,IX, which does not account for Galactic extinction (A$_i$ = 0 mag). There is no significant difference in results, but the SFH is shifted to earlier epochs when the Galactic extinction is ignored.

We estimate the total stellar mass of And\,IX in the lowest distance modulus ($23.89_{-0.08}^{+0.31}$ mag \citep{2019MNRAS.489..763W}) $\sim 3.50\pm0.50\times10^5$ M$\textsubscript{\(\odot\)}$ with A$_i$ = 0.128 mag, and $\sim 3.10\pm0.30\times10^5$ M$\textsubscript{\(\odot\)}$ with A$_i$ = 0 mag, at $Z = 0.0001$. In metallicity of $Z = 0.0002$, the total stellar mass is $\sim 2.70\pm0.40\times10^5$ M$\textsubscript{\(\odot\)}$ with A$_i$ = 0.128 mag and $\sim 2.60\pm0.10\times10^5$ M$\textsubscript{\(\odot\)}$ with A$_i$ = 0.
The total stellar mass of And\,IX at $Z = 0.0003$ is $\sim 2.50\pm0.40\times10^5$ M$\textsubscript{\(\odot\)}$ and $\sim 2.46\pm0.20\times10^5$ M$\textsubscript{\(\odot\)}$ considering A$_i$ = 0.128 mag and A$_i$ = 0 mag, respectively.

\begin{figure}
\includegraphics[width=.50\textwidth]{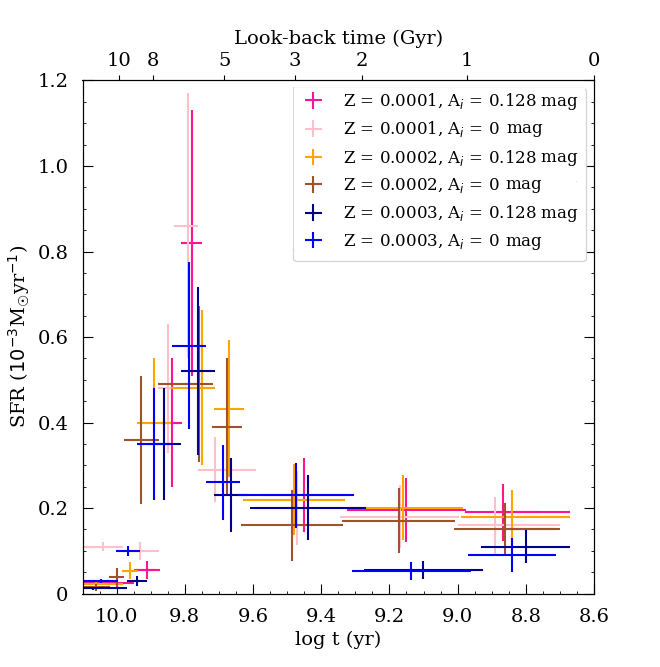}
\centering 
\caption{SFHs of And\,IX for three metallicities of $Z = 0.0001$, $Z = 0.0002$, and $Z = 0.0003$ within two half-light radii ($\sim$ 0.022 deg$^2$) with applying the Galactic extinction in filter $i$ (A$_i$ = 0.128 mag) and without considering the extinction (A$_i$ = 0 mag).}
\end{figure}

\begin{figure}
\includegraphics[width=.50\textwidth]{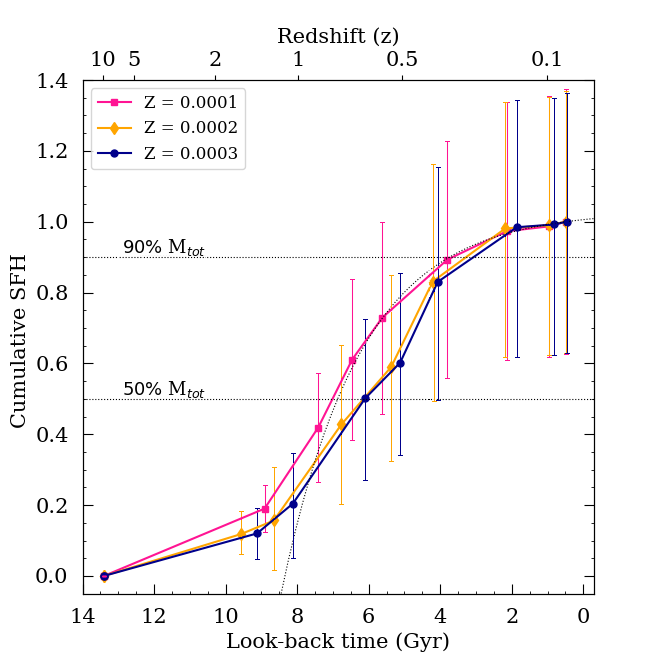}
\centering
\caption{Cumulative SFH as a function of look-back time and redshift within two half-light radii of And\,IX for three different adopted metallicities. Each bin is accompanied by a vertical error bar showing the statistical error of the SFR. The pink curve is also fitted by a $\tau$-model exponential function (the dotted black curve). Statistical errors in SFRs indicate the cumulative SFH errors for metallicities of $Z = 0.0001$ (pink error bars), $Z = 0.0002$ (orange error bars), and $Z = 0.0003$ (blue error bars).}
\end{figure}

As a function of look-back time and redshift, Fig.\ 23 shows the cumulative star formation (colors based on Fig.\ 20). The best exponential fit for our star formation model was obtained with the $\tau$-model with $\tau$ = $5$. The equation describes SFH with declining e-folding time $"\tau"$, which is initiated at $"t_i"$ with amplitude "$A$" \citep{Simha2014ParametrisingSF};

\begin{equation}
SFR (t) \propto A e^{-(t-t_i)/\tau}.
\end{equation}

According to Fig.\ 23, the horizontal dashed line labeled 90\% M$_{tot}$ indicates the time it took to assemble $90\%$ of the stellar mass (known quenching time). In addition, Fig.\ 23 shows the epoch by which $50\%$ of the stellar mass had been formed with a horizontal dashed line indicating 50\% M$_{tot}$. Table 4 summarizes the aggregation time for half ($t_{50}$) and $90\%$ ($t_{90}$) of the total stellar mass in different metallicities of And\,IX. The first line in $t_{50}$ and $t_{90}$ of Table 4 is based on SFHs in Fig.\ 20 and distance modulus of $24.56_{-0.15}^{+0.05}$, the second line is based on the SFH with distance modulus $23.89_{-0.08}^{+0.31}$ mag in Fig.\ 21. According to Table 4, the aggregation time for 50\% and 90\% of the total stellar mass move to the recent times in three metallicities by increasing the distance modulus. Additionally, increasing metallicity has a similar effect on t$_{50}$ and t$_{90}$ at the same distance modulus.
Our results are consistent with \cite{2019ApJ...885L...8W}, who estimated the quenching time as $5.1_{-2.0}^{+1.8}$ Gyr ago and the aggregation of half of the total stellar mass as $7.2_{-0.3}^{+2.5}$ Gyr ago.

\begin{table*}[t]
\begin{center}
\centering
\caption{Time to aggregation for half of the total stellar mass ($t_{50}$) and $90\%$ of the total stellar mass ($t_{90}$) in different metallicities.}
\begin{tabular}{c c c c c}
\hline\hline

&  $\mu (mag)$ &  $Z = 0.0001$ & $Z = 0.0002$ & $Z = 0.0003$ \\[1ex]

\hline
 $t_{50}$ (Gyr) & $24.56_{-0.15}^{+0.05}$ \footnote{Estimates based on the distance modulus calculated in this paper} & $7.02_{-0.56}^{+0.39}$ & $6.15_{-1.02}^{+0.77}$ & $6.10_{-0.97}^{+0.36}$\\[1ex]
 & $23.89_{-0.08}^{+0.31}$ \footnote{Estimates based on the distance modulus calculated in \cite{2019MNRAS.489..763W}} & $8.00_{-1.30}^{+0.10}$ & $7.20_{-0.70}^{+0.80}$ & $7.10_{-1.40}^{+0.30}$\\[1ex] 

 \tableline
$t_{90}$ (Gyr) & $24.56_{-0.15}^{+0.05}$ & $3.65_{-1.52}^{+0.13}$ &  $3.29_{-1.16}^{+0.97}$ & $3.07_{-1.39}^{+1.00}$ \\[1ex]
& $23.89_{-0.08}^{+0.31}$ & $3.95_{-0.50}^{+2.00}$ &  $3.40_{-0.60}^{+0.40}$ & $3.10_{-0.40}^{+0.30}$ \\[1ex]

\tableline
\centering
\end{tabular}
\end{center}
\end{table*}

\begin{figure}
\includegraphics[width=.52\textwidth]{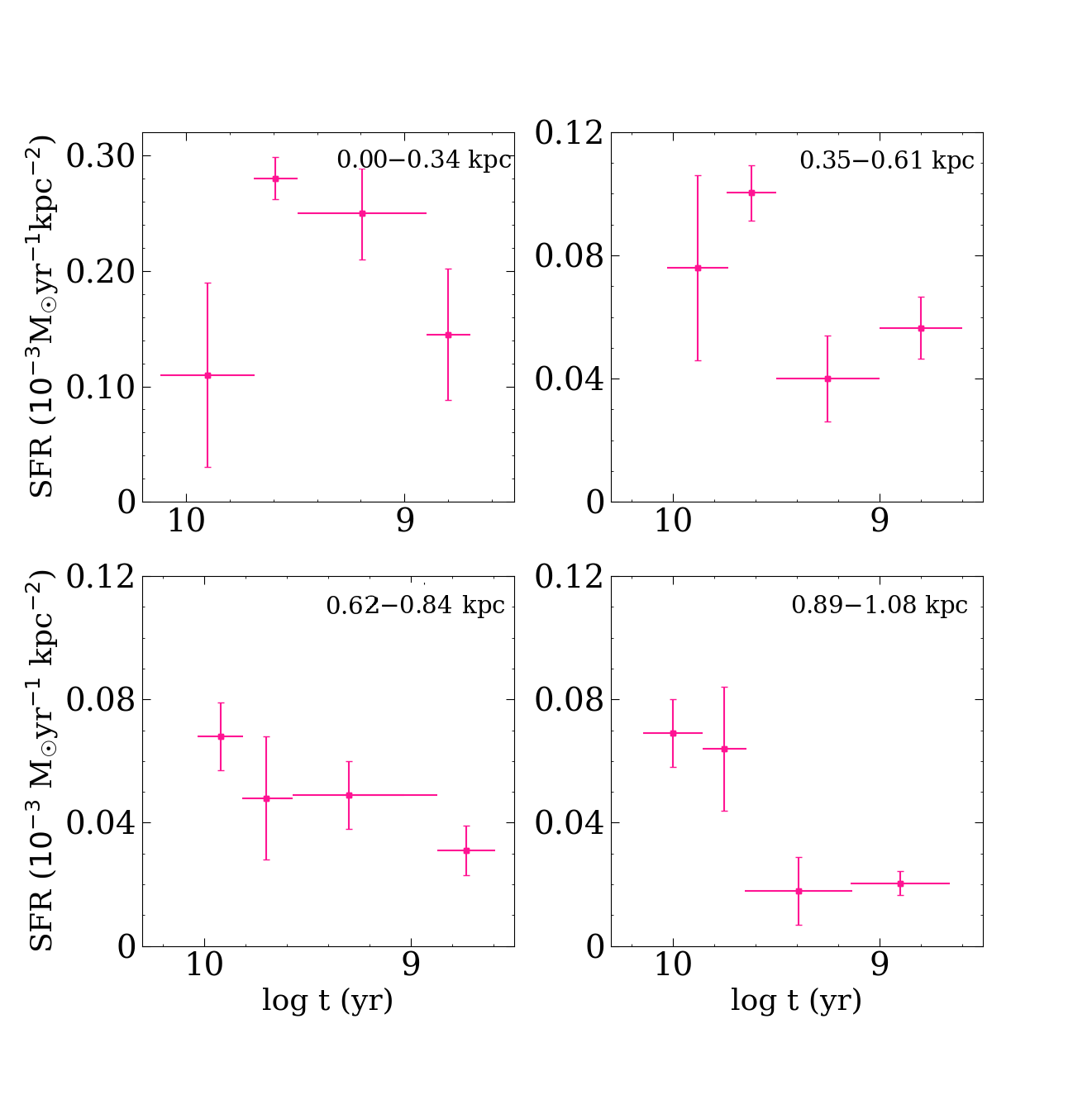}
\centering
\caption{SFR of And\,IX per unit area within four regions at galactocentric radii for a constant metallicity of $Z = 0.0001$.}
\end{figure}
 
\subsection{Radial star formation history} 

SFHs in different radial regions within a galaxy may reveal important information regarding the galaxy's formation history. For this, we divided the area of And\,IX into different radius bins with an equal number of stars to derive the radial gradient of SFH. In Fig.\ 24, the density of the SFR is plotted as a function of logarithmic time in four circular regions for metallicity $Z = 0.0001$. Due to the four annuls that make up And\,IX, each region may have a different population of stars.
The ratio of the SFR for $t > 3.16$ Gyr ago ($log\ (t) > 9.5$) to that for $t < 3.16$ Gyr ago ($log\ (t) < 9.5$) in the innermost region is $0.990_{-0.010}^{+0.005}$, while in the second region ($\sim$ $0.35-0.61$ kpc) it is $1.82_{-0.04}^{+0.07}$. The fraction reaches $1.45_{-0.06}^{+0.03}$ in the third region and $3.48_{-0.40}^{+0.91}$ in the outermost region. When moving toward central regions, this ratio decreases, suggesting that younger populations tend to be concentrated there. As a result, star formation began in the outer part and gradually spread inward. In these regions, $20\%$ of the total mass per unit area is formed at the outermost radius, $23\%$ at the middle regions, and the rest at the innermost radius. Therefore, the total stellar mass per unit area is more concentrated in the innermost region.

\begin{figure}
\includegraphics[width=.515\textwidth]{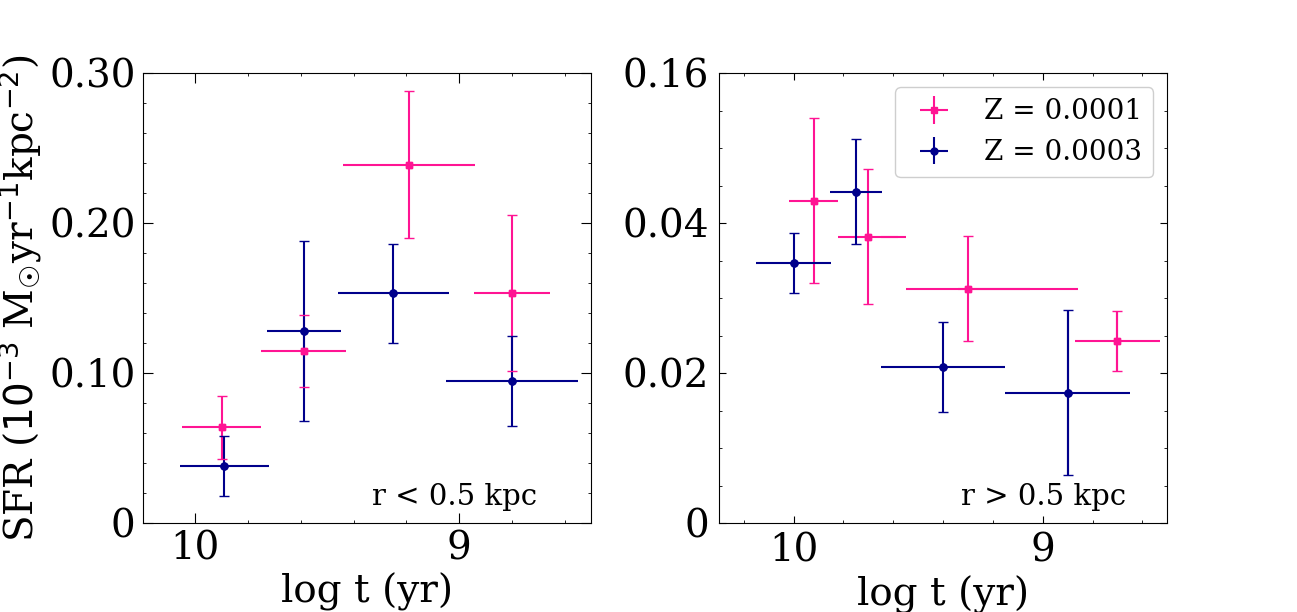}
\centering
\caption{SFR of And\,IX per unit area with equal numbers of LPV candidates within two regions at galactocentric radii for metallicities of $Z = 0.0001$ (pink) and $Z = 0.0003$ (blue).}
\end{figure}

In Fig.\ 25, we divided our sample into two regions ($r < 0.5$ and $r > 0.5$ kpc) to examine the radial gradient of SFH for metallicities $Z = 0.0001$ (in pink) and $Z = 0.0003$ (in blue). The star formation pattern suggests the highest SFRs in central regions at later times, which is the opposite of the outer regions.
In support of this, the fractions of $0.720_{-0.002}^{+0.003}$ in $r < 0.5$, and $2.13_{-0.29}^{+0.77}$ in $r > 0.5$ are obtained by estimating the ratio of the older population at $Z = 0.0001$ compared to the younger population at $Z = 0.0003$. When $r > 0.5$ kpc (estimated ratio greater than unity), older populations form earlier than when $r < 0.5$ kpc (estimated ratio less than unity). Obviously, this supports the outside-in formation scenario of And\,IX based on the different age gradients of the population in the inner and outer parts of the galaxy \citep{hidalgo2008disc, 2013ApJ...778..103H, 2015Llambay}. The dynamical effect could be another scenario for the distribution of stars in And\,IX, where stars migrate outwards after forming in more central regions. This is not unexpected as star formation only occurs if gas cools and falls deeply into the gravitational potential well of a small halo such as And\,IX; this is a highly non-equilibrium state, and internal dynamics would gradually cause the stars to fill the gravitational potential well (i.e., migrate outward); tidal stress would exacerbate that.

\subsection{Quenching mechanisms in And\,IX}

Several mechanisms lead to the quenching of a dwarf galaxy. Depletion of cold gas in the re-ionization era is supposed to affect the shutting-down of star formation in ultra-faint dwarfs (M$_V$ > $-6$ mag) and low-mass galaxies ($M$ < $10^5$ M$\textsubscript{\(\odot\)}$) \citep{2014ApJ...793...29G, 2016ApJ...833...84X, 2019MNRAS.490.4447W, 2021ApJ...906...96A}. At $z = 10$, low-mass galaxies are expected to be quenched by cosmic re-ionization, while for more massive galaxies, environmental processes have a more significant effect on the cessation of star formation at $z = 6$ \citep{2015ApJ...808L..27W}. It can be ruled out with greater certainty that the epoch of re-ionization will not affect And\,IX's quiescent since the SFRs are shut-down by $z = 6$. Environmental effects, such as ram-pressure stripping, tidal effects, and dwarf-galaxy interactions, may quench dwarfs with stellar masses of $10^5-10^7$ M$\textsubscript{\(\odot\)}$. In particular, ram-pressure stripping is a noticeable mechanism to stop star formation in galaxies with $M < 10^7$ M$\textsubscript{\(\odot\)}$ \citep{2018quenching}.

Another factor in the shutting-down of star formation of a satellite can be the fall in the virial radius of its host galaxy. For the Milky Way dwarf satellite galaxies, the infall time can be calculated using positions, line-of-sight velocities, and proper motions (if measured) \citep{2012Rocha}.
Since these data are not available for satellites of Andromeda, cosmological simulation is used to estimate the infall time. In a study by \cite{2015ApJ...808L..27W}, all Milky Way and Andromeda satellites with $M < 10^8$ M$\textsubscript{\(\odot\)}$ have been quenched after falling into their host galaxy's virial radius of fewer than $2$ Gyr, and quenching is more rapid at lower stellar masses. This study estimated infall time using N-body simulations. Furthermore, \cite{2021MNRAS.504.5270D} proposed a correlation between quenching time and the time when satellites enter the virial halo of their hosts (accretion time). 
In M31, massive accretion occurred around $5.5$ Gyr ago, around the time most M31 satellites quenched and also our estimate. Using 20 satellites of M31, \cite{2019ApJ...885L...8W} determined a look-back time of $3-6$ Gyr for the assembly of $90\%$ of stellar mass and $6-9$ Gyr for the assembly of $50\%$ of stellar mass. Our study also confirms these results. For metallicities of $Z = 0.0004$ and $Z = 0.0007$, \cite{2021ApJ...923..164S} estimated the quenching time of And\,I about 4 Gyr ago. Similarly, And\,VII, another satellite of M31, was quenched 5 Gyr ago at $Z = 0.0007$ and $5.7$ Gyr ago at $Z = 0.0004$ \citep{2021}. It is also consistent with the quenching time reported in \cite{2019ApJ...885L...8W}.

Environmental processes play an essential role in quenching And\;IX, a satellite with a stellar mass of $M \leq$ 10$^{8}$ M$\textsubscript{\(\odot\)}$. 
These processes include tidal effects and the depletion of cold gas through M31 (due to proximity). Internal processes also quench star formation, such as galactic winds, supernovae, and stellar feedback, specifically in low-mass dwarfs \citep{2020MNRAS.493..638L}.

\section{Probing of dust in And\,IX}

\subsection{C-rich and O-rich circumstellar envelopes}

The carbon abundance in the atmosphere of AGB stars increases after the third dredge-up process, despite the abundance of oxygen before it. Based on carbon and oxygen abundances in the atmosphere, AGBs are generally classified as carbon-rich (C/O > 1) or oxygen-rich (C/O < 1) \citep{2022Univ....8..465R}. As metal-poor environments have less oxygen, carbon stars form more easily since less carbon must be dredged-up to achieve C/O > 1 \citep{2008, 2022Univ....8..465R}. AGB star models estimate different thresholds for the third dredge-up, some of which are summarised below. In Magellanic Clouds (MCs), the mass range for C-rich stars is between $1.3-4$ M$\textsubscript{\(\odot\)}$ \citep{2005A&A...442..597V}; \cite{2007A&A...462..237G} estimate the mass ranges of C- and M-type stars using $1.5-2.8$ M$\textsubscript{\(\odot\)}$ for C-rich stars in MCs clusters. According to \citet{2012MNRAS.427.2647M}, the condition of C/O > 1 was first achieved in stars around 1 M$\textsubscript{\(\odot\)}$ in the Sgr dSph galaxy with metallicity Z $\sim 4\times10^{-3}$. In LMC, \cite{2017MNRAS.465..403G} estimations of $M \leq$ 1.5 M$\textsubscript{\(\odot\)}$ and $M \geq$ 4 M$\textsubscript{\(\odot\)}$ for O-rich stars are based on 1612 MHz circumstellar OH maser emissions from AGB stars and RSGs.

Given the low oxygen abundance in And\,IX, a range of $1.5-4$ M$\textsubscript{\(\odot\)}$ is adopted for the carbon stars \citep{2021ApJ...923..164S}. In our survey, no LPVs were found in the mass range of $1-1.5$ M$\textsubscript{\(\odot\)}$, thus substantial differences are not seen in the number of C- and O-rich LPVs based on these assumptions. As a result, the SFRs are almost the same regardless of the precise choice of the lower end of the C-rich star mass range. The chemical type of stars is determined by their masses, as described in Section 6.

\subsection{SED modelling through {\sc dusty}}

\begin{figure}
\includegraphics[width=.52\textwidth]{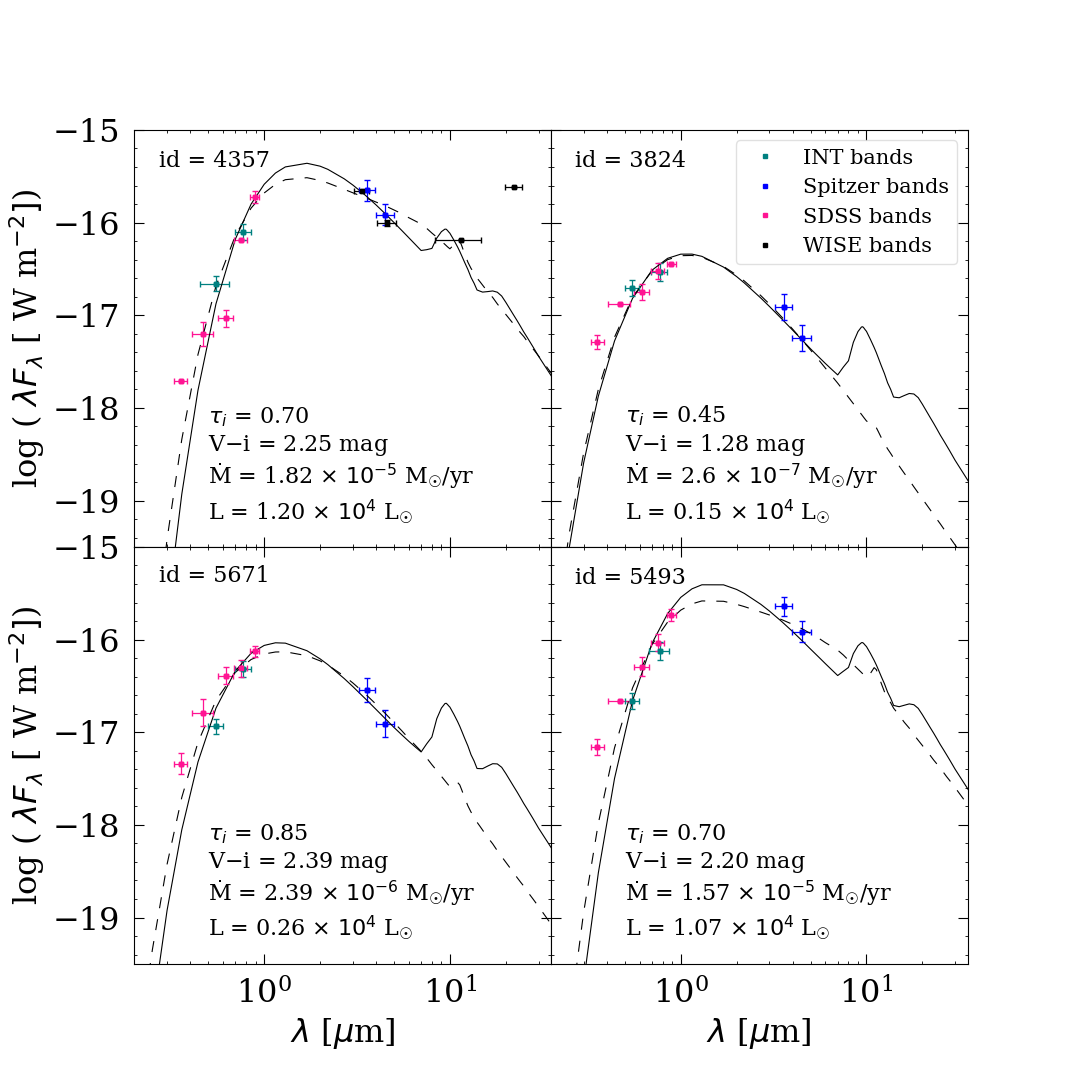}
\centering
\caption{Example SEDs of LPVs with the best fit for the C- and O-rich type of AGBs (dashed and solid black lines). Fluxes are modelled by {\sc dusty} as a function of wavelength. Fluxes observed in different bands with the INT ($i$ and $V$), {\it Spitzer} (3.6 and 4.5 $\mu$m), SDSS ($u$, $g$, $r$, $i$, and $z$), and WISE ($W_1$, $W_2$, $W_3$, and $W_4$) are shown by green, blue, pink, and black squares, respectively. Vertical and horizontal error bars show photometric uncertainty in the magnitude and the difference between the $\lambda_{max}$ and $\lambda_{min}$ around each filter's central wavelength, respectively.}
\end{figure}
In this paper, we modelled the SEDs of variables through the {\sc dusty} code, which was written by \cite{1997}. Modelling the SED requires input data such as stellar temperature, external radiation characteristics, dust properties (e.g., temperature, chemical compositions, grain size distribution), optical depth, and envelope density distribution. We put the star's and dust's temperatures at the inner edge of circumstellar envelope $3000-3500$ K and $500-1200$ K, respectively \citep{1999A&A...347..594G, 2018A&ARv..26....1H}. The C-rich type is made through the mixture of $85\%$ amorphous carbon \citep{hanner1988infrared} and $15\%$ silicon carbide \citep{pegourie1988optical}, while the O-rich type is made through the use of astronomical silicates \citep{1984ApJ...285...89D}. In solving hydrodynamic equations for AGB stars, {\sc dusty} assumes radiatively driven wind. When solving the equation, the default parameters (gas-to-dust mass ratio $\psi\textsubscript{\(\odot\)}$ = 200, $L = 10^4$ L$\textsubscript{\(\odot\)}$, and $\rho_{dust} = 3$ \; g cm$^{-3}$) are adopted. In this case, $\psi$ is scaled by the relation $\psi = \psi\textsubscript{\(\odot\)} 10^{-[Fe/H]}$ assuming the reverse metallicity relation with the gas-to-dust mass ratio \citep{2005A&A...442..597V}. Based on trial and error, the optical depth is estimated by comparing the simulated SED to the observed SED. As a result of the scaling relations, the mass-loss rates of LPVs are determined using the following equation \citep{1999LPICo.969...20N}:

\begin{equation}
    \dot {M} = \dot M_{DUSTY} \left(\frac{L}{10^4}\right)^{3/4} \left(\frac{\psi}{\psi_\textsubscript{\(\odot\)}}\right)^{1/2} 
\end{equation}

Examples of SEDs obtained with INT ($i$ and $V$), {\it Spitzer} (3.6 and 4.5 $\mu$m), SDSS ($u$, $g$, $r$, $i$, and $z$), and WISE ($W_1$, $W_2$, $W_3$, and $W_4$) are shown in Fig.\ 26. The best fit curves are constructed for two types of dust (solid and dashed black curves for O- and C-rich LPVs, respectively). Besides the best fit for the preferred dust species, we also show the best fit for the alternative dust species. Table 6 in the Appendix describes the physical properties of LPVs, which contains information such as star id, coordinates, magnitude ($V$) and error of magnitude ($\delta V$) in Harris $V$ filter, magnitude ($i$) and error of magnitude ($\delta i$) in Sloan $i'$ filter, amplitude in Sloan $i'$ filter (amplitude$_i$), birth mass (M$_{Birth}$), optical depth in Sloan $i'$ ($\tau_i$), mass-loss rate ($\dot{M}$), luminosity (L), and chemical type of LPVs. 

\begin{figure}
\includegraphics[width=.52\textwidth]{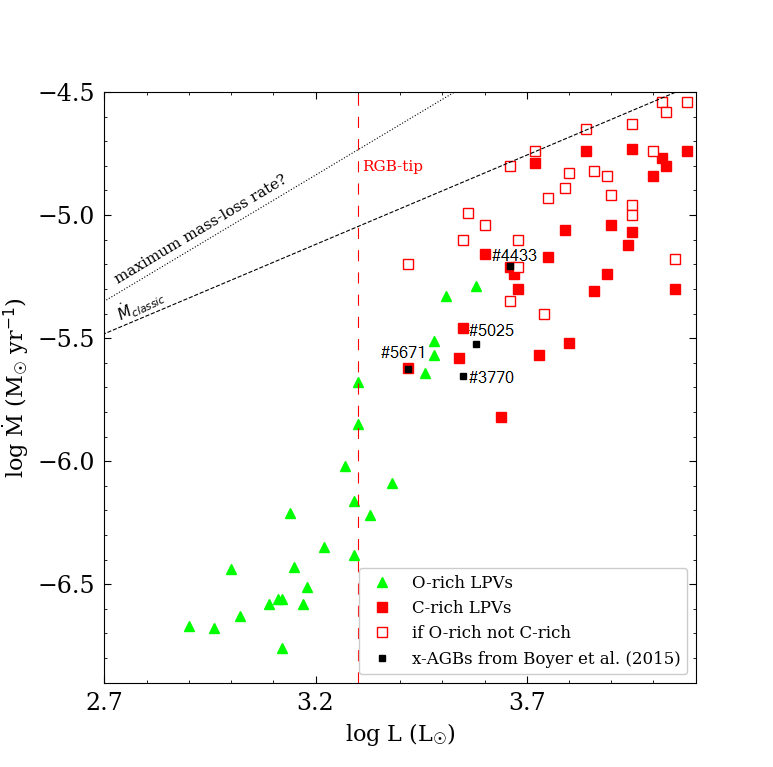}
\centering
\caption{Mass-loss rate as a function of luminosity for C-rich (red squares) and O-rich (green triangles) LPVs within two half-light radii of And\,IX. The open red squares would show the results if the carbon stars were assumed instead to be O-rich.
A dotted line represents the classical single scattering mass-loss limits $\dot M_{classic}$ $\propto$ L$^{0.75}$, and the mass-loss limits when multiple scattering of photons becomes important (maximum mass-loss rate?) is represented by the dashed line \citep{1999A&A...351..559V}. The tip luminosity of the RGB is marked with the red vertical dashed line. Four x-AGBs of \cite{2015ApJ...800...51B} are highlighted with their ids, two of which are mutual in {\it Spitzer} and INT surveys.}
\end{figure}

Fig.\ 27 shows mass-loss rates for C-rich (red squares) and O-rich (green triangles) LPVs as a function of luminosity. In order to assess the effect of chemical types of dust on optical depth and mass-loss ratio, carbon stars were assumed to have oxygen dust (open red squares). Changing optical depth under the effect of dust species alters the mass-loss ratio and luminosity of this sample.
Based on \cite{1999A&A...351..559V}'s paper, this figure shows the maximum and classical limit of mass-loss rates. The maximum mass-loss rate ridgeline is an extreme envelope of rates that were measured once in the past \citep{1999A&A...351..559V}, and given the uncertainties in those data and modelling assumptions, the actual limit may be somewhat lower; hence we added a question mark to it.
The black squares in this figure indicate \cite{2015ApJ...800...51B}'s x-AGBs with specific ids. We have used SED modeling to estimate all four x-AGBs' mass-loss (and luminosity). 
As shown in this figure, the two x-AGBs are mutual with the INT and {\it Spitzer} surveys, and two other \cite{2015ApJ...800...51B}'s variables were identified in the INT survey as non-variables. 
This plot shows that LPVs in And\,IX have mass-loss rates of $1.7\times10^{-7}$ $\leq$ $\dot{M}$ $\leq$ $1.9\times10^{-5}$ M$\textsubscript{\(\odot\)}$ yr$^{-1}$ and luminosity of $8.0\times10^{2}$ $\leq L \leq$ $1.2\times10^{4}$ L$\textsubscript{\(\odot\)}$. It is estimated that the mass-loss rate of the two mutual x-AGBs is about $3.6\%$ of the total mass return rate, while if four x-AGBs were identified, this rate would be $5.7\%$.
The maximum mass-loss rate of our LPVs ($\sim$ $10^{-5}$ M$\textsubscript{\(\odot\)}$ yr$^{-1}$) is around 2 dex less than the upper limit of maximum mass-loss rates of AGBs in the LMC ($10^{-7}-10^{-3}$ M$\textsubscript{\(\odot\)}$ yr$^{-1}$) \citep{1999A&A...351..559V}. 
In more metal-poor environments, such as in WLM galaxy with metallicity of [Fe/H] = $1.13$, the mass-loss rate of AGBs is calculated to be $10^{-10}-10^{-4}$ M$\textsubscript{\(\odot\)}$ yr$^{-1}$ for O-rich AGBs and $10^{-10}-10^{-5}$ M$\textsubscript{\(\odot\)}$ yr$^{-1}$ for C-rich ones \citep{2007ApJ...656..818J}.

\begin{table*} [t]
\begin{center}
\caption{Estimations of the mass-loss ratios of And\,IX LPVs in different metallicities.}
\begin{tabular}{c c c c c}
\tableline
\tableline
Metallicity  & $\dot{M}$ \footnote{Total mass return rate} ($10^{-4}$ M$\textsubscript{\(\odot\)}$ yr$^{-1}$) & $\dot{M}$ ($10^{-4}$ M$\textsubscript{\(\odot\)}$ yr$^{-1}$) (if all O-rich) & $\dot{M}$/$M_{LPV}$ \footnote{Total stellar mass of LPVs} ($10^{-6}$ yr$^{-1}$) & Specific mass-loss rate ($10^{-10}$ yr$^{-1}$)\\[1ex]

\tableline
$Z = 0.0001$ & $2.4$ & $3.7$ & $3.0$ & $8.0$\\[1ex] 

$Z = 0.0002$ & $1.5$ & $2.4$ & $2.0$ & $6.0$\\[1ex]

$Z = 0.0003$ &  $1.0$ & $1.7$ & $1.0$ & $4.0$ \\[1ex]

\tableline
\end{tabular}
\end{center}
\end{table*}

Mass-loss rate increases during stellar evolution along the AGB phase synchronously with luminosity (and hence birth mass) \citep{2018A&ARv..26....1H}. 
Fig.\ 27 illustrates the same point, as most luminous stars generally reach higher mass-loss rates. If LPVs with masses greater than 1.5 M$\textsubscript{\(\odot\)}$ are assumed to be O-rich, the mass-loss rate would be higher. Due to the lower opacity of silicates compared to amorphous carbon grains, fitting carbon stars with silicates usually results in higher mass-loss rates. In this diagram, the carbon star with luminosity $0.89\times10^{4}$ L$\textsubscript{\(\odot\)}$ has the most mass-loss $1.87\times10^{-5}$ M$\textsubscript{\(\odot\)}$ yr$^{-1}$. With silicon dust, the mass-loss of this star increases by $25\%$. As illustrated in Fig.\ 27, mass-loss variance is evident around the RGB-tip in a given luminosity (birth mass), indicating star evolution throughout the galaxy \citep{2013MNRAS.432.2824J}. When a star climbs the AGB, its mass-loss increases gradually with luminosity, allowing different mass-losses to be observed at the same luminosity. Carbon stars tend to have higher luminosities and mass-losses because of their higher mass; however, some carbon stars have lower luminosities ($log\ L$ < 3.6 L$\textsubscript{\(\odot\)}$ in Fig.\ 27) and mass-losses than oxygen stars. 
These carbon stars may be in the inter-thermal pulse luminosity dip and experience lower luminosity and mass-loss rate, whereas, after thermal pulses, they experience higher luminosity and mass-loss rate \citep{1993ApJ...413..641V, mattsson2007mass}. On the other hand, the metallicity gradient in galaxies can also lead to overlaps between carbon and oxygen stars, which seem unlikely here. Even if there is a metallicity gradient in this galaxy, since we did not detect any stars in the range of 1 to 1.5 M$\textsubscript{\(\odot\)}$, changing this limit due to metallicity does not affect our results. The limit at which oxygen stars turn into carbon stars is affected by metallicity. Oxygen stars with higher mass-loss and luminosity than carbon stars have spent more time evolving in the AGB phase, the luminosity of these carbon stars can also increase with further evolution in this phase \citep{2008A&A...482..883M}.

The results of the total mass return rate and the ratio of the total mass return to the total stellar mass at three different metallicities are summarized in Table 5. Moreover, we estimated the ratio of the total mass return rate to the total stellar mass of LPVs. This ratio is a measure of the duration of the dominant mass-loss phase or rather the inverse of it. In fact, it sets an upper limit to the duration, as the stars do not completely vanish but leave remnants (white dwarfs, neutron stars, or black holes). With $Z = 0.0001$, the timescale is $3.0\times 10^5$ yr, which is a few times shorter than the radial pulsation-phase timescale of $\sim 10^6$ yr as predicted by the models \citep{2021ApJ...923..164S}. Based on this, the more extreme phase of the LPV and heavy mass-loss seems to last for a shorter period of time than the whole TP-AGB lifetime.
Also, we estimated the specific mass-loss rate as the total mass return by the total stellar mass in each metallicity. It is estimated that the mass-loss of LPVs in about a billion years could enrich the ISM and revive star formation in the galaxy. Despite this, the mass of the ISM may be impacted by the interaction with the M31 galaxy. We can refer to \cite{2019MNRAS.483.1314B}'s simulation for more information on the possibility of mass-loss in satellites in LG, such as And\,IX, via stripping. Based on this simulation, it was found that satellite galaxies have a lower present-day stellar mass ($z = 0$) than the maximum total mass they have ever reached during their evolution. Accordingly, satellite galaxies with present-day stellar masses of order $\sim 10^6$ M$\textsubscript{\(\odot\)}$ had a maximum stellar mass of order $\sim 10^9$ M$\textsubscript{\(\odot\)}$ \citep{2019MNRAS.483.1314B}.

The total stellar mass and the total mass return rate in And\,IX are decreased with increasing metallicity. So the specific mass-loss rate follows this pattern. C-rich and O-rich LPVs have total mass-loss rates of $2.1\times10^{-4}$ M$\textsubscript{\(\odot\)}$ yr$^{-1}$ and $2.9\times10^{-5}$ M$\textsubscript{\(\odot\)}$ yr$^{-1}$ at $Z = 0.0001$, respectively. As carbon stars account for about 80\% of the total mass return rate in three metallicities, most of the dust that enters the ISM by LPVs is carbon dust.

\section{Summary}

From June 21, 2015, to October 6, 2017, nine observations were undertaken to study the properties of And\,IX, the dwarf spheroidal satellite of the M31. Observations were conducted using the 2.5-m wide-field camera (WFC) of the INT with the Sloan $i'$ and Harris $V$ filters. We detected 54 AGB candidates within two half-light radii ( $\sim$ 5 arcmin) of And\,IX by employing {\sc daophot/allstar} software \citep{1987PASP...99..191S, 1990PASP..102..932S, 1996PASP..108..851S}. About 50 of them are LPVs with amplitude$_i>0.2$ mag. We calculated the SFRs as a galaxy evolution tracer in two half-light radii for metallicities of $Z = 0.0001, 0.0002, 0.0003$. Due to the temperature and radius variations, LPVs experience significant mass-loss in the form of a stellar wind. We measure mass-loss rates using multi-wavelength data from INT, {\it Spitzer}, SDSS, and WISE. Our primary conclusions are as follows:

\begin{itemize}
\item The maximum rate of star formation $\sim$ $8.2\pm3.1\times10^{-4}$ M$\textsubscript{\(\odot\)}$ yr$^{-1}$ occurred $6$ Gyr ago at $Z = 0.0001$.
Compared to the peak of SFR in the more metal-rich estimation ($\sim 5.2\pm2.0\times10^{-4}$ M$\textsubscript{\(\odot\)}$ yr$^{-1}$ at $5.7$ Gyr ago), the peak of SFR in the more metal-poor estimation is $57\%$ higher.
\item The total stellar mass is estimated $\sim 3.0\times10^5$ M$\textsubscript{\(\odot\)}$ ($Z = 0.0001$), which decreased up to $2.3\times10^5$ M$\textsubscript{\(\odot\)}$ by increasing metallicity to $Z = 0.0003$. Furthermore, based on And IX's cumulative SFH, 90\% of its total stellar mass was formed until $\sim 3.65_{-1.52}^{+0.13}$ Gyr ago at $Z = 0.0001$, indicating that this galaxy had an extended SFH. Furthermore, half of the mass of the And\,IX was formed about $7.02_{-0.56}^{+0.39}$ Gyr ago in more metal-poor estimation. Consequently, our results imply that this dSph satellite was quenched late, possibly due to late infall.
\item And\,IX shows a late epoch of star formation, peaking
around 630 Myr ago, but the SFR is low $1.1\pm0.4\times10^{-4}$ M$\textsubscript{\(\odot\)}$ yr$^{-1}$. In this quenched galaxy, dusty stellar winds at earlier times may have contributed to this late
epoch of star formation.
\item According to the different age gradients of the population in the inner and outer parts of the galaxy, the
outside-in star formation scenario could be a galaxy
evolution scenario. Furthermore, the separation of population ages might result from stars migrating outward
after forming in more central regions.
\item We estimated the mass-loss rate of LPVs ($10^{-7}-10^{-5}$ M$\textsubscript{\(\odot\)}$ yr$^{-1}$) employing {\sc dusty} code \citep{1997}. We have shown a correlation between mass-loss rates and luminosity for AGB stars. However, there is also an evolution term for stars of a given mass that should be considered. In addition, the carbon stars contribute much to the replenishment of the ISM with a timescale of $\sim 3.0\times10^5$ yr, a few times shorter than the TP-AGB duration. Also, we calculated the total mass returned rate to the ISM by LPVs $\sim$ $1.0-2.4\times10^{-4}$ M$\textsubscript{\(\odot\)}$ yr$^{-1}$ depending on the adopted metallicity. The mass-loss of LPVs could enrich the ISM in about a billion years if external or internal processes do not remove the gas.
\end{itemize}

\section{Acknowledgements}
The observing time for this survey was provided by the Iranian National Observatory (INO) and the UK-PATT allocation of time to programs I/2016B/09 and I/2017B/04 (PI: J. van Loon). We thank the INO and the School of Astronomy (IPM) for the financial support of this project. We thank
the referee for their comments which helped enhance
the analysis. HA is grateful to Peter Stetson for sharing his photometry routines. We thank Alireza Molaeinezhad, Arash Danesh, Mojtaba Raouf, Ghassem Gozaliasl, James Bamber, Philip Short, Lucia Su\'arez-Andr\'es, and Rosa Clavero for their help with the observations.

\bibliography{bibliography}

\begin{thebibliography}{}
\makeatletter
\relax
\def\mn@urlcharsother{\let\do\@makeother \do\$\do\&\do\#\do\^\do\_\do\%\do\~}
\def\mn@doi{\begingroup\mn@urlcharsother \@ifnextchar [ {\mn@doi@}
  {\mn@doi@[]}}
\def\mn@doi@[#1]#2{\def\@tempa{#1}\ifx\@tempa\@empty \href
  {http://dx.doi.org/#2} {doi:#2}\else \href {http://dx.doi.org/#2} {#1}\fi
  \endgroup}
\def\mn@eprint#1#2{\mn@eprint@#1:#2::\@nil}
\def\mn@eprint@arXiv#1{\href {http://arxiv.org/abs/#1} {{\tt arXiv:#1}}}
\def\mn@eprint@dblp#1{\href {http://dblp.uni-trier.de/rec/bibtex/#1.xml}
  {dblp:#1}}
\def\mn@eprint@#1:#2:#3:#4\@nil{\def\@tempa {#1}\def\@tempb {#2}\def\@tempc
  {#3}\ifx \@tempc \@empty \let \@tempc \@tempb \let \@tempb \@tempa \fi \ifx
  \@tempb \@empty \def\@tempb {arXiv}\fi \@ifundefined
  {mn@eprint@\@tempb}{\@tempb:\@tempc}{\expandafter \expandafter \csname
  mn@eprint@\@tempb\endcsname \expandafter{\@tempc}}}

\bibitem[\protect\citeauthoryear{{Applebaum}, {Brooks}, {Christensen},
  {Munshi}, {Quinn}, {Shen}  \& {Tremmel}}{{Applebaum}
  et~al.}{2021}]{2021ApJ...906...96A}
{Applebaum} E.,  {Brooks} A.~M.,  {Christensen} C.~R.,  {Munshi} F.,  {Quinn}
  T.~R.,  {Shen} S.,   {Tremmel} M.,  2021, \mn@doi [\apj]
  {10.3847/1538-4357/abcafa}, \href
  {https://ui.adsabs.harvard.edu/abs/2021ApJ...906...96A} {906, 96}

\bibitem[\protect\citeauthoryear{{Battinelli} \& {Demers}}{{Battinelli} \&
  {Demers}}{2013}]{2013POBeo..92..117B}
{Battinelli} P.,  {Demers} S.,  2013, Publications de l'Observatoire
  Astronomique de Beograd, \href
  {https://ui.adsabs.harvard.edu/abs/2013POBeo..92..117B} {92, 117}

\bibitem[\protect\citeauthoryear{{Ben{\'\i}tez-Llambay}, {Navarro}, {Abadi},
  {Gottl{\"o}ber}, {Yepes}, {Hoffman}  \& {Steinmetz}}{{Ben{\'\i}tez-Llambay}
  et~al.}{2016}]{2015Llambay}
{Ben{\'\i}tez-Llambay} A.,  {Navarro} J.~F.,  {Abadi} M.~G.,  {Gottl{\"o}ber}
  S.,  {Yepes} G.,  {Hoffman} Y.,   {Steinmetz} M.,  2016, \mn@doi [\mnras]
  {10.1093/mnras/stv2722}, \href
  {https://ui.adsabs.harvard.edu/abs/2016MNRAS.456.1185B} {456, 1185}

\bibitem[\protect\citeauthoryear{{Boyer}, {Skillman}, {van Loon}, {Gehrz}  \&
  {Woodward}}{{Boyer} et~al.}{2009}]{boyer2009spitzer}
{Boyer} M.~L.,  {Skillman} E.~D.,  {van Loon} J.~T.,  {Gehrz} R.~D.,
  {Woodward} C.~E.,  2009, \mn@doi [\apj] {10.1088/0004-637X/697/2/1993}, \href
  {https://ui.adsabs.harvard.edu/abs/2009ApJ...697.1993B} {697, 1993}

\bibitem[\protect\citeauthoryear{{Boyer} et~al.,}{{Boyer}
  et~al.}{2012}]{2012ApJ...748...40B}
{Boyer} M.~L.,  et~al., 2012, \mn@doi [\apj] {10.1088/0004-637X/748/1/40},
  \href {https://ui.adsabs.harvard.edu/abs/2012ApJ...748...40B} {748, 40}

\bibitem[\protect\citeauthoryear{{Boyer} et~al.,}{{Boyer}
  et~al.}{2015a}]{2015ApJS..216...10B}
{Boyer} M.~L.,  et~al., 2015a, \mn@doi [\apjs] {10.1088/0067-0049/216/1/10},
  \href {https://ui.adsabs.harvard.edu/abs/2015ApJS..216...10B} {216, 10}

\bibitem[\protect\citeauthoryear{{Boyer} et~al.,}{{Boyer}
  et~al.}{2015b}]{2015ApJ...800...51B}
{Boyer} M.~L.,  et~al., 2015b, \mn@doi [\apj] {10.1088/0004-637X/800/1/51},
  \href {https://ui.adsabs.harvard.edu/abs/2015ApJ...800...51B} {800, 51}

\bibitem[\protect\citeauthoryear{{Boylan-Kolchin}, {Bullock}  \&
  {Kaplinghat}}{{Boylan-Kolchin} et~al.}{2011}]{2011toobigtoofail}
{Boylan-Kolchin} M.,  {Bullock} J.~S.,   {Kaplinghat} M.,  2011, \mn@doi
  [\mnras] {10.1111/j.1745-3933.2011.01074.x}, \href
  {https://ui.adsabs.harvard.edu/abs/2011MNRAS.415L..40B} {415, L40}

\bibitem[\protect\citeauthoryear{{Boylan-Kolchin}, {Bullock}  \&
  {Kaplinghat}}{{Boylan-Kolchin} et~al.}{2012}]{2012MW}
{Boylan-Kolchin} M.,  {Bullock} J.~S.,   {Kaplinghat} M.,  2012, \mn@doi
  [\mnras] {10.1111/j.1365-2966.2012.20695.x}, \href
  {https://ui.adsabs.harvard.edu/abs/2012MNRAS.422.1203B} {422, 1203}

\bibitem[\protect\citeauthoryear{{Bressan}, {Marigo}, {Girardi}, {Salasnich},
  {Dal Cero}, {Rubele}  \& {Nanni}}{{Bressan}
  et~al.}{2012}]{2012MNRAS.427..127B}
{Bressan} A.,  {Marigo} P.,  {Girardi} L.,  {Salasnich} B.,  {Dal Cero} C.,
  {Rubele} S.,   {Nanni} A.,  2012, \mn@doi [\mnras]
  {10.1111/j.1365-2966.2012.21948.x}, \href
  {https://ui.adsabs.harvard.edu/abs/2012MNRAS.427..127B} {427, 127}

\bibitem[\protect\citeauthoryear{{Buck}, {Macci{\`o}}, {Dutton}, {Obreja}  \&
  {Frings}}{{Buck} et~al.}{2019}]{2019MNRAS.483.1314B}
{Buck} T.,  {Macci{\`o}} A.~V.,  {Dutton} A.~A.,  {Obreja} A.,   {Frings} J.,
  2019, \mn@doi [\mnras] {10.1093/mnras/sty2913}, \href
  {https://ui.adsabs.harvard.edu/abs/2019MNRAS.483.1314B} {483, 1314}

\bibitem[\protect\citeauthoryear{{Cignoni} \& {Tosi}}{{Cignoni} \&
  {Tosi}}{2010}]{2010AdAst2010E...3C}
{Cignoni} M.,  {Tosi} M.,  2010, \mn@doi [Advances in Astronomy]
  {10.1155/2010/158568}, \href
  {https://ui.adsabs.harvard.edu/abs/2010AdAst2010E...3C} {2010, 158568}

\bibitem[\protect\citeauthoryear{{Collins} et~al.,}{{Collins}
  et~al.}{2010}]{2010MNRAS.407.2411C}
{Collins} M.~L.~M.,  et~al., 2010, \mn@doi [\mnras]
  {10.1111/j.1365-2966.2010.17069.x}, \href
  {https://ui.adsabs.harvard.edu/abs/2010MNRAS.407.2411C} {407, 2411}

\bibitem[\protect\citeauthoryear{{Conn} et~al.,}{{Conn}
  et~al.}{2012}]{2012ApJ...758...11C}
{Conn} A.~R.,  et~al., 2012, \mn@doi [\apj] {10.1088/0004-637X/758/1/11}, \href
  {https://ui.adsabs.harvard.edu/abs/2012ApJ...758...11C} {758, 11}

\bibitem[\protect\citeauthoryear{{Cutri} et~al.,}{{Cutri}
  et~al.}{2021}]{2014yCat.2328....0C}
{Cutri} R.~M.,  et~al., 2021, VizieR Online Data Catalog, \href
  {https://ui.adsabs.harvard.edu/abs/2014yCat.2328....0C} {p. II/328}

\bibitem[\protect\citeauthoryear{{D'Souza} \& {Bell}}{{D'Souza} \&
  {Bell}}{2021}]{2021MNRAS.504.5270D}
{D'Souza} R.,  {Bell} E.~F.,  2021, \mn@doi [\mnras] {10.1093/mnras/stab1283},
  \href {https://ui.adsabs.harvard.edu/abs/2021MNRAS.504.5270D} {504, 5270}

\bibitem[\protect\citeauthoryear{Del~Popolo \& Le~Delliou}{Del~Popolo \&
  Le~Delliou}{2017}]{2017CDM}
Del~Popolo A.,  Le~Delliou M.,  2017, \mn@doi [Galaxies]
  {10.3390/galaxies5010017}, 5, 17

\bibitem[\protect\citeauthoryear{{Draine} \& {Lee}}{{Draine} \&
  {Lee}}{1984}]{1984ApJ...285...89D}
{Draine} B.~T.,  {Lee} H.~M.,  1984, \mn@doi [\apj] {10.1086/162480}, \href
  {https://ui.adsabs.harvard.edu/abs/1984ApJ...285...89D} {285, 89}

\bibitem[\protect\citeauthoryear{{Fillingham} et~al.,}{{Fillingham}
  et~al.}{2019}]{fillingham2019characterizing}
{Fillingham} S.~P.,  et~al., 2019, arXiv e-prints, \href
  {https://ui.adsabs.harvard.edu/abs/2019arXiv190604180F} {p. arXiv:1906.04180}

\bibitem[\protect\citeauthoryear{{Gaia Collaboration} et~al.,}{{Gaia
  Collaboration} et~al.}{2021}]{GaiaDR3}
{Gaia Collaboration} et~al., 2021, \mn@doi [\aap]
  {10.1051/0004-6361/202039657}, \href
  {https://ui.adsabs.harvard.edu/abs/2021A&A...649A...1G} {649, A1}

\bibitem[\protect\citeauthoryear{{Gail} \& {Sedlmayr}}{{Gail} \&
  {Sedlmayr}}{1999}]{1999A&A...347..594G}
{Gail} H.~P.,  {Sedlmayr} E.,  1999, \aap, \href
  {https://ui.adsabs.harvard.edu/abs/1999A&A...347..594G} {347, 594}

\bibitem[\protect\citeauthoryear{{Girardi} \& {Marigo}}{{Girardi} \&
  {Marigo}}{2007}]{2007A&A...462..237G}
{Girardi} L.,  {Marigo} P.,  2007, \mn@doi [\aap] {10.1051/0004-6361:20065249},
  \href {https://ui.adsabs.harvard.edu/abs/2007A&A...462..237G} {462, 237}

\bibitem[\protect\citeauthoryear{{Girardi}, {Groenewegen}, {Hatziminaoglou}  \&
  {da Costa}}{{Girardi} et~al.}{2005}]{2005A&A...436..895G}
{Girardi} L.,  {Groenewegen} M.~A.~T.,  {Hatziminaoglou} E.,   {da Costa} L.,
  2005, \mn@doi [\aap] {10.1051/0004-6361:20042352}, \href
  {https://ui.adsabs.harvard.edu/abs/2005A&A...436..895G} {436, 895}

\bibitem[\protect\citeauthoryear{{Gnedin}}{{Gnedin}}{2014}]{2014ApJ...793...29G}
{Gnedin} N.~Y.,  2014, \mn@doi [\apj] {10.1088/0004-637X/793/1/29}, \href
  {https://ui.adsabs.harvard.edu/abs/2014ApJ...793...29G} {793, 29}

\bibitem[\protect\citeauthoryear{{Goldman} et~al.,}{{Goldman}
  et~al.}{2017}]{2017MNRAS.465..403G}
{Goldman} S.~R.,  et~al., 2017, \mn@doi [\mnras] {10.1093/mnras/stw2708}, \href
  {https://ui.adsabs.harvard.edu/abs/2017MNRAS.465..403G} {465, 403}

\bibitem[\protect\citeauthoryear{{Goldman} et~al.,}{{Goldman}
  et~al.}{2019}]{goldman2019infrared}
{Goldman} S.~R.,  et~al., 2019, \mn@doi [\apj] {10.3847/1538-4357/ab0965},
  \href {https://ui.adsabs.harvard.edu/abs/2019ApJ...877...49G} {877, 49}

\bibitem[\protect\citeauthoryear{{Hamedani Golshan}, {Javadi}, {van Loon},
  {Khosroshahi}  \& {Saremi}}{{Hamedani Golshan}
  et~al.}{2017}]{2017MNRAS.466.1764H}
{Hamedani Golshan} R.,  {Javadi} A.,  {van Loon} J.~T.,  {Khosroshahi} H.,
  {Saremi} E.,  2017, \mn@doi [\mnras] {10.1093/mnras/stw3174}, \href
  {https://ui.adsabs.harvard.edu/abs/2017MNRAS.466.1764H} {466, 1764}

\bibitem[\protect\citeauthoryear{Hanner}{Hanner}{1988}]{hanner1988infrared}
Hanner M.~S.,  1988, Infrared Observations of Comets Halley and Wilson and
  Properties of the Grains

\bibitem[\protect\citeauthoryear{{Harbeck}, {Gallagher}, {Grebel}, {Koch}  \&
  {Zucker}}{{Harbeck} et~al.}{2005}]{2005ApJ...623..159H}
{Harbeck} D.,  {Gallagher} J.~S.,  {Grebel} E.~K.,  {Koch} A.,   {Zucker}
  D.~B.,  2005, \mn@doi [\apj] {10.1086/428650}, \href
  {https://ui.adsabs.harvard.edu/abs/2005ApJ...623..159H} {623, 159}

\bibitem[\protect\citeauthoryear{{Hashemi}, {Javadi}  \& {van Loon}}{{Hashemi}
  et~al.}{2019}]{2019MNRAS.483.4751H}
{Hashemi} S.~A.,  {Javadi} A.,   {van Loon} J.~T.,  2019, \mn@doi [\mnras]
  {10.1093/mnras/sty3450}, \href
  {https://ui.adsabs.harvard.edu/abs/2019MNRAS.483.4751H} {483, 4751}

\bibitem[\protect\citeauthoryear{Hidalgo, Aparicio  \& Gallart}{Hidalgo
  et~al.}{2008}]{hidalgo2008disc}
Hidalgo S.~L.,  Aparicio A.,   Gallart C.,  2008, The Astronomical Journal,
  136, 2332

\bibitem[\protect\citeauthoryear{{Hidalgo} et~al.,}{{Hidalgo}
  et~al.}{2013}]{2013ApJ...778..103H}
{Hidalgo} S.~L.,  et~al., 2013, \mn@doi [\apj] {10.1088/0004-637X/778/2/103},
  \href {https://ui.adsabs.harvard.edu/abs/2013ApJ...778..103H} {778, 103}

\bibitem[\protect\citeauthoryear{{H{\"o}fner} \& {Olofsson}}{{H{\"o}fner} \&
  {Olofsson}}{2018}]{2018A&ARv..26....1H}
{H{\"o}fner} S.,  {Olofsson} H.,  2018, \mn@doi [\aapr]
  {10.1007/s00159-017-0106-5}, \href
  {https://ui.adsabs.harvard.edu/abs/2018A&ARv..26....1H} {26, 1}

\bibitem[\protect\citeauthoryear{{Ivezic} \& {Elitzur}}{{Ivezic} \&
  {Elitzur}}{1997}]{1997}
{Ivezic} Z.,  {Elitzur} M.,  1997, \mn@doi [\mnras] {10.1093/mnras/287.4.799},
  \href {https://ui.adsabs.harvard.edu/abs/1997MNRAS.287..799I} {287, 799}

\bibitem[\protect\citeauthoryear{{Jackson}, {Skillman}, {Gehrz}, {Polomski}  \&
  {Woodward}}{{Jackson} et~al.}{2007}]{2007ApJ...656..818J}
{Jackson} D.~C.,  {Skillman} E.~D.,  {Gehrz} R.~D.,  {Polomski} E.,
  {Woodward} C.~E.,  2007, \mn@doi [\apj] {10.1086/510354}, \href
  {https://ui.adsabs.harvard.edu/abs/2007ApJ...656..818J} {656, 818}

\bibitem[\protect\citeauthoryear{{Javadi}, {van Loon}  \& {Mirtorabi}}{{Javadi}
  et~al.}{2011a}]{javadia}
{Javadi} A.,  {van Loon} J.~T.,   {Mirtorabi} M.~T.,  2011a, \mn@doi [\mnras]
  {10.1111/j.1365-2966.2010.17678.x}, \href
  {https://ui.adsabs.harvard.edu/abs/2011MNRAS.411..263J} {411, 263}

\bibitem[\protect\citeauthoryear{{Javadi}, {van Loon}  \& {Mirtorabi}}{{Javadi}
  et~al.}{2011b}]{javadib}
{Javadi} A.,  {van Loon} J.~T.,   {Mirtorabi} M.~T.,  2011b, \mn@doi [\mnras]
  {10.1111/j.1365-2966.2011.18638.x}, \href
  {https://ui.adsabs.harvard.edu/abs/2011MNRAS.414.3394J} {414, 3394}

\bibitem[\protect\citeauthoryear{{Javadi}, {van Loon}, {Khosroshahi}  \&
  {Mirtorabi}}{{Javadi} et~al.}{2013}]{2013MNRAS.432.2824J}
{Javadi} A.,  {van Loon} J.~T.,  {Khosroshahi} H.,   {Mirtorabi} M.~T.,  2013,
  \mn@doi [\mnras] {10.1093/mnras/stt640}, \href
  {https://ui.adsabs.harvard.edu/abs/2013MNRAS.432.2824J} {432, 2824}

\bibitem[\protect\citeauthoryear{{Javadi}, {Saberi}, {van Loon}, {Khosroshahi},
  {Golabatooni}  \& {Mirtorabi}}{{Javadi} et~al.}{2015}]{2015MNRAS.447.3973J}
{Javadi} A.,  {Saberi} M.,  {van Loon} J.~T.,  {Khosroshahi} H.,  {Golabatooni}
  N.,   {Mirtorabi} M.~T.,  2015, \mn@doi [\mnras] {10.1093/mnras/stu2637},
  \href {https://ui.adsabs.harvard.edu/abs/2015MNRAS.447.3973J} {447, 3973}

\bibitem[\protect\citeauthoryear{{Javadi}, {van Loon}, {Khosroshahi},
  {Tabatabaei}, {Hamedani Golshan}  \& {Rashidi}}{{Javadi}
  et~al.}{2017}]{2017MNRAS.464.2103J}
{Javadi} A.,  {van Loon} J.~T.,  {Khosroshahi} H.~G.,  {Tabatabaei} F.,
  {Hamedani Golshan} R.,   {Rashidi} M.,  2017, \mn@doi [\mnras]
  {10.1093/mnras/stw2463}, \href
  {https://ui.adsabs.harvard.edu/abs/2017MNRAS.464.2103J} {464, 2103}

\bibitem[\protect\citeauthoryear{{Jordi}, {Grebel}  \& {Ammon}}{{Jordi}
  et~al.}{2006}]{2006A&A...460..339J}
{Jordi} K.,  {Grebel} E.~K.,   {Ammon} K.,  2006, \mn@doi [\aap]
  {10.1051/0004-6361:20066082}, \href
  {https://ui.adsabs.harvard.edu/abs/2006A&A...460..339J} {460, 339}

\bibitem[\protect\citeauthoryear{{Kirby}, {Cohen}, {Guhathakurta}, {Cheng},
  {Bullock}  \& {Gallazzi}}{{Kirby} et~al.}{2013}]{2013ApJ...779..102K}
{Kirby} E.~N.,  {Cohen} J.~G.,  {Guhathakurta} P.,  {Cheng} L.,  {Bullock}
  J.~S.,   {Gallazzi} A.,  2013, \mn@doi [\apj] {10.1088/0004-637X/779/2/102},
  \href {https://ui.adsabs.harvard.edu/abs/2013ApJ...779..102K} {779, 102}

\bibitem[\protect\citeauthoryear{{Klypin}, {Kravtsov}, {Valenzuela}  \&
  {Prada}}{{Klypin} et~al.}{1999}]{1999missing}
{Klypin} A.,  {Kravtsov} A.~V.,  {Valenzuela} O.,   {Prada} F.,  1999, \mn@doi
  [\apj] {10.1086/307643}, \href
  {https://ui.adsabs.harvard.edu/abs/1999ApJ...522...82K} {522, 82}

\bibitem[\protect\citeauthoryear{{Kroupa}}{{Kroupa}}{2001}]{2001MNRAS.322..231K}
{Kroupa} P.,  2001, \mn@doi [\mnras] {10.1046/j.1365-8711.2001.04022.x}, \href
  {https://ui.adsabs.harvard.edu/abs/2001MNRAS.322..231K} {322, 231}

\bibitem[\protect\citeauthoryear{{Ledinauskas} \& {Zubovas}}{{Ledinauskas} \&
  {Zubovas}}{2020}]{2020MNRAS.493..638L}
{Ledinauskas} E.,  {Zubovas} K.,  2020, \mn@doi [\mnras]
  {10.1093/mnras/staa298}, \href
  {https://ui.adsabs.harvard.edu/abs/2020MNRAS.493..638L} {493, 638}

\bibitem[\protect\citeauthoryear{{Lee}, {Freedman}  \& {Madore}}{{Lee}
  et~al.}{1993}]{lee1993tip}
{Lee} M.~G.,  {Freedman} W.~L.,   {Madore} B.~F.,  1993, \mn@doi [\apj]
  {10.1086/173334}, \href
  {https://ui.adsabs.harvard.edu/abs/1993ApJ...417..553L} {417, 553}

\bibitem[\protect\citeauthoryear{{Leisenring}, {Kemper}  \&
  {Sloan}}{{Leisenring} et~al.}{2008}]{2008}
{Leisenring} J.~M.,  {Kemper} F.,   {Sloan} G.~C.,  2008, \mn@doi [\apj]
  {10.1086/588378}, \href
  {https://ui.adsabs.harvard.edu/abs/2008ApJ...681.1557L} {681, 1557}

\bibitem[\protect\citeauthoryear{Lupton}{Lupton}{2005}]{Lupton(2005)}
Lupton R.~H.,  2005, {MS Windows NT} Kernel Description, \url
  {https://www.sdss3.org/dr8/algorithms/sdssUBVRITransform.php}

\bibitem[\protect\citeauthoryear{{Marigo}, {Girardi}, {Bressan}, {Groenewegen},
  {Silva}  \& {Granato}}{{Marigo} et~al.}{2008}]{2008A&A...482..883M}
{Marigo} P.,  {Girardi} L.,  {Bressan} A.,  {Groenewegen} M.~A.~T.,  {Silva}
  L.,   {Granato} G.~L.,  2008, \mn@doi [\aap] {10.1051/0004-6361:20078467},
  \href {https://ui.adsabs.harvard.edu/abs/2008A&A...482..883M} {482, 883}

\bibitem[\protect\citeauthoryear{{Marigo} et~al.,}{{Marigo}
  et~al.}{2017}]{2017ApJ...835...77M}
{Marigo} P.,  et~al., 2017, \mn@doi [\apj] {10.3847/1538-4357/835/1/77}, \href
  {https://ui.adsabs.harvard.edu/abs/2017ApJ...835...77M} {835, 77}

\bibitem[\protect\citeauthoryear{Mattsson, H{\"o}fner  \& Herwig}{Mattsson
  et~al.}{2007}]{mattsson2007mass}
Mattsson L.,  H{\"o}fner S.,   Herwig F.,  2007, Astronomy \& Astrophysics,
  470, 339

\bibitem[\protect\citeauthoryear{{McConnachie}}{{McConnachie}}{2012}]{2012AJ....144....4M}
{McConnachie} A.~W.,  2012, \mn@doi [\aj] {10.1088/0004-6256/144/1/4}, \href
  {https://ui.adsabs.harvard.edu/abs/2012AJ....144....4M} {144, 4}

\bibitem[\protect\citeauthoryear{{McConnachie}, {Irwin}, {Ferguson}, {Ibata},
  {Lewis}  \& {Tanvir}}{{McConnachie} et~al.}{2004}]{2004MNRAS.350..243M}
{McConnachie} A.~W.,  {Irwin} M.~J.,  {Ferguson} A.~M.~N.,  {Ibata} R.~A.,
  {Lewis} G.~F.,   {Tanvir} N.,  2004, \mn@doi [\mnras]
  {10.1111/j.1365-2966.2004.07637.x}, \href
  {https://ui.adsabs.harvard.edu/abs/2004MNRAS.350..243M} {350, 243}

\bibitem[\protect\citeauthoryear{{McConnachie}, {Irwin}, {Ferguson}, {Ibata},
  {Lewis}  \& {Tanvir}}{{McConnachie} et~al.}{2005}]{2005MNRAS.356..979M}
{McConnachie} A.~W.,  {Irwin} M.~J.,  {Ferguson} A.~M.~N.,  {Ibata} R.~A.,
  {Lewis} G.~F.,   {Tanvir} N.,  2005, \mn@doi [\mnras]
  {10.1111/j.1365-2966.2004.08514.x}, \href
  {https://ui.adsabs.harvard.edu/abs/2005MNRAS.356..979M} {356, 979}

\bibitem[\protect\citeauthoryear{{McDonald} \& {Trabucchi}}{{McDonald} \&
  {Trabucchi}}{2019}]{2019MNRAS.484.4678M}
{McDonald} I.,  {Trabucchi} M.,  2019, \mn@doi [\mnras] {10.1093/mnras/stz324},
  \href {https://ui.adsabs.harvard.edu/abs/2019MNRAS.484.4678M} {484, 4678}

\bibitem[\protect\citeauthoryear{{McDonald} \& {Zijlstra}}{{McDonald} \&
  {Zijlstra}}{2016}]{2016ApJ...823L..38M}
{McDonald} I.,  {Zijlstra} A.~A.,  2016, \mn@doi [\apjl]
  {10.3847/2041-8205/823/2/L38}, \href
  {https://ui.adsabs.harvard.edu/abs/2016ApJ...823L..38M} {823, L38}

\bibitem[\protect\citeauthoryear{{McDonald}, {White}, {Zijlstra}, {Guzman
  Ramirez}, {Szyszka}, {van Loon}, {Lagadec}  \& {Jones}}{{McDonald}
  et~al.}{2012}]{2012MNRAS.427.2647M}
{McDonald} I.,  {White} J.~R.,  {Zijlstra} A.~A.,  {Guzman Ramirez} L.,
  {Szyszka} C.,  {van Loon} J.~T.,  {Lagadec} E.,   {Jones} O.~C.,  2012,
  \mn@doi [\mnras] {10.1111/j.1365-2966.2012.22109.x}, \href
  {https://ui.adsabs.harvard.edu/abs/2012MNRAS.427.2647M} {427, 2647}

\bibitem[\protect\citeauthoryear{{M{\"u}ller} et~al.,}{{M{\"u}ller}
  et~al.}{2021}]{2021A&A...645A..92M}
{M{\"u}ller} O.,  et~al., 2021, \mn@doi [\aap] {10.1051/0004-6361/202039359},
  \href {https://ui.adsabs.harvard.edu/abs/2021A&A...645A..92M} {645, A92}

\bibitem[\protect\citeauthoryear{{Navabi} et~al.,}{{Navabi}
  et~al.}{2021}]{2021}
{Navabi} M.,  et~al., 2021, \mn@doi [\apj] {10.3847/1538-4357/abdec1}, \href
  {https://ui.adsabs.harvard.edu/abs/2021ApJ...910..127N} {910, 127}

\bibitem[\protect\citeauthoryear{Navarro}{Navarro}{2018}]{2018dg}
Navarro J.~F.,  2018, \mn@doi [Proceedings of the International Astronomical
  Union] {10.1017/s1743921318005963}, 14, 455–463

\bibitem[\protect\citeauthoryear{{Nenkova}, {Ivezic}  \& {Elitzur}}{{Nenkova}
  et~al.}{1999}]{1999LPICo.969...20N}
{Nenkova} M.,  {Ivezic} Z.,   {Elitzur} M.,  1999, in {Sprague} A.,  {Lynch}
  D.~K.,   {Sitko} M.,  eds,  LPI Contributions Vol. 969, Thermal Emission
  Spectroscopy and Analysis of Dust, Disks, and Regoliths. p.~20

\bibitem[\protect\citeauthoryear{{Parto} et~al.,}{{Parto}
  et~al.}{2023}]{2023ApJ...942...33P}
{Parto} T.,  et~al., 2023, \mn@doi [\apj] {10.3847/1538-4357/aca471}, \href
  {https://ui.adsabs.harvard.edu/abs/2023ApJ...942...33P} {942, 33}

\bibitem[\protect\citeauthoryear{P{\'e}gouri{\'e}}{P{\'e}gouri{\'e}}{1988}]{pegourie1988optical}
P{\'e}gouri{\'e} B.,  1988, Astronomy and Astrophysics, 194, 335

\bibitem[\protect\citeauthoryear{{Planck Collaboration} et~al.,}{{Planck
  Collaboration} et~al.}{2014}]{2014A&A...571A..16P}
{Planck Collaboration} et~al., 2014, \mn@doi [\aap]
  {10.1051/0004-6361/201321591}, \href
  {https://ui.adsabs.harvard.edu/abs/2014A&A...571A..16P} {571, A16}

\bibitem[\protect\citeauthoryear{{Ren}, {Jiang}, {Ren}  \& {Yang}}{{Ren}
  et~al.}{2022}]{2022Univ....8..465R}
{Ren} T.,  {Jiang} B.,  {Ren} Y.,   {Yang} M.,  2022, \mn@doi [Universe]
  {10.3390/universe8090465}, \href
  {https://ui.adsabs.harvard.edu/abs/2022Univ....8..465R} {8, 465}

\bibitem[\protect\citeauthoryear{{Rezaeikh}, {Javadi}, {Khosroshahi}  \& {van
  Loon}}{{Rezaeikh} et~al.}{2014}]{2014MNRAS.445.2214R}
{Rezaeikh} S.,  {Javadi} A.,  {Khosroshahi} H.,   {van Loon} J.~T.,  2014,
  \mn@doi [\mnras] {10.1093/mnras/stu1807}, \href
  {https://ui.adsabs.harvard.edu/abs/2014MNRAS.445.2214R} {445, 2214}

\bibitem[\protect\citeauthoryear{{Rocha-Pinto}, {Majewski}, {Skrutskie},
  {Crane}  \& {Patterson}}{{Rocha-Pinto} et~al.}{2004}]{2004ApJ...615..732R}
{Rocha-Pinto} H.~J.,  {Majewski} S.~R.,  {Skrutskie} M.~F.,  {Crane} J.~D.,
  {Patterson} R.~J.,  2004, \mn@doi [\apj] {10.1086/424585}, \href
  {https://ui.adsabs.harvard.edu/abs/2004ApJ...615..732R} {615, 732}

\bibitem[\protect\citeauthoryear{{Rocha}, {Peter}  \& {Bullock}}{{Rocha}
  et~al.}{2012}]{2012Rocha}
{Rocha} M.,  {Peter} A. H.~G.,   {Bullock} J.,  2012, \mn@doi [\mnras]
  {10.1111/j.1365-2966.2012.21432.x}, \href
  {https://ui.adsabs.harvard.edu/abs/2012MNRAS.425..231R} {425, 231}

\bibitem[\protect\citeauthoryear{{Rosenfield} et~al.,}{{Rosenfield}
  et~al.}{2014}]{2014ApJ...790...22R}
{Rosenfield} P.,  et~al., 2014, \mn@doi [\apj] {10.1088/0004-637X/790/1/22},
  \href {https://ui.adsabs.harvard.edu/abs/2014ApJ...790...22R} {790, 22}

\bibitem[\protect\citeauthoryear{{Sakai}, {Madore}  \& {Freedman}}{{Sakai}
  et~al.}{1996}]{1996ApJ...461..713S}
{Sakai} S.,  {Madore} B.~F.,   {Freedman} W.~L.,  1996, \mn@doi [\apj]
  {10.1086/177096}, \href
  {https://ui.adsabs.harvard.edu/abs/1996ApJ...461..713S} {461, 713}

\bibitem[\protect\citeauthoryear{{Saremi} et~al.,}{{Saremi}
  et~al.}{2020}]{2020ApJ...894..135S}
{Saremi} E.,  et~al., 2020, \mn@doi [\apj] {10.3847/1538-4357/ab88a2}, \href
  {https://ui.adsabs.harvard.edu/abs/2020ApJ...894..135S} {894, 135}

\bibitem[\protect\citeauthoryear{{Saremi}, {Javadi}, {Navabi}, {van Loon},
  {Khosroshahi}, {Bojnordi Arbab}  \& {McDonald}}{{Saremi}
  et~al.}{2021}]{2021ApJ...923..164S}
{Saremi} E.,  {Javadi} A.,  {Navabi} M.,  {van Loon} J.~T.,  {Khosroshahi}
  H.~G.,  {Bojnordi Arbab} B.,   {McDonald} I.,  2021, \mn@doi [\apj]
  {10.3847/1538-4357/ac2d96}, \href
  {https://ui.adsabs.harvard.edu/abs/2021ApJ...923..164S} {923, 164}

\bibitem[\protect\citeauthoryear{{Schlafly} \& {Finkbeiner}}{{Schlafly} \&
  {Finkbeiner}}{2011}]{2011ApJ...737..103S}
{Schlafly} E.~F.,  {Finkbeiner} D.~P.,  2011, \mn@doi [\apj]
  {10.1088/0004-637X/737/2/103}, \href
  {https://ui.adsabs.harvard.edu/abs/2011ApJ...737..103S} {737, 103}

\bibitem[\protect\citeauthoryear{{Schlegel}, {Finkbeiner}  \&
  {Davis}}{{Schlegel} et~al.}{1998}]{1998ApJ...500..525S}
{Schlegel} D.~J.,  {Finkbeiner} D.~P.,   {Davis} M.,  1998, \mn@doi [\apj]
  {10.1086/305772}, \href
  {https://ui.adsabs.harvard.edu/abs/1998ApJ...500..525S} {500, 525}

\bibitem[\protect\citeauthoryear{{Shi} et~al.,}{{Shi} et~al.}{2020}]{2020SFH}
{Shi} J.,  et~al., 2020, \mn@doi [\apj] {10.3847/1538-4357/ab8464}, \href
  {https://ui.adsabs.harvard.edu/abs/2020ApJ...893..139S} {893, 139}

\bibitem[\protect\citeauthoryear{Simha, Weinberg, Conroy, Dav{\'e}, Fardal,
  Katz  \& Oppenheimer}{Simha et~al.}{2014}]{Simha2014ParametrisingSF}
Simha V.,  Weinberg D.~H.,  Conroy C.,  Dav{\'e} R.,  Fardal M.~A.,  Katz N.,
  Oppenheimer B.~D.,  2014, arXiv: Astrophysics of Galaxies

\bibitem[\protect\citeauthoryear{{Simpson}, {Grand}, {G{\'o}mez}, {Marinacci},
  {Pakmor}, {Springel}, {Campbell}  \& {Frenk}}{{Simpson}
  et~al.}{2018}]{2018quenching}
{Simpson} C.~M.,  {Grand} R. J.~J.,  {G{\'o}mez} F.~A.,  {Marinacci} F.,
  {Pakmor} R.,  {Springel} V.,  {Campbell} D. J.~R.,   {Frenk} C.~S.,  2018,
  \mn@doi [\mnras] {10.1093/mnras/sty774}, \href
  {https://ui.adsabs.harvard.edu/abs/2018MNRAS.478..548S} {478, 548}

\bibitem[\protect\citeauthoryear{{Stetson}}{{Stetson}}{1987}]{1987PASP...99..191S}
{Stetson} P.~B.,  1987, \mn@doi [\pasp] {10.1086/131977}, \href
  {https://ui.adsabs.harvard.edu/abs/1987PASP...99..191S} {99, 191}

\bibitem[\protect\citeauthoryear{{Stetson}}{{Stetson}}{1990}]{1990PASP..102..932S}
{Stetson} P.~B.,  1990, \mn@doi [\pasp] {10.1086/132719}, \href
  {https://ui.adsabs.harvard.edu/abs/1990PASP..102..932S} {102, 932}

\bibitem[\protect\citeauthoryear{Stetson}{Stetson}{1994}]{stetson1994center}
Stetson P.~B.,  1994, Publications of the Astronomical Society of the Pacific,
  106, 250

\bibitem[\protect\citeauthoryear{{Stetson}}{{Stetson}}{1996}]{1996PASP..108..851S}
{Stetson} P.~B.,  1996, \mn@doi [\pasp] {10.1086/133808}, \href
  {https://ui.adsabs.harvard.edu/abs/1996PASP..108..851S} {108, 851}

\bibitem[\protect\citeauthoryear{{Tollerud} et~al.,}{{Tollerud}
  et~al.}{2012}]{2012ApJ...752...45T}
{Tollerud} E.~J.,  et~al., 2012, \mn@doi [\apj] {10.1088/0004-637X/752/1/45},
  \href {https://ui.adsabs.harvard.edu/abs/2012ApJ...752...45T} {752, 45}

\bibitem[\protect\citeauthoryear{{Tolstoy}}{{Tolstoy}}{2003}]{2003Ap&SS.284..579T}
{Tolstoy} E.,  2003, \mn@doi [\apss] {10.1023/A:1024006006003}, \href
  {https://ui.adsabs.harvard.edu/abs/2003Ap&SS.284..579T} {284, 579}

\bibitem[\protect\citeauthoryear{{Valiante}, {Schneider}, {Bianchi}  \&
  {Andersen}}{{Valiante} et~al.}{2009}]{2009MNRAS.397.1661V}
{Valiante} R.,  {Schneider} R.,  {Bianchi} S.,   {Andersen} A.~C.,  2009,
  \mn@doi [\mnras] {10.1111/j.1365-2966.2009.15076.x}, \href
  {https://ui.adsabs.harvard.edu/abs/2009MNRAS.397.1661V} {397, 1661}

\bibitem[\protect\citeauthoryear{{Vassiliadis} \& {Wood}}{{Vassiliadis} \&
  {Wood}}{1993}]{1993ApJ...413..641V}
{Vassiliadis} E.,  {Wood} P.~R.,  1993, \mn@doi [\apj] {10.1086/173033}, \href
  {https://ui.adsabs.harvard.edu/abs/1993ApJ...413..641V} {413, 641}

\bibitem[\protect\citeauthoryear{{Weisz}, {Dolphin}, {Skillman}, {Holtzman},
  {Gilbert}, {Dalcanton}  \& {Williams}}{{Weisz}
  et~al.}{2014}]{2014ApJ...789..147W}
{Weisz} D.~R.,  {Dolphin} A.~E.,  {Skillman} E.~D.,  {Holtzman} J.,  {Gilbert}
  K.~M.,  {Dalcanton} J.~J.,   {Williams} B.~F.,  2014, \mn@doi [\apj]
  {10.1088/0004-637X/789/2/147}, \href
  {https://ui.adsabs.harvard.edu/abs/2014ApJ...789..147W} {789, 147}

\bibitem[\protect\citeauthoryear{{Weisz} et~al.,}{{Weisz}
  et~al.}{2019a}]{2019MNRAS.489..763W}
{Weisz} D.~R.,  et~al., 2019a, \mn@doi [\mnras] {10.1093/mnras/stz1984}, \href
  {https://ui.adsabs.harvard.edu/abs/2019MNRAS.489..763W} {489, 763}

\bibitem[\protect\citeauthoryear{{Weisz} et~al.,}{{Weisz}
  et~al.}{2019b}]{2019ApJ...885L...8W}
{Weisz} D.~R.,  et~al., 2019b, \mn@doi [\apjl] {10.3847/2041-8213/ab4b52},
  \href {https://ui.adsabs.harvard.edu/abs/2019ApJ...885L...8W} {885, L8}

\bibitem[\protect\citeauthoryear{{Welch} \& {Stetson}}{{Welch} \&
  {Stetson}}{1993}]{1993AJ....105.1813W}
{Welch} D.~L.,  {Stetson} P.~B.,  1993, \mn@doi [\aj] {10.1086/116556}, \href
  {https://ui.adsabs.harvard.edu/abs/1993AJ....105.1813W} {105, 1813}

\bibitem[\protect\citeauthoryear{{Wetzel}, {Tollerud}  \& {Weisz}}{{Wetzel}
  et~al.}{2015}]{2015ApJ...808L..27W}
{Wetzel} A.~R.,  {Tollerud} E.~J.,   {Weisz} D.~R.,  2015, \mn@doi [\apjl]
  {10.1088/2041-8205/808/1/L27}, \href
  {https://ui.adsabs.harvard.edu/abs/2015ApJ...808L..27W} {808, L27}

\bibitem[\protect\citeauthoryear{{Wheeler} et~al.,}{{Wheeler}
  et~al.}{2019}]{2019MNRAS.490.4447W}
{Wheeler} C.,  et~al., 2019, \mn@doi [\mnras] {10.1093/mnras/stz2887}, \href
  {https://ui.adsabs.harvard.edu/abs/2019MNRAS.490.4447W} {490, 4447}

\bibitem[\protect\citeauthoryear{{Whitelock}, {Feast}  \&
  {Catchpole}}{{Whitelock} et~al.}{1991}]{1991MNRAS.248..276W}
{Whitelock} P.,  {Feast} M.,   {Catchpole} R.,  1991, \mn@doi [\mnras]
  {10.1093/mnras/248.2.276}, \href
  {https://ui.adsabs.harvard.edu/abs/1991MNRAS.248..276W} {248, 276}

\bibitem[\protect\citeauthoryear{{Whitelock}, {Feast}, {van Loon}  \&
  {Zijlstra}}{{Whitelock} et~al.}{2003}]{whitelock2003obscured}
{Whitelock} P.~A.,  {Feast} M.~W.,  {van Loon} J.~T.,   {Zijlstra} A.~A.,
  2003, \mn@doi [\mnras] {10.1046/j.1365-8711.2003.06514.x}, \href
  {https://ui.adsabs.harvard.edu/abs/2003MNRAS.342...86W} {342, 86}

\bibitem[\protect\citeauthoryear{Whitelock, Kasliwal  \& Boyer}{Whitelock
  et~al.}{2017}]{2017}
Whitelock P.~A.,  Kasliwal M.,   Boyer M.,  2017, \mn@doi [EPJ Web of
  Conferences] {10.1051/epjconf/201715201009}, 152, 01009

\bibitem[\protect\citeauthoryear{{Wojno}, {Gilbert}, {Kirby}, {Escala},
  {Beaton}, {Tollerud}, {Majewski}  \& {Guhathakurta}}{{Wojno}
  et~al.}{2020}]{2020ApJ...895...78W}
{Wojno} J.,  {Gilbert} K.~M.,  {Kirby} E.~N.,  {Escala} I.,  {Beaton} R.~L.,
  {Tollerud} E.~J.,  {Majewski} S.~R.,   {Guhathakurta} P.,  2020, \mn@doi
  [\apj] {10.3847/1538-4357/ab8ccb}, \href
  {https://ui.adsabs.harvard.edu/abs/2020ApJ...895...78W} {895, 78}

\bibitem[\protect\citeauthoryear{{Wood}}{{Wood}}{1998}]{1998A&A...338..592W}
{Wood} P.~R.,  1998, \aap, \href
  {https://ui.adsabs.harvard.edu/abs/1998A&A...338..592W} {338, 592}

\bibitem[\protect\citeauthoryear{{Wood}, {Whiteoak}, {Hughes}, {Bessell},
  {Gardner}  \& {Hyland}}{{Wood} et~al.}{1992}]{1992ApJ...397..552W}
{Wood} P.~R.,  {Whiteoak} J.~B.,  {Hughes} S.~M.~G.,  {Bessell} M.~S.,
  {Gardner} F.~F.,   {Hyland} A.~R.,  1992, \mn@doi [\apj] {10.1086/171812},
  \href {https://ui.adsabs.harvard.edu/abs/1992ApJ...397..552W} {397, 552}

\bibitem[\protect\citeauthoryear{{Xu}, {Wise}, {Norman}, {Ahn}  \&
  {O'Shea}}{{Xu} et~al.}{2016}]{2016ApJ...833...84X}
{Xu} H.,  {Wise} J.~H.,  {Norman} M.~L.,  {Ahn} K.,   {O'Shea} B.~W.,  2016,
  \mn@doi [\apj] {10.3847/1538-4357/833/1/84}, \href
  {https://ui.adsabs.harvard.edu/abs/2016ApJ...833...84X} {833, 84}

\bibitem[\protect\citeauthoryear{{Zijlstra}, {Loup}, {Waters}, {Whitelock},
  {van Loon}  \& {Guglielmo}}{{Zijlstra} et~al.}{1996}]{1996MNRAS.279...32Z}
{Zijlstra} A.~A.,  {Loup} C.,  {Waters} L.~B.~F.~M.,  {Whitelock} P.~A.,  {van
  Loon} J.~T.,   {Guglielmo} F.,  1996, \mn@doi [\mnras]
  {10.1093/mnras/279.1.32}, \href
  {https://ui.adsabs.harvard.edu/abs/1996MNRAS.279...32Z} {279, 32}

\bibitem[\protect\citeauthoryear{{Zucker} et~al.,}{{Zucker}
  et~al.}{2004}]{2004ApJ...612L.121Z}
{Zucker} D.~B.,  et~al., 2004, \mn@doi [\apjl] {10.1086/424691}, \href
  {https://ui.adsabs.harvard.edu/abs/2004ApJ...612L.121Z} {612, L121}

\bibitem[\protect\citeauthoryear{{van Loon}, {Groenewegen}, {de Koter},
  {Trams}, {Waters}, {Zijlstra}, {Whitelock}  \& {Loup}}{{van Loon}
  et~al.}{1999}]{1999A&A...351..559V}
{van Loon} J.~T.,  {Groenewegen} M.~A.~T.,  {de Koter} A.,  {Trams} N.~R.,
  {Waters} L.~B.~F.~M.,  {Zijlstra} A.~A.,  {Whitelock} P.~A.,   {Loup} C.,
  1999, \aap, \href {https://ui.adsabs.harvard.edu/abs/1999A&A...351..559V}
  {351, 559}

\bibitem[\protect\citeauthoryear{{van Loon}, {Cioni}, {Zijlstra}  \&
  {Loup}}{{van Loon} et~al.}{2005a}]{2005A&A...438..273V}
{van Loon} J.~T.,  {Cioni} M. R.~L.,  {Zijlstra} A.~A.,   {Loup} C.,  2005a,
  \mn@doi [\aap] {10.1051/0004-6361:20042555}, \href
  {https://ui.adsabs.harvard.edu/abs/2005A&A...438..273V} {438, 273}

\bibitem[\protect\citeauthoryear{{van Loon}, {Marshall}  \& {Zijlstra}}{{van
  Loon} et~al.}{2005b}]{2005A&A...442..597V}
{van Loon} J.~T.,  {Marshall} J.~R.,   {Zijlstra} A.~A.,  2005b, \mn@doi [\aap]
  {10.1051/0004-6361:20053528}, \href
  {https://ui.adsabs.harvard.edu/abs/2005A&A...442..597V} {442, 597}

\bibitem[\protect\citeauthoryear{{van Loon}, {Cohen}, {Oliveira}, {Matsuura},
  {McDonald}, {Sloan}, {Wood}  \& {Zijlstra}}{{van Loon}
  et~al.}{2008}]{2008A&A...487.1055V}
{van Loon} J.~T.,  {Cohen} M.,  {Oliveira} J.~M.,  {Matsuura} M.,  {McDonald}
  I.,  {Sloan} G.~C.,  {Wood} P.~R.,   {Zijlstra} A.~A.,  2008, \mn@doi [\aap]
  {10.1051/0004-6361:200810036}, \href
  {https://ui.adsabs.harvard.edu/abs/2008A&A...487.1055V} {487, 1055}

\makeatother
\end{thebibliography}

\begin{table}
\begin{center}
\caption{Characterizing of variable candidates.}
\begin{tabular}{c c c c c c c c c c c c c}

\tableline
\tableline

id & R.A. & Dec & $V$ &  $\delta V$  & $i$ & $\delta i$ & amplitude$_i$ & M$_{Birth}$ & $\tau_i$  &  $\dot{M}$  &  $L$  & chemical type \\
& (J2000) & (J2000) & (mag) & (mag) & (mag) & (mag) & (mag) & (M$\textsubscript{\(\odot\)}$) & & ($10^{-6}$ M$\textsubscript{\(\odot\)}$ yr$^{-1}$) & (10$^{4}$ L$\textsubscript{\(\odot\)}$) &  \\
\tableline
4957 & 00 52 40.39 & +43 06 37.20 & 23.26 & 0.08 & 22.68 & 0.06 & 0.56 & 0.95 & 0.30 &  $2.11$ & $0.19$  & O\\[1ex]
2686 & 00 53 25.47 & +43 15 19.35 & 21.35 & 0.03 & 20.77 & 0.01 & 0.32 & 1.73 & 0.13 &  $7.61$ & $0.88$  & C\\[1ex]
4837 & 00 52 49.79 & +43 10 51.66 & 21.89 & 0.02 & 21.29 & 0.04 & 0.30 & 0.93 & 0.25 &  $1.42$ & $0.19$  & O\\[1ex]
4799 & 00 52 49.13 & +43 13 21.94 & 21.93 & 0.03 & 21.32 & 0.02 & 0.30 & 0.97 & 0.20 &  $5.09$ & $0.38$  & O\\[1ex]
4601 & 00 53 18.14 & +43 15 05.66 & 22.34 & 0.03 & 21.70 & 0.01 & 0.64 & 0.80 & 0.36 &  $0.17$ & $0.13$ & O\\[1ex]
3543 & 00 53 13.45 & +43 13 31.77 & 22.91 & 0.08 & 22.27 & 0.07 & 0.67 & 0.89 & 0.48 &  $0.37$ & $0.10$  & O\\[1ex]
3481 & 00 52 13.34 & +43 13 02.46 & 21.40 & 0.03 & 20.70 & 0.01 & 0.29 & 2.10 & 0.38 &  $9.06$ & $0.79$  &  C\\[1ex]
4973 & 00 52 52.45 & +43 11 18.77 & 22.22 & 0.06 & 21.44 & 0.01 & 0.32 & 0.87 & 0.60 &  $0.37$ & $0.14$ &  O\\[1ex]
4901 & 00 53 11.49 & +43 16 21.95 & 22.56 & 0.07 & 21.77 & 0.03 & 0.54 & 0.83 & 0.50 & $0.31$ & $0.15$  &  O\\[1ex]
4448 & 00 52 55.29 & +43 10 27.09 & 21.99 & 0.03 & 21.20 & 0.00 & 0.51 & 2.10 & 0.22 &  $8.71$ & $0.62$  &  C\\[1ex]
4190 & 00 52 33.54 & +43 12 17.17 & 21.60 & 0.01 & 20.77 & 0.00 & 0.72 & 2.90 & 0.34 &  $18.74$ & $0.89$ &  C\\[1ex]
3798 & 00 52 48.01 & +43 07 33.08 & 21.35 & 0.01 & 20.50 & 0.02 & 0.65 & 1.53 & 0.15 &  $1.53$ & $0.43$  &  C\\[1ex]
6338 & 00 52 46.09 & +43 09 05.12 & 21.90 & 0.02 & 21.02 & 0.03 & 1.11 & 0.93 & 0.40 &  $0.81$ & $0.24$  &  O\\[1ex]
3236 & 00 52 36.14 & +43 10 11.38 & 21.74 & 0.09 & 20.85 & 0.04 & 0.29 & 1.72 & 0.20 &  $5.73$ & $0.47$ &  C\\[1ex]
4757 & 00 53 00.79 & +43 10 03.84 & 22.11 & 0.02 & 21.16 & 0.04 & 1.30 & 0.97 & 0.65 &  $4.69$ & $0.32$  &  O\\[1ex]
3919 & 00 53 23.70 & +43 14 25.45 & 21.56 & 0.07 & 20.61 & 0.01 & 0.43 & 1.91 & 0.28 &  $8.57$ & $0.89$  &  C\\[1ex]
4951 & 00 52 39.94 & +43 09 55.47 & 22.15 & 0.05 & 21.20 & 0.05 & 0.41 & 0.93 & 0.51 &  $0.95$ & $0.19$  &  O\\[1ex]
5284 & 00 52 59.36 & +43 12 04.35 & 22.32 & 0.06 & 21.36 & 0.05 & 0.56 & 0.93 & 0.65 &  $0.70$ & $0.19$  &  O\\[1ex]
5073 & 00 53 29.77 & +43 09 25.70 & 22.67 & 0.04 & 21.69 & 0.03 & 1.01 & 0.81 & 0.66 &  $0.28$ & $0.13$  & O\\[1ex]
4770 & 00 52 53.68 & +43 13 41.74 & 22.33 & 0.07 & 21.33 & 0.04 & 1.07 & 0.81 & 0.60 &  $0.23$ & $0.10$ &  O\\[1ex]
4435 & 00 52 54.35 & +43 13 43.06 & 22.33 & 0.06 & 21.33 & 0.03 & 1.20 & 0.95 & 0.65 &  $2.30$ & $0.29$  &  O\\[1ex]
4127 & 00 52 34.93 & +43 11 22.88 & 21.81 & 0.01 & 20.80 & 0.01 & 0.82 & 2.10 & 0.52 &  $5.01$ & $0.48$ &  C\\[1ex]
4250 & 00 53 00.93 & +43 11 21.29 & 22.15 & 0.07 & 21.14 & 0.05 & 0.91 & 1.53 & 0.14 &  $2.68$ & $0.54$  & C\\[1ex]
4512 & 00 52 59.41 & +43 10 35.64 & 22.27 & 0.04 & 21.16 & 0.04 & 1.22 & 0.91 & 0.66 &  $0.60$ & $0.21$  &  O\\[1ex]
5339 & 00 53 01.96 & +43 09 00.93 & 21.56 & 0.06 & 20.42 & 0.01 & 0.62 & 2.20 & 0.29 &  $14.56$ & $0.99$ &  C\\[1ex]
4349 & 00 52 06.19 & +43 13 30.85 & 21.80 & 0.06 & 20.64 & 0.01 & 0.46 & 0.97 & 0.65 &  $3.11$ & $0.30$ &  O\\[1ex]
4573 & 00 52 20.74 & +43 09 21.40 & 21.91 & 0.06 & 20.75 & 0.02 & 0.32 & 1.93 & 0.37 &  $5.72$ & $0.77$  &  C\\[1ex]
3703 & 00 52 36.50 & +43 09 43.58 & 22.03 & 0.07 & 20.86 & 0.01 & 1.02 & 2.00 & 0.34 &  $6.69$ & $0.56$ &  C\\[1ex]
3824 & 00 52 51.91 & +43 12 15.05 & 22.72 & 0.04 & 21.44 & 0.06 & 1.12 & 0.83 & 0.45 &  $0.26$ & $0.15$  &  O\\[1ex]
3010 & 00 53 03.48 & +43 09 08.18 & 21.76 & 0.03 & 20.40 & 0.02 & 0.68 & 1.70 & 0.35 &  $4.85$ & $0.72$  &  C\\[1ex]
5503 & 00 52 14.24 & +43 08 31.38 & 21.85 & 0.01 & 20.50 & 0.01 & 0.40 & 1.90 & 0.55 &  $5.01$ & $1.12$  &  C\\[1ex]
3416 & 00 52 10.98 & +43 13 22.30 & 22.15 & 0.05 & 20.71 & 0.05 & 0.22 & 1.50 & 0.41 &  $2.65$ & $0.35$  &  C\\[1ex]
3587 & 00 52 23.64 & +43 08 45.66 & 22.16 & 0.01 & 20.71 & 0.04 & 1.14 & 2.20 & 0.74 & $16.41$ & $0.52$  & C\\[1ex]
\tableline

\end{tabular}

\end{center}
\end{table}

\begin{table}
\begin{center}

\begin{tabular}{c c c c c c c c c c c c c}

\tableline
\tableline
id & R.A. & Dec & $V$ &  $\delta V$  & $i$ & $\delta i$ & amplitude$_i$ & M$_{Birth}$ & $\tau_i$  &  $\dot{M}$  &  $L$  & chemical type \\

& (J2000) & (J2000) & (mag) & (mag) & (mag) & (mag) & (mag) & (M$\textsubscript{\(\odot\)}$) & & ($10^{-6}$ M$\textsubscript{\(\odot\)}$ yr$^{-1})$ & (10$^{4}$ L$\textsubscript{\(\odot\)})$ & \\
\tableline

5437 & 00 52 54.34 & +43 12 40.17 & 23.00 & 0.08 & 21.46 & 0.07 & 0.90 & 0.91 & 0.98 &  $0.61$ & $0.14$ &  O\\[1ex]
5325 & 00 52 47.61 & +43 11 08.23 & 22.91 & 0.06 & 21.31 & 0.05 & 1.90 & 0.90 & 0.67 &  $0.45$ & $0.16$  &  O\\[1ex]
4388 & 00 52 45.19 & +43 11 49.30 & 22.97 & 0.07 & 21.32 & 0.06 & 1.43 & 1.55 & 0.50 &  $3.02$ & $0.63$  &  C\\[1ex]
4381 & 00 53 01.70 & +43 06 45.22 & 22.16 & 0.05 & 20.42 & 0.01 & 0.75 & 1.70 & 0.50 &  $3.49$ & $0.35$ &  C\\[1ex]
4550 & 00 52 56.15 & +43 12 22.27 & 23.31 & 0.09 & 21.57 & 0.07 & 1.32 & 0.85 & 0.99 &  $0.27$ & $0.13$  &  O\\[1ex]
5546 & 00 53 00.38 & +43 14 10.44 & 23.38 & 0.10 & 21.62 & 0.06 & 0.96 & 0.83 & 0.95 &  $0.26$ & $0.12$  &  O\\[1ex]
4925 & 00 52 55.79 & +43 12 38.16 & 23.30 & 0.07 & 21.49 & 0.06 & 1.72 & 2.80 & 0.70 &  $18.26$ & $0.69$  &  C\\[1ex]
4541 & 00 52 52.66 & +43 15 12.78 & 23.61 & 0.07 & 21.80 & 0.02 & 1.41 & 0.89 & 0.97 &  $0.42$ & $0.19$  &  O\\[1ex]
5848 & 00 53 25.79 & +43 10 20.78 & 22.18 & 0.01 & 20.35 & 0.00 & 1.31 & 2.60 & 0.60 &  $17.04$ & $1.04$  & C\\[1ex]
5030 & 00 52 43.08 & +43 14 32.50 & 23.83 & 0.05 & 21.86 & 0.04 & 1.00 & 2.00 & 1.02 &  $0.21$ & $0.08$  &  C\\[1ex]
4151 & 00 52 48.26 & +43 08 49.87 & 22.98 & 0.01 & 21.00 & 0.01 & 1.23 & 0.83 & 0.67 &  $6.90$ & $0.40$  &  O\\[1ex]
3243 & 00 52 22.35 & +43 06 58.08 & 23.11 & 0.07 & 20.95 & 0.00 & 1.16 & 0.90 & 0.87 &  $2.67$ & $0.30$  &  O\\[1ex]
5493 & 00 53 00.08 & +43 07 44.17 & 22.63 & 0.06 & 20.42 & 0.02 & 2.20 & 2.20 & 0.70 &  $15.67$ & $1.07$  &  C\\[1ex]
4357 & 00 53 09.67 & +43 10 44.04 & 22.61 & 0.05 & 20.36 & 0.00 & 1.41 & 2.60 & 0.70 &  $18.18$ & $1.20$ &  C\\[1ex]
4839 & 00 53 02.26 & +43 14 47.48 & 23.89 & 0.06 & 21.59 & 0.01 & 1.62 & 0.83 & 1.20 &  $0.21$ & $0.09$  &  O\\[1ex]
5671 & 00 52 54.62 & +43 10 01.29 & 23.30 & 0.08 & 20.91 & 0.06 & 1.55 & 1.55 & 0.85 &  $2.39$ & $0.26$  & C\\[1ex]
4433 & 00 53 08.08 & +43 15 01.26 & 22.91 & 0.06 & 20.36 & 0.04 & 1.76 & 1.82 & 0.87 &  $6.16$ & $0.46$  &  C\\[1ex]

\tableline
\end{tabular}

\end{center}
\end{table}

\begin{figure}
\centering
\includegraphics[height=\textheight]{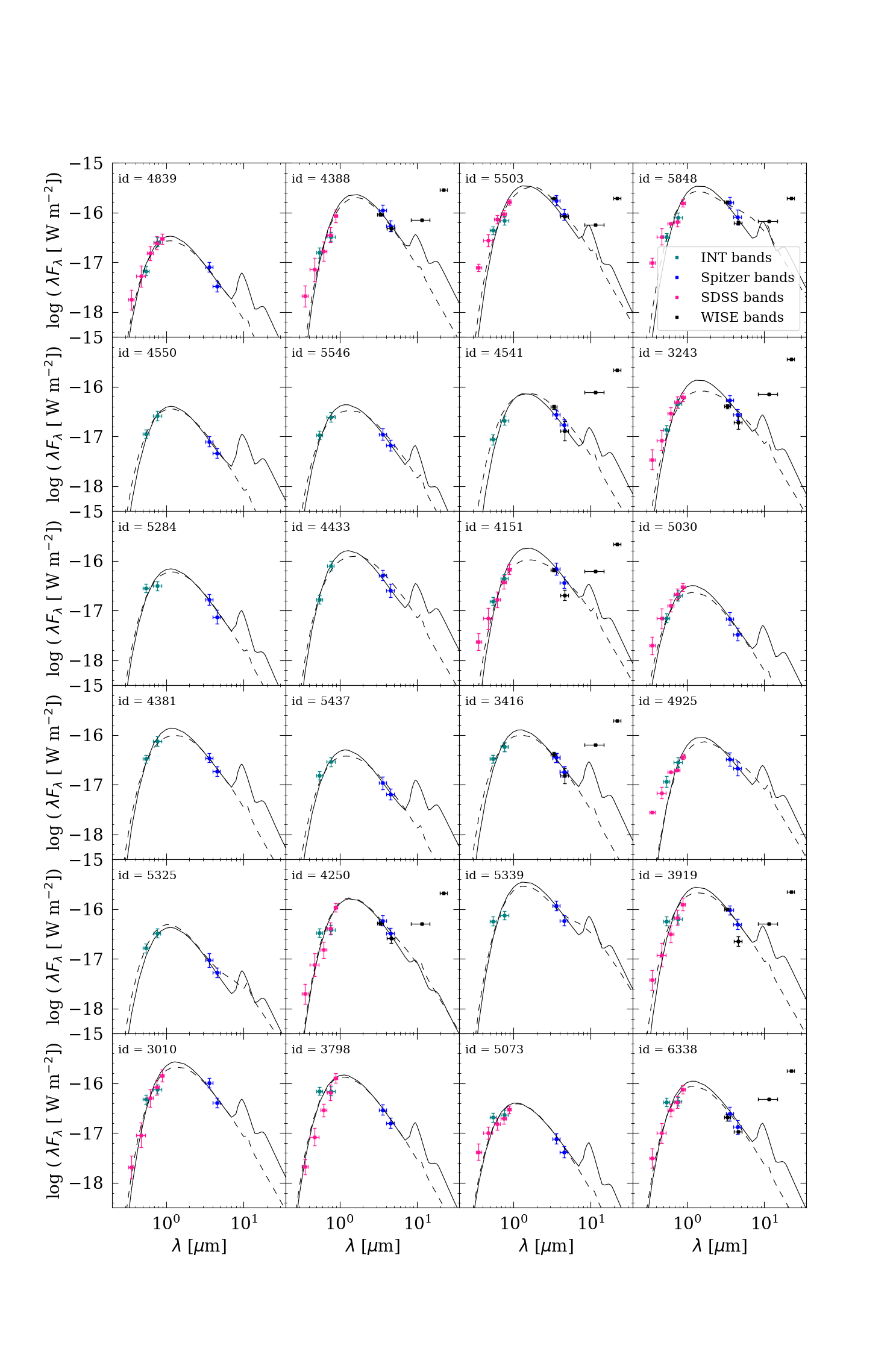}
\end{figure} 
\begin{figure}
\centering
\includegraphics[width=.85\textwidth]{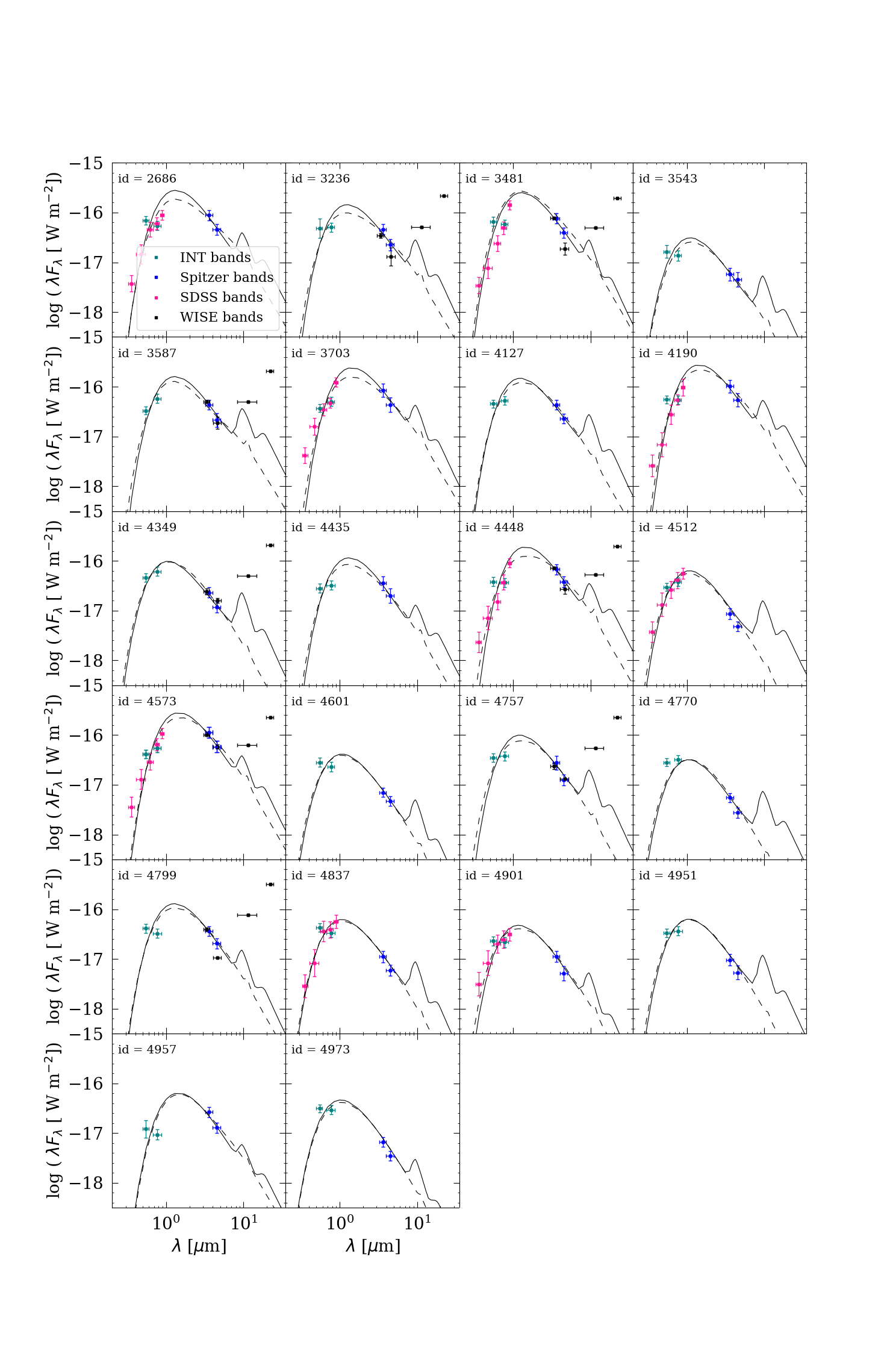}

\caption{SEDs of LPVs with the best fit for the C- and O-rich type of AGBs (dashed and solid black lines). Fluxes are modelled by {\sc dusty} as a function of wavelength. Fluxes observed in different bands with the INT ($i$ and $V$), {\it Spitzer} (3.6 and 4.5 $\mu$m), SDSS ($u$, $g$, $r$, $i$, and $z$), and WISE ($W_1$, $W_2$, $W_3$, and $W_4$) are shown by green, blue, pink, and black squares, respectively. Vertical and horizontal error bars show photometric uncertainty in the magnitude and the difference between the $\lambda_{max}$ and $\lambda_{min}$ around each filter's central wavelength, respectively.}
\end{figure}

\label{lastpage}

\end{document}